\documentclass[fleqn,usenatbib]{mnras}

\usepackage{newtxtext,newtxmath}
\usepackage{ulem}
\usepackage{longtable}
\usepackage{lscape}
\usepackage{hyperref}

\usepackage[T1]{fontenc}

\DeclareRobustCommand{\VAN}[3]{#2}
\let\VANthebibliography\thebibliography
\def\thebibliography{\DeclareRobustCommand{\VAN}[3]{##3}\VANthebibliography}

\usepackage{graphicx}	
\usepackage{amsmath}	

\def\flux{erg s$^{-1}$ cm$^{-2}$}
\def\phflux{ph\,cm$^{-2}$\,s$^{-1}$}

\def\integral{\textit{INTEGRAL}}

\def\logn{Log$N$-Log$S$}

\def\rosat{\textit{ROSAT}}
\def\nustar{\textit{NuSTAR}}
\def\swift{\textit{Swift}}
\def\xmm{\textit{XMM--Newton}}

\title[INTEGRAL 17-yr all-sky survey]{INTEGRAL/IBIS 17-yr hard X-ray all-sky survey}

\author[Krivonos et al.]{
Roman A. Krivonos,$^{1}$\thanks{E-mail: krivonos@cosmos.ru (IKI)}
Sergey Yu. Sazonov,$^{1,2}$ 
Ekaterina A. Kuznetsova,$^{1}$ \newauthor
Alexander A. Lutovinov,$^{1}$  
Ilya A. Mereminskiy,$^{1}$ 
and Sergey S. Tsygankov$^{1,3}$\\
$^{1}$Space Research Institute (IKI), Profsoyuznaya 84/32, Moscow 117997, Russia\\
$^{2}$Moscow Institute of Physics and Technology, Institutsky per. 9, 141700 Dolgoprudny, Russia\\
$^{3}$Department of Physics and Astronomy, FI-20014 University of Turku, Finland
}

\date{Accepted 2021 December 21. Received 2021 December 5; in original form 2021 October 25}

\pubyear{2021}

\begin{document}
\label{firstpage}
\pagerange{\pageref{firstpage}--\pageref{lastpage}}
\maketitle

\begin{abstract}
The International Gamma-Ray Astrophysics Laboratory (\integral), launched in 2002, continues its successful work in observing the sky at energies $E>20$~keV. The legacy of the mission already includes a large number of discovered or previously poorly studied hard X-ray sources. The growing \integral\ archive allows one to conduct an all-sky survey including a number of deep extragalactic fields and the deepest ever hard X-ray survey of the Galaxy. Taking advantage of the data gathered over 17 years with the IBIS coded-mask telescope of \integral, we conducted survey of hard X-ray sources, providing flux information from 17 to 290~keV. The catalog includes 929 objects, 890 of which exceed a detection threshold of $4.5\sigma$ and the rest are detected at $4.0-4.5\sigma$ and belong to known cataloged hard X-ray sources. Among the identified sources of known or suspected nature, 376 are associated with the Galaxy and Magellanic clouds, including 145 low-mass and 115 high-mass X-ray binaries, 79 cataclysmic variables, and 37 of other types; and 440 are extragalactic, including 429 active galactic nuclei (AGNs), 2 ultra-luminous sources, one supernova (AT2018cow) and 8 galaxy clusters. 113 sources remain unclassified. 46 objects are detected in the hard X-ray band for the first time. The \logn\ distribution of 356 non-blazar AGNs is measured down to a flux of $2\times10^{-12}$~\flux\ and can be described by a power law with a slope of $1.44 \pm 0.09$ and normalization $8\times10^{-3}$  deg$^{-2}$ at $10^{-11}$~\flux. The \logn\ distribution of unclassified sources indicates that the majority of them are of extragalactic origin.
\end{abstract}




\section{Introduction}

The \integral\ \citep{2003A&A...411L...1W} observatory has been successfully operating in orbit since its launch in 2002 October. The high sensitivity in the hard X-ray band (${\sim}20-100$~keV) and relatively good angular resolution of the IBIS coded-mask telescope \citep{2003A&A...411L.131U} makes surveying the hard X-ray sky one of the primary goals of \integral. Over the past years, \integral\ has conducted many surveys, starting with sky areas of a few hundred squared degrees and gradually covering the entire celestial sphere \citep[see][for a review]{2021NewAR..9201612K}. 

The \integral\ hard X-ray surveys have been used as a basis for studies of various classes of objects, from cataclysmic variables (CVs) and symbiotic stars \citep[see][for a review]{2020NewAR..9101547L}, through low mass X-ray binaries \citep[LMXB; see][for a review]{2020NewAR..8801536S} and high mass X-ray binaries \citep[HMXB; see][for a review]{2019NewAR..8601546K}, to extragalactic objects, mainly active galactic nuclei \citep[AGNs; see][for a review]{2020NewAR..9001545M}.

Since 2004, the sky has also been surveyed in hard X-rays by the Burst Alert Telescope \citep[BAT;][]{2005SSRv..120..143B} of the {\it Neil Gehrels Swift Observatory}  \citep[\swift;][]{2004ApJ...611.1005G}, which provides a nearly uniform all-sky coverage \citep[see][and references therein]{2018ApJS..235....4O} with somewhat longer exposures at high Galactic latitudes. In contrast to \swift, the \integral\ observatory provides a sky survey with exposures that are deeper near the Galactic plane, which renders the \swift/BAT and \integral/IBIS hard X-ray surveys complementary to each other.

\cite{2007AandA...475..775K} presented the first \integral/IBIS all-sky survey based on 3.5~years of observations at the beginning of the mission. The catalog of sources detected in the 17--60~keV energy band comprises 403 objects, 316 of which were found above a threshold of $5\sigma$ on the time-averaged sky map, and the rest were detected in various subsamples of observations. The most recent data release of the \integral/IBIS catalog with all-sky coverage, presented by \cite{2016ApJS..223...15B}, was based on the first 1000 orbits of \integral, i.e. the data acquired from the launch at the end of 2002 until the end of 2010. This catalog includes 939 sources detected above a significance threshold of $4.5\sigma$ in the $17-100$~keV energy band on sky maps with different exposures.

In this work, we extend the \integral/IBIS all-sky survey to 17 years, using the whole set of observations carried out between 2002 December and 2020 January. We obtain a statistically clean catalog of hard X-ray sources detected above the $4.5\sigma$ threshold along with their classification, if available. This list includes 46 newly detected sources. We also provide a sample of sub-threshold, $4.0-4.5\sigma$, \integral\ sources that were already known from previous \integral\ and \swift\ hard X-ray catalogs. In total, the final catalog comprises 929 hard X-ray sources.

\section{Data analysis}

For the current hard X-ray survey, we utilized all publicly available \integral\ data acquired by the ISGRI low-energy detector layer \citep{2003A&A...411L.141L} of the IBIS telescope between 2002 December and 2020 January (spacecraft revolutions 26--2180). This corresponds to a total nominal exposure time of 406~Ms. We reduced the IBIS/ISGRI data with the \integral\ data analysis software developed at IKI\footnote{Space Research Institute of the Russian Academy of Sciences, Moscow, Russia} \citep[see e.g.,][and references therein]{2010A&A...519A.107K,2012AandA...545A..27K,2014Natur.512..406C}. Below we describe details of the data analysis that are relevant for the current work.

We first applied the latest energy calibration \citep{2013arXiv1304.1349C} for the registered IBIS/ISGRI detector events with the INTEGRAL Offline Scientific Analysis (OSA) provided by the INTEGRAL Science Data Centre (ISDC) Data Centre for Astrophysics up to the COR level. Because the latest OSA version 11.0 is only applicable to the data obtained since 26 December 2015 (revolution 1626), we calibrated ISGRI events obtained within the orbit ranges of  26--1625 and 1626--2180 with OSA versions 10.2 and 11.0, respectively, 

We then produced a sky image of every individual \integral\ observation with a typical exposure time of 2~ks (referred as a {\it Science Window}, or {\it ScW}). The flux scale in each {\it ScW} sky image was adjusted to the flux of the Crab nebula measured in the nearest observation. This procedure was applied to account for the loss of sensitivity at low energies $E\lesssim25$~keV caused by ongoing detector degradation. 

\subsection{Energy bands}


To be consistent with our previous works, we chose 17--60~keV as a working energy band for detection of sources in the present hard X-ray survey. This also allows us to utilize the time period of spacecraft orbits before $\sim1000$ when ISGRI had good sensitivity in the ${\sim}17-30$~keV band. To determine an energy range that is not affected by detector aging, we analyzed the long-term (2003-2019) light curve of the persistent X-ray source Ophiuchus cluster in different energy bands and found that at energies above $30$~keV ISGRI demonstrates stable efficiency over the whole period of observations \citep{2022MNRAS.509.1605K}. We selected 30--80~keV as an additional energy range for studying source light curves. Finally, in order to provide rough broad-band spectral information, we analysed the sky in logarithmically spaced energy bands of 17--26, 26--38, 38--57, 57--86, 86--129, 129--194 and 194--290~keV. Table~\ref{tab:ebands} contains the list of the used energy bands and conversion factors between physical and mCrab units assuming the Crab spectrum in the form of $10(E/{\rm 1\ keV})^{-2.1}$~photons cm$^{-2}$~s$^{-1}$~keV$^{-1}$.

\begin{table}
\noindent
\centering
\caption{Working energy bands for the current \integral\ all-sky survey.}\label{tab:ebands}
\centering
\vspace{1mm}
  \begin{tabular}{|c|c|c|c|c|c|c|}
\hline\hline
Name & Range & Width &  \multicolumn{2}{c}{1~mCrab}  \\
     & [ keV ] & [ keV ] & [ \flux ] & [ \phflux ]  \\
\hline
E1 & $17-60$& 43 & $1.43\times10^{-11}$  & $3.02\times10^{-4}$  \\
E2 & $30-80$& 50 & $1.06\times10^{-11}$ & $1.42\times10^{-4}$  \\
E3 & $17-26$& 9 & $5.02\times10^{-12}$ & $1.50\times10^{-4}$  \\
E4 & $26-38$& 12 & $4.31\times10^{-12}$ & $8.61\times10^{-5}$  \\
E5 & $38-57$& 19 & $4.42\times10^{-12}$  & $5.98\times10^{-5}$  \\
E6 & $57-86$& 29 & $4.31\times10^{-12}$ & $3.87\times10^{-5}$  \\
E7 & $86-129$& 43 & $4.08\times10^{-12}$  & $2.43\times10^{-5}$  \\
E8 & $129-194$& 65 & $3.94\times10^{-12}$  & $1.57\times10^{-5}$  \\
E9 & $194-290$& 96 & $3.73\times10^{-12}$  & $9.89\times10^{-6}$  \\
\hline
\end{tabular}
\vspace{3mm}
\end{table}

\subsection{Sky map mosaics}

After applying selection criteria over the list of reconstructed {\it ScW} sky images, as described in \cite{2007AandA...475..775K}, we obtained $157,200$ {\it ScWs} in each energy band, which comprises 273~Ms of dead-time corrected exposure. Finally, we projected the sky images of all {\it ScWs} on to $25^{\circ}\times25^{\circ}$ sky frames covering the whole sky in the HEALPIX reference grid \citep{2005ApJ...622..759G}, with 192 frames in total.

The procedure of sky reconstruction for IBIS/ISGRI suffers from systematic noise, which is mainly caused by the presence of bright sources in the field of view (FOV) \citep[see e.g.,][]{2007AA...475..775K,2012AandA...545A..27K}. As a result, some sky artefacts are still present around persistent bright sources, such as Crab, Sco~X--1, Cyg~X--1, Cyg~X--3, Vela~X--1, GX~301--2 and GRS 1915+105, which requires manual inspection of the list of source candidates. Apart from the bright persistent sources, sky regions can be seriously contaminated by bright Galactic transients, which, however, can be excluded from the analysis. To produce sky maps that are as clean as possible, we avoided observations with angular offsets closer than $20^{\circ}$ from the following bright transients: IGR~J17480--2446 \citep[1519 -- 1530,][]{2010ATel.2919....1B}, GRS~1716--249 \citep[1780 -- 1806,][]{2019MNRAS.482.1587B}, MAXI J1535--571 \citep[1860 -- 1865,][]{2019ApJ...883..198R,2019ApJ...878L..28P}, Swift~J174510.8--262411 \citep[1212 -- 1264,][]{2016MNRAS.456.3585D}, 
MAXI~J1820+070 \citep[1931 -- 1952,][]{2018ApJ...868...54S} and MAXI~J1348--630 \citep[2050 -- 2061,][]{2019ATel12425....1Y}, where the range in parenthesis means the interval of excluded spacecraft orbits.

\subsection{Detection of sources}
\label{sec:detection}
Sources were searched as excesses on $25^{\circ}\times25^{\circ}$ ISGRI sky frames, convolved with a Gaussian with $\sigma=5'$, which approximates the effective point spread function (PSF) of IBIS/ISGRI. The signal-to-noise (S/N) distribution of pixels is dominated by the statistical noise and can be described by a Gaussian with zero mean and unit variance, as demonstrated in our previous works \citep[e.g.,][]{2007AandA...475..775K}. However, in sky regions that contain very bright point sources or the crowded field near the Galactic center, the root-mean-square (RMS) scatter of the S/N distribution increases \citep[e.g.,][]{2010A&A...519A.107K}, which requires manual inspection of detected excesses, as was already noted above.

We considered two criteria for detecting sources. First, a source is considered to be detected if the significance of its detection exceeds $4.5\sigma$. Given the IBIS/ISGRI angular resolution of $12'$, the all-sky map may be regarded as consisting of $\sim10^{6}$ statistically independent pixels, which implies that less than 10 false detections are expected to be registered due to statistical fluctuations. The second, alternative, criterion, is intended to increase the completeness of the survey with respect to previously known hard X-ray sources that fall slightly below the $4.5\sigma$ detection threshold. Specifically, if the spatial position of an \integral\ source candidate detected with a significance of $4.0-4.5\sigma$ is consistent with that of a known hard X-ray source from an external catalog it is considered to be detected. To this end, we cross-correlated the list of $S/N>4.0\sigma$ source candidates with previous \integral/IBIS and \swift/BAT hard X-ray source catalogs that are listed in Table~\ref{tab:surveys}. The search was done within a radius of $10'$. Table~\ref{tab:surveys} also provides the number of cross-matches with a given catalog. 

As a result, all $4.0<S/N<4.5$ sources included in the final catalog are known hard X-ray sources. Consequently, we mark $S/N>4.5\sigma$ sources that do not have a counterpart in any of the hard X-ray surveys listed in Table~\ref{tab:surveys} as newly detected in hard X-rays, except for a number of known X-ray transients and historical X-ray sources that are not listed in the mentioned hard X-ray catalogs for some reason. There are in total 46 newly detected hard X-ray sources.

\subsection{Survey sensitivity}

\begin{figure}
  \includegraphics[width=0.99\columnwidth]{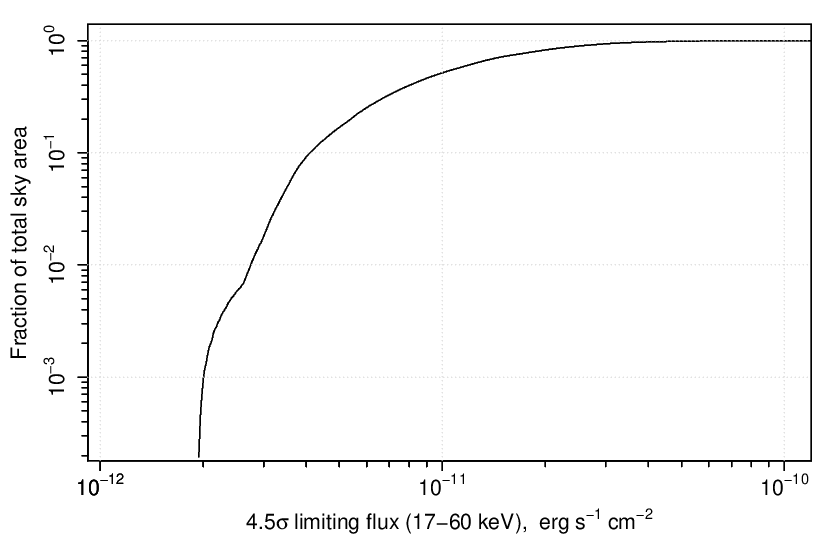}
\caption
   { \label{fig:area} 
Fraction of the total sky area surveyed as a function of the limiting flux for source detection with $4.5\sigma$ significance in the 17--60~keV energy band.}
\end{figure}

Figure~\ref{fig:area} shows the survey sky coverage as a function of the flux corresponding to the $4.5\sigma$ significance limit. The peak sensitivity, reached in the Galactic Center region, is about 0.14~mCrab ($2\times10^{-12}$~\flux) in the $17-60$~keV energy band. Compared to our previous work based on 14 years of data \citep{2017MNRAS.470..512K} the peak sensitivity in this band has improved by nearly $10\%$. The present survey covers 10\% of the sky down to a flux limit of ${\sim}0.3$~mCrab ($4.3\times10^{-12}$~\flux) and 90\% of the sky down to ${\sim}1.8$~mCrab ($2.6\times10^{-11}$~\flux) in the $17-60$~keV energy band. A limiting ($4.5\sigma$) flux of 1~mCrab or better is achieved for ${\sim}70\%$ of the sky.

\section{Catalog of sources}

The catalog of detected sources was compiled in the reference $17-60$~keV energy band following the detection criteria described in Section~\ref{sec:detection}. In addition, Table~\ref{tab:ebands:stats} provides source counts statistics for all energy bands considered in this work. The list of sources detected in the $17-60$~keV band is presented in Table~A1 in Appendix~\ref{sec:catalog}, and its content is described below.

{\it Column (1) ``Id''} -- sequence number of the source in the current catalog.

{\it Column (2) ``Name''} -- name of the source. Common names are given for sources which were known before their detection with \integral. Sources discovered by \integral\ or those whose nature was established thanks to the \integral\ observations are named as ``IGR''.

{\it Columns (3,4) ``RA, Dec''} -- source equatorial (J2000) coordinates. The positional accuracy depends on the significance of detection by IBIS/ISGRI \citep{ 2003A&A...411L.179G,2007ApJS..170..175B}. According to \cite{2007AandA...475..775K}, the estimated 68\% confidence intervals for sources detected at $5-6$, 10, and $>20\sigma$ are $2.1'$, $1.5'$, and $<0.8'$, respectively.

{\it Column (5) ``Flux''} -- time-averaged flux of the source in the $17-60$~keV energy band. The online version of the table also contains source fluxes in all energy bands listed in Table~\ref{tab:ebands}.         

{\it Column (6) ``Type''} -- general astrophysical type of the object. LMXB (HMXB) -- low- (high-) mass X-ray binary; X-RAY BINARY -- X-ray binary of uncertain type; CV -- cataclysmic variable or symbiotic binary; SNR -- supernova remnant; SNR/Pulsar -- supernova remnant with a central pulsar (when both may contribute to the hard X-ray emission); MAGNETAR -- magnetar (anomalous X-ray pulsars and soft gamma-ray repeaters); SUPERNOVA -- supernova; STAR -- active star (of various types, excluding the previously listed types of stellar objects); ULX -- ultraluminous X-ray source; SEYFERT -- AGN of the Seyfert or LINER type; BLAZAR -- beamed AGN (BL Lac objects and flat-spectrum radio quasars); AGN -- unclassified AGN {(i.e., the object is known to be an AGN in a general sense but detailed optical classification is missing)}; CLUSTER -- a cluster of galaxies; and UNIDENT -- an unclassified object. A question mark indicates that the specified type is not firmly established. {These classifications are mainly based on the information contained in the Simbad\footnote{Simbad Astronomical Database \url{http://simbad.u-strasbg.fr}} and NED\footnote{NASA/IPAC Extragalactic Database \url{http://ned.ipac.caltech.edu}} databases as well as in the previous versions of the \integral/IBIS and \swift/BAT hard X-ray source catalogs (see Table~\ref{tab:surveys}). In addition, we strove to select the most recent and reliable source identifications from the literature if necessary.}

{\it Column (7) ``Notes''} -- additional notes such as references, redshift information, alternative source names, spatial confusion and transient flags. The references are mainly provided for recently discovered sources and are related to determination of their nature. The redshifts of the extragalactic sources were adopted from SIMBAD, NED and the {\it Swift}/BAT AGN Spectroscopic Survey (BASS) \citep[see][and references therein]{2017ApJS..233...17R}.

\begin{table*}
\noindent
\centering
\caption{List of selected hard X-ray catalogs used for cross-matching.}\label{tab:surveys}
\centering
\vspace{1mm}
\begin{tabular}{ccclr}
\hline\hline
X-ray survey & Band (keV) & Sources & Cross-matches$^{\rm a}$ & Reference \\
\hline
\multicolumn{5}{c}{\swift/BAT}\\
\hline
all-sky, 70 months & 14--195 & 1210 & 617 (+20)  &\citet{2013ApJS..207...19B}  \\
all-sky, 105 months & 14--195 & 1632 & 693 (+18) & \citet{2018ApJS..235....4O}  \\
\hline
\multicolumn{5}{c}{\integral/IBIS}\\
\hline
1st catalog & 20--100 & 123 & 122 (+6)  & \citet{2004ApJ...607L..33B}  \\
${\sim}50$\% of sky, 2nd cat. & 20--100 & 209 & 197 (+8)  & \citet{2006ApJ...636..765B}  \\
${\sim}70$\% of sky, 3rd cat. & 17--100 & 421 & 358 (+12) & \citet{2007ApJS..170..175B}  \\
all-sky, 4th catalog & 17--100 & 723 & 485 (+12)  & \citet{2010ApJS..186....1B}  \\
all-sky, 1000 orbits & 17--100 & 939 &  629 (+14) & \citet{2016ApJS..223...15B}  \\
all-sky 3.5 years & 17--60 & 403 & 369 (+13) & \citet{2007AandA...475..775K}  \\
all-sky, 7 years & 17--60 & 521 & 486 (+15)  & \citet{2010AandA...523A..61K}  \\
$|b|<17.5^\circ$, 9 years & 17--80 & 402 & 387 (+13)  & \citet{2012AandA...545A..27K}  \\
$|b|<17.5^\circ$, 14 years & 17--60 & 72 & 70 (+2)  & \citet{2017MNRAS.470..512K}  \\
deep extragal. fields & 17--60 & 147 & 140 (+4)  & \citet{2016MNRAS.459..140M} \\
\hline

\end{tabular}\\
\begin{flushleft}
$^{\rm a}$ {The result of cross-correlation is regarded as positive if at least one source from the \integral/IBIS 17-year catalog is found within $10'$ from a source in the external catalog. The number of cases where there is more than one \integral\ source within $10'$ from the same source in the external catalog is given in brackets.} 
\end{flushleft}
\vspace{3mm}
\end{table*}

\begin{table}
\noindent
\centering
\caption{Source detection statistics in different energy bands (see Table~\ref{tab:ebands} for reference) of the current \integral/IBIS 17-year all-sky survey.}\label{tab:ebands:stats}
\centering
\vspace{1mm}
  \begin{tabular}{|c|c|c|c|c|c|c|}
\hline\hline
Band & S/N > 4.5 & 4.0 > S/N > 4.5$^{\rm a}$ &  Total  \\
\hline
E1 & 890 & 39 & 929  \\
E2 & 693 & 40 & 733  \\
E3 & 554 & 45 & 599  \\
E4 & 700 & 55 & 755 \\
E5 & 597 & 48 & 645  \\
E6 & 314 & 33 & 347  \\
E7 & 208 & 29 & 237  \\
E8 & 63 & 9  & 72  \\
E9 & 17 & 5  & 22  \\
\hline
\end{tabular}\\
\begin{flushleft}
$^{\rm a}$ Such sources are required to have been previously detected in any of the all-sky hard X-ray surveys listed in Table~\ref{tab:surveys}.
\end{flushleft}
\vspace{3mm}
\end{table}

\section{Discussion and summary}

In this work we present the all-sky survey of hard X-ray sources in the reference $17-60$~keV energy band, based on the data archive of the IBIS coded-mask telescope on board the \integral\ observatory accumulated over 17 years, from 2002 December to 2020 January. The combined catalog of sources includes 929 objects, 890 of which exceed a detection threshold of $4.5\sigma$ and the rest are known hard X-ray sources detected at $4.0\sigma < S/N <4.5\sigma$.

\subsection{Properties of the catalog}

Table~\ref{tab:types} presents source statistics by astrophysical type and Table~\ref{tab:source_category} lists source counts over three categories: Galactic objects, sources located in the Large and Small Magellanic Clouds, and extragalactic objects. 113 sources from the catalog remain unclassified. 

\begin{table}
\noindent
\centering
\caption{Statistics of sources in the catalog.}\label{tab:types}
\centering
\vspace{1mm}
\begin{tabular}{clr}
\hline\hline
Type & Count (Notes) \\
\hline
LMXB &  145 \\
HMXB &  115 \\
X-ray binary (unclassified) &  1 (SWIFT J1858.6-0814)\\
CV & 79 \\
Star  & 4  \\
Magnetar & 5 \\
SNR, SNR/Pulsar & 25 \\
Molecular cloud & 1 (Sgr B2) \\
Galactic Center & 1 (Sgr A*) \\
Supernova &  1 (AT2018cow) \\
ULX & 2 \\
Seyfert galaxy & 336 \\
AGN (unclassified) & 39 \\
Blazar & 54 \\
Galaxy cluster & 8 \\
Unidentified & 113 \\
\hline
\end{tabular}
\end{table}

\begin{table}
\caption{Statistics of identified sources by category.}
\label{tab:source_category}
\centering

\begin{tabular}{cl}
\hline
Category & Count \\
\hline
Galactic & 356 \\
Magellanic Clouds & 20 \\
Extragalactic & 440 \\
\hline
\end{tabular}
\end{table}


We mark 46 sources detected at $S/N>4.5\sigma$ as newly detected in hard X-rays since they do not have a counterpart in any of the selected hard X-ray catalogs (Table~\ref{tab:surveys}). We cross-correlated these sources with a number of catalogs in the soft X-ray band, namely the \rosat\ all-sky bright source catalog \citep[1RXS;][]{voges99}, the \xmm\ slew survey  \citep[XMMSL2;][]{2008A&A...480..611S}, the \xmm\ serendipitous survey \citep[][]{2009A&A...493..339W,2020A&A...641A.136W}, the \swift/XRT point-source catalog \citep[2SXPS;][]{2020ApJS..247...54E}, and the {\it Chandra} source catalog 2.0 \citep{2010ApJS..189...37E}. Specifically, we searched for soft X-ray counterparts within $5'$ of the positions of the \integral\ sources and only considered sources with a significant detection above 2 keV and the corresponding X-ray flux $F_{X}\gtrsim10^{-13}$ \flux. We then searched for a positionally coincident optical or radio source using the VizieR \citep{vizier} service, in order to determine the source class. As a result, we have found soft X-ray counterparts for 8 out of the 46 new \integral\, sources and classified some of these objects (see Table~\ref{tab:softxray}). Most of them are likely of extragalactic nature.

\begin{table*}
\noindent
\centering
\caption{Soft X-ray ($E>2$~keV) counterparts of \integral\ sources discovered in the hard X-ray ($17-60$~keV) band.}\label{tab:softxray}
\centering
\vspace{1mm}
  \begin{tabular}{|c|l|c|c|l|c|c|}
\hline\hline
Name &  X-ray & Offset & Type & Notes  \\
\hline
IGR J04085-6546 & 2RXS J040840.0-654545, XMMSL2 J040839.1-654600 & $0.3'$ & AGN & LEDA 310383, z=0.125 \citep{2013ATel.5537....1S}\\ 
IGR J07328-4640 & 2RXS J073245.8-464006, 4XMM J073244.3-464017   & $1.4'$ & BLAZAR? &  PKS 0731-465\\
IGR J10595-5125 & 2RXS J105920.3-512644, 2SXPS J105918.9-512632  & $2.3'$ & SEYFERT & ESO 215-14, z=0.019\\
IGR J16005-4645 & 2RXS J160019.6-464802, 2SXPS J160020.4-464841 & $3.6'$ & AGN?  & \cite{2012ApJ...751...52E} \\
IGR J17227+3411 & 4XMM J172230.8+341339, 2SXPS J172230.9+341341 & $3.5'$ & AGN? & z=0.425 \citep{2008AandA...477..735C} \\
IGR J17342-4049 & XMMSL2 J173425.6-405121 & $2.2'$ & & \\
IGR J17449-3037 & 4XMM J174507.9-303905 & $3.4'$ & &  \\
IGR J18006-3426 & 2SXPS J180050.6-342322 &$4.1'$ & & \\
\hline
\end{tabular}\\
\begin{flushleft}
\end{flushleft}
\vspace{3mm}
\end{table*}

The identification completeness of the survey, i.e. the fraction of identified sources, is ($N_{\rm Tot} - N_{\rm NotID})/N_{\rm Tot} = 1-113/929 \sim 0.88$. Interestingly, it is the same for both the Galactic ($|b|<5^{\circ}$) and extragalactic ($|b|>5^{\circ}$) parts of the sky. As expected, due to the continuously increasing depth of the \integral\ all-sky survey, the fraction of identified sources sources has decreased compared to our previous compilation of the all-sky catalog, where it was 93\% and 96\% at $|b|<5^{\circ}$ and $|b|>5^{\circ}$, respectively \citep{2010AandA...523A..61K}. 

{Most of the \integral\ sources in the current survey have X-ray counterparts in the \swift/BAT 70- and 105-month catalogs (\citealt{2013ApJS..207...19B,2018ApJS..235....4O}, respectively). Taking the different sky coverage by \integral/IBIS and \swift/BAT into account, it is interesting to investigate the properties of the \integral\ sources that are not present in the \swift/BAT catalogs. According to Table~\ref{tab:surveys}, 711 \integral\ sources have counterparts in the \swift/BAT 105-month survey and 218 (929-711) sources are missed. The majority of the unmatched \integral\ sources (165 out of 218) prove to be located near the Galactic plane at $|b|<17.5^{\circ}$, which is the characteristic latitude span of the \integral\ Galactic surveys \citep{2012AandA...545A..27K}. This indicates that the \integral/IBIS survey has higher sensitivity near the Galactic plane compared to the \swift/BAT surveys. Similar evidence has been demonstrated by \cite{2016ApJS..223...15B} in comparing their \integral\ catalog with the \swift/BAT 70-month survey by \cite{2013ApJS..207...19B}. Among the 53 \integral\ sources at $|b|>17.5^{\circ}$ that are absent in the \swift/BAT 105-month catalog there are 5~HMXBs, 11~non-blazar AGN, 4~blazars, 32 unidentified objects, and the supernova AT2018cow; 17 are newly discovered sources in hard X-ray domain (see Sect.~\ref{sec:detection}).}

{\cite{2016ApJS..223...15B} (hereafter B16) presented a hard X-ray source catalog based on \integral/IBIS observations performed in the first 1000 orbits of \integral\ (up to the end of 2010). This catalog includes 939 sources detected in the 17--100~keV band above a $4.5\sigma$ significance threshold. The final catalog was constructed using the ``burstcity'' method, based on finding the time window wherein the significance of detection of a given source is highest. As a result, 342 sources (out of 939) are transients found on different timescales, labeled with a variability flag in the B16 catalog. We have cross-correlated our source list with the B16 catalog and found 643 matches within a $10'$ radius. Among them, 140 sources are marked with a `Y' flag by B16, which implies moderate variability, and 12 are strongly variable sources, labeled as `YY'. The remaining 491 sources are not labeled as variable in the B16 catalog. Therefore, the majority (76\%) of our 643 sources that have matches in the B16 catalog are presumed to be persistent ones, whereas the ${\sim}300$ B16 sources that are not confirmed by the present survey are mainly transients detected on \integral/IBIS sky maps of various timescales and not present on the all-time maps (see B16 for details). However, ${\sim}100$ persistent B16 sources have not been found in our catalog either. These sources are characterized by sub-mCrab fluxes and have a median detection significance ${\sim}6\sigma$, i.e. they were found close to the detection threshold in the B16 survey. Possible explanations for their absence in the present catalog may lie in the different data analysis methods and slightly different energy bands, substantial source variability, and some of the weak B16 sources being false detections. Because B16 is the most recent previous \integral/IBIS all-sky survey, we added a special flag for all 286 new sources in our catalog that were not detected by B16.}

\subsection{Extragalactic LogN--LogS}
\label{sec:lognlogs}

\begin{figure}
  \includegraphics[width=0.99\columnwidth]{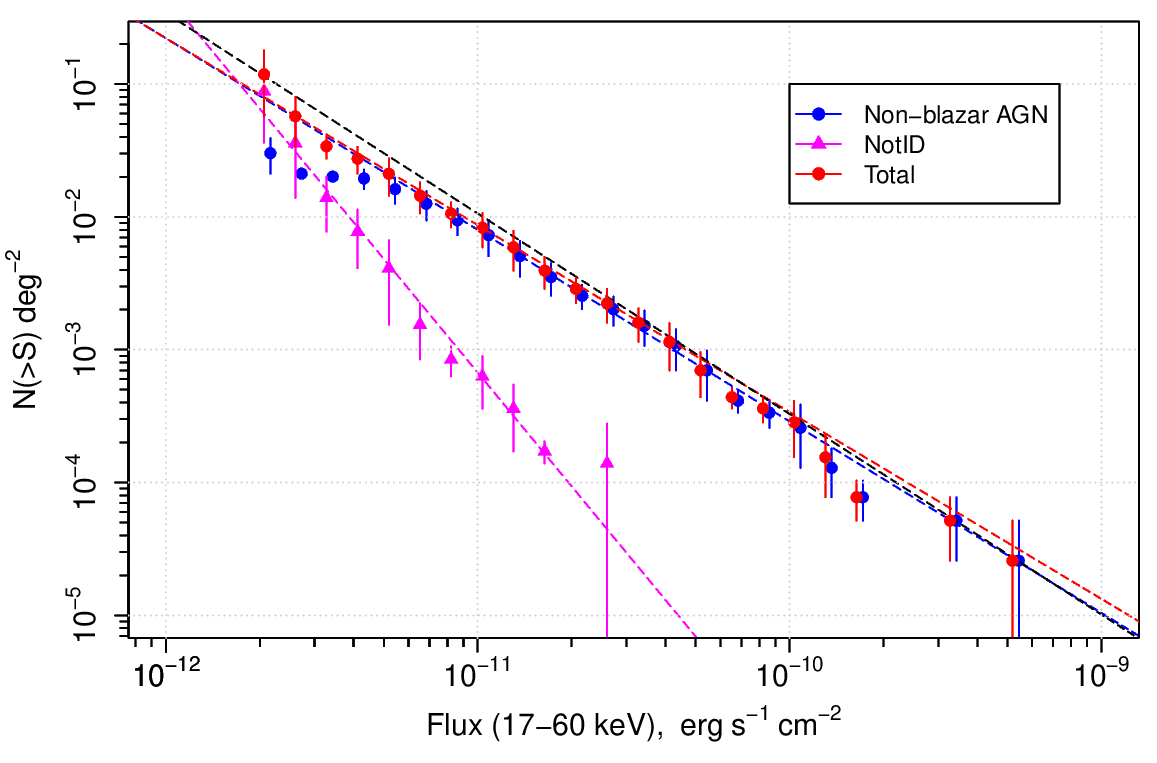}
\caption
   { \label{fig:lognlogs} 
Binned number-flux relation for the 356 non-blazar AGNs (Seyferts and unclassified AGNs, blue points), 104 unidentified sources (magenta triangles), and 460 objects in total (red points) detected above the $4.5\sigma$ detection level. The binning of the blue and red points is slightly shifted for better visibility. The best-fitting power laws for the non-blazar AGNs and unidentified sources are shown by the dashed lines in the corresponding colors. {For comparison, the dashed black line shows the AGN number-flux relation from  \citet{2019AandA...625A.131A} based on the \swift/BAT 105-months source catalog \citep{2018ApJS..235....4O}.}}
\end{figure}

Assuming that AGNs are uniformly distributed over the sky, we can construct their number--flux function in the hard X-ray band. As \integral\ observations cover the sky inhomogeneously, we must take the sensitivity map into account. To this end, we divided the observed source counts by the sky coverage at the $4.5\sigma$ level as a function of flux (Fig.~\ref{fig:area}). Figure~\ref{fig:lognlogs} shows the resulting cumulative \logn\ distribution based on the 356 non-blazar AGNs detected at $S/N>4.5\sigma$ over the whole sky. AGN \logn\ distributions are usually approximated by a power law $N(>S) = AS^{-\alpha}$. Using a maximum-likelihood estimator \citep[][]{1967Natur.216..877J,1970ApJ...162..405C}, we determined the best-fit value of the slope $\alpha = 1.44 \pm 0.09$ using the source counts at fluxes above $10^{-11}$~\flux, where our source sample is expected to be highly complete. We fixed the normalization of the power law at the observed value of the \logn\ distribution at $S = 10^{-11}$~\flux, $A=8\times10^{-3}$~deg$^{-2}$. The inferred \logn\ slope $\alpha$ is consistent with a homogeneous distribution of sources in space ($\alpha = 3/2$).

At fluxes below ${\sim}5\times10^{-12}$~\flux, the measured \logn\ demonstrates a significant deficit of observed source counts relative to the aforementioned power-law dependence. This occurs far above both the known flattening of AGN number counts (in the 2--10~keV energy band) at $\sim 10^{-14}$~\flux\ \citep{2008MNRAS.388.1205G} and the upturn at fluxes below ${\sim}2\times10^{-13}$~\flux\ observed between the \nustar\ and \swift/BAT number-flux relations \citep[][see also below]{2012ApJ...749...21A,2016ApJ...831..185H,2019AandA...625A.131A}. Thus, no deviations from the canonical $\alpha = 3/2$ are expected at fluxes $\lesssim 10^{-11}$\flux, and the observed slope flattening is likely caused by incompleteness of the source sample. To check this, we constructed an all-sky \logn\ relation for the 104 unclassified sources at $S/N>4.5\sigma$, which is shown in Fig.~\ref{fig:lognlogs}. This \logn\ is well approximated with a power law and characterized by a steep $\alpha=2.8\pm0.2$ and normalization $A=4.9\times10^{-3}$~deg$^{-2}$ at $S=5\times10^{-12}$\flux. 
{It turns out that the combined \logn\ distribution of unidentified sources and non-blazar AGNs follows a power law with $\alpha  {\sim}3/2$ ($\alpha = 1.40 \pm 0.05$;  $A=8.7\times10^{-3}$~deg$^{-2}$ at $S = 10^{-11}$~\flux; calculated over the whole range of fluxes)}, as demonstrated in Fig.~\ref{fig:lognlogs}. This suggests that the majority of the unidentified sources are of extragalactic origin.

{Over the past two decades, number counts of extragalactic objects in hard X-ray domain have been independently measured with \integral/IBIS \citep[e.g.,][]{2005ApJ...625...89K,2007AandA...475..775K,2015MNRAS.448.3766K,2006ApJ...652..126B,2006ApJ...649L...9B} and \swift/BAT \citep[e.g.,][]{2008ApJ...681..113T,2012ApJ...749...21A}. \cite{2010AA...523A..61K} compared the AGN \logn\ distribution determined with \integral/IBIS in the 17--60~keV band with the one derived in the 15--55~keV band from the \swift/BAT AGN sample by \cite{2009ApJ...699..603A} and found an excellent agreement \citep[see also][]{2010A&A...510A..48C}. On the other hand, \cite{2016ApJ...831..185H} presented \nustar\ 8--24~keV AGN number counts at lower fluxes that lay significantly above a simple extrapolation with a Euclidean slope of the \swift/BAT counts. However, \cite{2019AandA...625A.131A} demonstrated that the bright part of the \nustar\ AGN number counts was in agreement with the \swift/BAT counts.}

{In light of the updated \integral\ AGN number counts, we compare them in Fig.~\ref{fig:lognlogs} with the \swift/BAT 105-month AGN \logn\ modelled by \cite{2019AandA...625A.131A}. To this end, we converted the 14--195~keV \swift/BAT number counts to the 17--60~keV band following \cite{2010AA...523A..61K}. As seen from Fig.~\ref{fig:lognlogs}, the \swift/BAT \logn\ has a slope \citep[$\alpha = 1.51\pm0.10$,][]{2019AandA...625A.131A}, which is better consistent with the canonical $\alpha = 3/2$ value than the slope obtained in the present work but is consistent with the latter as well. The small difference between \swift/BAT and \integral/IBIS may be partially caused by the fact that \cite{2019AandA...625A.131A} included beamed AGNs in their counts while we tried to select only non-blazar AGN (although some blazars may be present among our unclassified AGN and unidentified sources). Nevertheless, our \logn\ tends to be even less compatible with the \nustar\ number counts than \swift/BAT.}

{Finally,} we plan to maintain an extended online version of the \integral/IBIS 17-year hard X-ray source catalog at \url{http://integral.cosmos.ru} where additional information will be presented. Specifically, we plan to provide {soft X-ray and optical coordinates (where available),} spectral information, and multi-year light curves of the detected sources, as well as sky images in different energy bands up to 290~keV.


\section*{Acknowledgements}

{We thank the anonymous referee for many suggestions that helped us to improve the paper.} This work is based on observations with \integral, an ESA project with instruments and the science data centre funded by ESA member states (especially the PI countries: Denmark, France, Germany, Italy, Switzerland, Spain), and Poland, and with the participation of Russia and the USA. This research has made use of: data obtained from the High Energy Astrophysics Science Archive Research Center (HEASARC) provided by NASA’s Goddard Space Flight Center; the SIMBAD database operated at CDS, Strasbourg, France; the NASA/IPAC Extragalactic Database (NED) operated by the Jet Propulsion Laboratory, California Institute of Technology, under contract with the National Aeronautics and Space Administration. The data were obtained from the European\footnote{\url{http://isdc.unige.ch}} and Russian\footnote{\url{http://hea.iki.rssi.ru/rsdc}} INTEGRAL Science Data Centers. The authors are grateful to E.~M.~Churazov, who developed the INTEGRAL/IBIS data analysis methods and provided the software, and thank the Max Planck Institute for Astrophysics for computational support. This work was financially supported by grant 19-12-00396 from the Russian Science Foundation. 

\section*{Data Availability}

This work is based on publicly available data of the \integral\ observatory. The ISDC Data Centre for Astrophysics (\url{http://isdc.unige.ch/}) provides the scientific archive of the \integral\ data to the community.



\bibliographystyle{mnras}
\bibliography{main} 

\begin{thebibliography}{}
\makeatletter
\relax
\def\mn@urlcharsother{\let\do\@makeother \do\$\do\&\do\#\do\^\do\_\do\%\do\~}
\def\mn@doi{\begingroup\mn@urlcharsother \@ifnextchar [ {\mn@doi@}
  {\mn@doi@[]}}
\def\mn@doi@[#1]#2{\def\@tempa{#1}\ifx\@tempa\@empty \href
  {http://dx.doi.org/#2} {doi:#2}\else \href {http://dx.doi.org/#2} {#1}\fi
  \endgroup}
\def\mn@eprint#1#2{\mn@eprint@#1:#2::\@nil}
\def\mn@eprint@arXiv#1{\href {http://arxiv.org/abs/#1} {{\tt arXiv:#1}}}
\def\mn@eprint@dblp#1{\href {http://dblp.uni-trier.de/rec/bibtex/#1.xml}
  {dblp:#1}}
\def\mn@eprint@#1:#2:#3:#4\@nil{\def\@tempa {#1}\def\@tempb {#2}\def\@tempc
  {#3}\ifx \@tempc \@empty \let \@tempc \@tempb \let \@tempb \@tempa \fi \ifx
  \@tempb \@empty \def\@tempb {arXiv}\fi \@ifundefined
  {mn@eprint@\@tempb}{\@tempb:\@tempc}{\expandafter \expandafter \csname
  mn@eprint@\@tempb\endcsname \expandafter{\@tempc}}}

\bibitem[\protect\citeauthoryear{{Acero} et~al.,}{{Acero}
  et~al.}{2015}]{2015ApJS..218...23A}
{Acero} F.,  et~al., 2015, \mn@doi [\apjs] {10.1088/0067-0049/218/2/23}, \href
  {https://ui.adsabs.harvard.edu/abs/2015ApJS..218...23A} {218, 23}

\bibitem[\protect\citeauthoryear{{Aharonian} et~al.,}{{Aharonian}
  et~al.}{2005}]{2005A&A...442....1A}
{Aharonian} F.,  et~al., 2005, \mn@doi [\aap] {10.1051/0004-6361:20052983},
  \href {https://ui.adsabs.harvard.edu/abs/2005A&A...442....1A} {442, 1}

\bibitem[\protect\citeauthoryear{{Ajello} et~al.,}{{Ajello}
  et~al.}{2009}]{2009ApJ...699..603A}
{Ajello} M.,  et~al., 2009, \mn@doi [\apj] {10.1088/0004-637X/699/1/603}, \href
  {https://ui.adsabs.harvard.edu/abs/2009ApJ...699..603A} {699, 603}

\bibitem[\protect\citeauthoryear{{Ajello}, {Alexander}, {Greiner}, {Madejski},
  {Gehrels}  \& {Burlon}}{{Ajello} et~al.}{2012}]{2012ApJ...749...21A}
{Ajello} M.,  {Alexander} D.~M.,  {Greiner} J.,  {Madejski} G.~M.,  {Gehrels}
  N.,   {Burlon} D.,  2012, \mn@doi [\apj] {10.1088/0004-637X/749/1/21}, \href
  {https://ui.adsabs.harvard.edu/abs/2012ApJ...749...21A} {749, 21}

\bibitem[\protect\citeauthoryear{{Akylas} \& {Georgantopoulos}}{{Akylas} \&
  {Georgantopoulos}}{2019}]{2019AandA...625A.131A}
{Akylas} A.,  {Georgantopoulos} I.,  2019, \mn@doi [\aap]
  {10.1051/0004-6361/201834234}, \href
  {https://ui.adsabs.harvard.edu/abs/2019A&A...625A.131A} {625, A131}

\bibitem[\protect\citeauthoryear{{Altamirano} et~al.,}{{Altamirano}
  et~al.}{2008}]{2008ATel.1651....1A}
{Altamirano} D.,  et~al., 2008, The Astronomer's Telegram, \href
  {https://ui.adsabs.harvard.edu/abs/2008ATel.1651....1A} {1651, 1}

\bibitem[\protect\citeauthoryear{{Apparao}, {Bignami}, {Maraschi}, {Helmken},
  {Margon}, {Hjellming}, {Bradt}  \& {Dower}}{{Apparao}
  et~al.}{1978}]{1978Natur.273..450A}
{Apparao} K.~M.~V.,  {Bignami} G.~F.,  {Maraschi} L.,  {Helmken} H.,  {Margon}
  B.,  {Hjellming} R.,  {Bradt} H.~V.,   {Dower} R.~G.,  1978, \mn@doi [\nat]
  {10.1038/273450a0}, \href
  {https://ui.adsabs.harvard.edu/abs/1978Natur.273..450A} {273, 450}

\bibitem[\protect\citeauthoryear{{Armas Padilla}, {Mu{\~n}oz-Darias},
  {S{\'a}nchez-Sierras}, {De Marco}, {Jim{\'e}nez-Ibarra}, {Casares},
  {Corral-Santana}  \& {Torres}}{{Armas Padilla}
  et~al.}{2019}]{2019MNRAS.485.5235A}
{Armas Padilla} M.,  {Mu{\~n}oz-Darias} T.,  {S{\'a}nchez-Sierras} J.,  {De
  Marco} B.,  {Jim{\'e}nez-Ibarra} F.,  {Casares} J.,  {Corral-Santana} J.~M.,
   {Torres} M.~A.~P.,  2019, \mn@doi [\mnras] {10.1093/mnras/stz737}, \href
  {https://ui.adsabs.harvard.edu/abs/2019MNRAS.485.5235A} {485, 5235}

\bibitem[\protect\citeauthoryear{{Bahramian}, {Gladstone}, {Heinke},
  {Wijnands}, {Kaur}  \& {Altamirano}}{{Bahramian}
  et~al.}{2014}]{2014MNRAS.441..640B}
{Bahramian} A.,  {Gladstone} J.~C.,  {Heinke} C.~O.,  {Wijnands} R.,  {Kaur}
  R.,   {Altamirano} D.,  2014, \mn@doi [\mnras] {10.1093/mnras/stu611}, \href
  {https://ui.adsabs.harvard.edu/abs/2014MNRAS.441..640B} {441, 640}

\bibitem[\protect\citeauthoryear{{Barlow} et~al.,}{{Barlow}
  et~al.}{2005}]{2005A&A...437L..27B}
{Barlow} E.~J.,  et~al., 2005, \mn@doi [\aap] {10.1051/0004-6361:200500131},
  \href {https://ui.adsabs.harvard.edu/abs/2005A&A...437L..27B} {437, L27}

\bibitem[\protect\citeauthoryear{{Barlow}, {Knigge}, {Bird}, {J Dean}, {Clark},
  {Hill}, {Molina}  \& {Sguera}}{{Barlow} et~al.}{2006}]{2006MNRAS.372..224B}
{Barlow} E.~J.,  {Knigge} C.,  {Bird} A.~J.,  {J Dean} A.,  {Clark} D.~J.,
  {Hill} A.~B.,  {Molina} M.,   {Sguera} V.,  2006, \mn@doi [\mnras]
  {10.1111/j.1365-2966.2006.10836.x}, \href
  {https://ui.adsabs.harvard.edu/abs/2006MNRAS.372..224B} {372, 224}

\bibitem[\protect\citeauthoryear{{Barthelmy} et~al.,}{{Barthelmy}
  et~al.}{2005}]{2005SSRv..120..143B}
{Barthelmy} S.~D.,  et~al., 2005, \mn@doi [\ssr] {10.1007/s11214-005-5096-3},
  \href {https://ui.adsabs.harvard.edu/abs/2005SSRv..120..143B} {120, 143}

\bibitem[\protect\citeauthoryear{{Barthelmy} et~al.,}{{Barthelmy}
  et~al.}{2019}]{2019ATel12436....1B}
{Barthelmy} S.~D.,  et~al., 2019, The Astronomer's Telegram, \href
  {https://ui.adsabs.harvard.edu/abs/2019ATel12436....1B} {12436, 1}

\bibitem[\protect\citeauthoryear{{Bassani} et~al.,}{{Bassani}
  et~al.}{2004}]{2004ATel..232....1B}
{Bassani} L.,  et~al., 2004, The Astronomer's Telegram, \href
  {https://ui.adsabs.harvard.edu/abs/2004ATel..232....1B} {232, 1}

\bibitem[\protect\citeauthoryear{{Bassani} et~al.,}{{Bassani}
  et~al.}{2005}]{2005ApJ...634L..21B}
{Bassani} L.,  et~al., 2005, \mn@doi [\apjl] {10.1086/498718}, \href
  {https://ui.adsabs.harvard.edu/abs/2005ApJ...634L..21B} {634, L21}

\bibitem[\protect\citeauthoryear{{Bassani} et~al.,}{{Bassani}
  et~al.}{2006}]{2006ApJ...636L..65B}
{Bassani} L.,  et~al., 2006, \mn@doi [\apjl] {10.1086/500132}, \href
  {https://ui.adsabs.harvard.edu/abs/2006ApJ...636L..65B} {636, L65}

\bibitem[\protect\citeauthoryear{{Bassani} et~al.,}{{Bassani}
  et~al.}{2007}]{2007ApJ...669L...1B}
{Bassani} L.,  et~al., 2007, \mn@doi [\apjl] {10.1086/523757}, \href
  {https://ui.adsabs.harvard.edu/abs/2007ApJ...669L...1B} {669, L1}

\bibitem[\protect\citeauthoryear{{Bassani} et~al.,}{{Bassani}
  et~al.}{2009}]{2009MNRAS.395L...1B}
{Bassani} L.,  et~al., 2009, \mn@doi [\mnras]
  {10.1111/j.1745-3933.2009.00626.x}, \href
  {https://ui.adsabs.harvard.edu/abs/2009MNRAS.395L...1B} {395, L1}

\bibitem[\protect\citeauthoryear{{Bassani}, {Landi}, {Malizia}, {Stephen},
  {Bazzano}, {Bird}  \& {Ubertini}}{{Bassani}
  et~al.}{2014}]{2014A&A...561A.108B}
{Bassani} L.,  {Landi} R.,  {Malizia} A.,  {Stephen} J.~B.,  {Bazzano} A.,
  {Bird} A.~J.,   {Ubertini} P.,  2014, \mn@doi [\aap]
  {10.1051/0004-6361/201322292}, \href
  {https://ui.adsabs.harvard.edu/abs/2014A&A...561A.108B} {561, A108}

\bibitem[\protect\citeauthoryear{{Bassi} et~al.,}{{Bassi}
  et~al.}{2019}]{2019MNRAS.482.1587B}
{Bassi} T.,  et~al., 2019, \mn@doi [\mnras] {10.1093/mnras/sty2739}, \href
  {https://ui.adsabs.harvard.edu/abs/2019MNRAS.482.1587B} {482, 1587}

\bibitem[\protect\citeauthoryear{{Baumgartner}, {Tueller}, {Markwardt}  \&
  {Skinner}}{{Baumgartner} et~al.}{2010}]{2010HEAD...11.1305B}
{Baumgartner} W.~H.,  {Tueller} J.,  {Markwardt} C.,   {Skinner} G.,  2010, in
  AAS/High Energy Astrophysics Division \#11. p. 13.05

\bibitem[\protect\citeauthoryear{{Baumgartner}, {Tueller}, {Markwardt},
  {Skinner}, {Barthelmy}, {Mushotzky}, {Evans}  \& {Gehrels}}{{Baumgartner}
  et~al.}{2013}]{2013ApJS..207...19B}
{Baumgartner} W.~H.,  {Tueller} J.,  {Markwardt} C.~B.,  {Skinner} G.~K.,
  {Barthelmy} S.,  {Mushotzky} R.~F.,  {Evans} P.~A.,   {Gehrels} N.,  2013,
  \mn@doi [\apjs] {10.1088/0067-0049/207/2/19}, \href
  {https://ui.adsabs.harvard.edu/abs/2013ApJS..207...19B} {207, 19}

\bibitem[\protect\citeauthoryear{{Bazzano} et~al.,}{{Bazzano}
  et~al.}{2006}]{2006ApJ...649L...9B}
{Bazzano} A.,  et~al., 2006, \mn@doi [\apjl] {10.1086/508171}, \href
  {https://ui.adsabs.harvard.edu/abs/2006ApJ...649L...9B} {649, L9}

\bibitem[\protect\citeauthoryear{{Beckmann} et~al.,}{{Beckmann}
  et~al.}{2005}]{2005ApJ...631..506B}
{Beckmann} V.,  et~al., 2005, \mn@doi [\apj] {10.1086/432600}, \href
  {https://ui.adsabs.harvard.edu/abs/2005ApJ...631..506B} {631, 506}

\bibitem[\protect\citeauthoryear{{Beckmann}, {Soldi}, {Shrader}, {Gehrels}  \&
  {Produit}}{{Beckmann} et~al.}{2006}]{2006ApJ...652..126B}
{Beckmann} V.,  {Soldi} S.,  {Shrader} C.~R.,  {Gehrels} N.,   {Produit} N.,
  2006, \mn@doi [\apj] {10.1086/507510}, \href
  {https://ui.adsabs.harvard.edu/abs/2006ApJ...652..126B} {652, 126}

\bibitem[\protect\citeauthoryear{{B{\'e}langer} et~al.,}{{B{\'e}langer}
  et~al.}{2006}]{2006ApJ...636..275B}
{B{\'e}langer} G.,  et~al., 2006, \mn@doi [\apj] {10.1086/497629}, \href
  {https://ui.adsabs.harvard.edu/abs/2006ApJ...636..275B} {636, 275}

\bibitem[\protect\citeauthoryear{{Beri} et~al.,}{{Beri}
  et~al.}{2021}]{2020MNRAS.500..565B}
{Beri} A.,  et~al., 2021, \mn@doi [\mnras] {10.1093/mnras/staa3254}, \href
  {https://ui.adsabs.harvard.edu/abs/2021MNRAS.500..565B} {500, 565}

\bibitem[\protect\citeauthoryear{{Bernardini}, {de Martino}, {Falanga},
  {Mukai}, {Matt}, {Bonnet-Bidaud}, {Masetti}  \& {Mouchet}}{{Bernardini}
  et~al.}{2012}]{2012A&A...542A..22B}
{Bernardini} F.,  {de Martino} D.,  {Falanga} M.,  {Mukai} K.,  {Matt} G.,
  {Bonnet-Bidaud} J.~M.,  {Masetti} N.,   {Mouchet} M.,  2012, \mn@doi [\aap]
  {10.1051/0004-6361/201219233}, \href
  {https://ui.adsabs.harvard.edu/abs/2012A&A...542A..22B} {542, A22}

\bibitem[\protect\citeauthoryear{{Bernardini}, {de Martino}, {Mukai}, {Israel},
  {Falanga}, {Ramsay}  \& {Masetti}}{{Bernardini}
  et~al.}{2015}]{2015MNRAS.453.3100B}
{Bernardini} F.,  {de Martino} D.,  {Mukai} K.,  {Israel} G.,  {Falanga} M.,
  {Ramsay} G.,   {Masetti} N.,  2015, \mn@doi [\mnras] {10.1093/mnras/stv1673},
  \href {https://ui.adsabs.harvard.edu/abs/2015MNRAS.453.3100B} {453, 3100}

\bibitem[\protect\citeauthoryear{{Bernardini}, {de Martino}, {Mukai}  \&
  {Falanga}}{{Bernardini} et~al.}{2018}]{2018MNRAS.478.1185B}
{Bernardini} F.,  {de Martino} D.,  {Mukai} K.,   {Falanga} M.,  2018, \mn@doi
  [\mnras] {10.1093/mnras/sty1090}, \href
  {https://ui.adsabs.harvard.edu/abs/2018MNRAS.478.1185B} {478, 1185}

\bibitem[\protect\citeauthoryear{{Bikmaev}, {Sunyaev}, {Revnivtsev}  \&
  {Burenin}}{{Bikmaev} et~al.}{2006}]{2006AstL...32..221B}
{Bikmaev} I.~F.,  {Sunyaev} R.~A.,  {Revnivtsev} M.~G.,   {Burenin} R.~A.,
  2006, \mn@doi [Astronomy Letters] {10.1134/S1063773706040013}, \href
  {https://ui.adsabs.harvard.edu/abs/2006AstL...32..221B} {32, 221}

\bibitem[\protect\citeauthoryear{{Bikmaev}, {Burenin}, {Revnivtsev}, {Sazonov},
  {Sunyaev}, {Pavlinsky}  \& {Sakhibullin}}{{Bikmaev}
  et~al.}{2008a}]{2008AstL...34..653B}
{Bikmaev} I.~F.,  {Burenin} R.~A.,  {Revnivtsev} M.~G.,  {Sazonov} S.~Y.,
  {Sunyaev} R.~A.,  {Pavlinsky} M.~N.,   {Sakhibullin} N.~A.,  2008a, \mn@doi
  [Astronomy Letters] {10.1134/S1063773708100010}, \href
  {https://ui.adsabs.harvard.edu/abs/2008AstL...34..653B} {34, 653}

\bibitem[\protect\citeauthoryear{{Bikmaev}, {Revnivtsev}, {Burenin}, {Sazonov},
  {Sunyaev}, {Pavlinsky}, {Galeev}  \& {Sakhibullin}}{{Bikmaev}
  et~al.}{2008b}]{2008Atel.1363....1B}
{Bikmaev} I.,  {Revnivtsev} M.,  {Burenin} R.,  {Sazonov} S.,  {Sunyaev} R.,
  {Pavlinsky} M.,  {Galeev} A.,   {Sakhibullin} N.,  2008b, The Astronomer's
  Telegram, \href {https://ui.adsabs.harvard.edu/abs/2008ATel.1363....1B}
  {1363, 1}

\bibitem[\protect\citeauthoryear{{Bikmaev} et~al.,}{{Bikmaev}
  et~al.}{2017}]{2017ATel10968....1B}
{Bikmaev} I.,  et~al., 2017, The Astronomer's Telegram, \href
  {https://ui.adsabs.harvard.edu/abs/2017ATel10968....1B} {10968, 1}

\bibitem[\protect\citeauthoryear{{Bird} et~al.,}{{Bird}
  et~al.}{2004}]{2004ApJ...607L..33B}
{Bird} A.~J.,  et~al., 2004, \mn@doi [\apjl] {10.1086/421772}, \href
  {https://ui.adsabs.harvard.edu/abs/2004ApJ...607L..33B} {607, L33}

\bibitem[\protect\citeauthoryear{{Bird} et~al.,}{{Bird}
  et~al.}{2006}]{2006ApJ...636..765B}
{Bird} A.~J.,  et~al., 2006, \mn@doi [\apj] {10.1086/498090}, \href
  {https://ui.adsabs.harvard.edu/abs/2006ApJ...636..765B} {636, 765}

\bibitem[\protect\citeauthoryear{{Bird} et~al.,}{{Bird}
  et~al.}{2007}]{2007ApJS..170..175B}
{Bird} A.~J.,  et~al., 2007, \mn@doi [\apjs] {10.1086/513148}, \href
  {https://ui.adsabs.harvard.edu/abs/2007ApJS..170..175B} {170, 175}

\bibitem[\protect\citeauthoryear{{Bird} et~al.,}{{Bird}
  et~al.}{2010}]{2010ApJS..186....1B}
{Bird} A.~J.,  et~al., 2010, \mn@doi [\apjs] {10.1088/0067-0049/186/1/1}, \href
  {https://ui.adsabs.harvard.edu/abs/2010ApJS..186....1B} {186, 1}

\bibitem[\protect\citeauthoryear{{Bird} et~al.,}{{Bird}
  et~al.}{2016}]{2016ApJS..223...15B}
{Bird} A.~J.,  et~al., 2016, \mn@doi [\apjs] {10.3847/0067-0049/223/1/15},
  \href {https://ui.adsabs.harvard.edu/abs/2016ApJS..223...15B} {223, 15}

\bibitem[\protect\citeauthoryear{{Bodaghee}, {Walter}, {Zurita Heras}, {Bird},
  {Courvoisier}, {Malizia}, {Terrier}  \& {Ubertini}}{{Bodaghee}
  et~al.}{2006}]{2006A&A...447.1027B}
{Bodaghee} A.,  {Walter} R.,  {Zurita Heras} J.~A.,  {Bird} A.~J.,
  {Courvoisier} T.~J.~L.,  {Malizia} A.,  {Terrier} R.,   {Ubertini} P.,  2006,
  \mn@doi [\aap] {10.1051/0004-6361:20053809}, \href
  {https://ui.adsabs.harvard.edu/abs/2006A&A...447.1027B} {447, 1027}

\bibitem[\protect\citeauthoryear{{Bodaghee}, {Tomsick}  \&
  {Rodriguez}}{{Bodaghee} et~al.}{2012}]{2012ApJ...753....3B}
{Bodaghee} A.,  {Tomsick} J.~A.,   {Rodriguez} J.,  2012, \mn@doi [\apj]
  {10.1088/0004-637X/753/1/3}, \href
  {https://ui.adsabs.harvard.edu/abs/2012ApJ...753....3B} {753, 3}

\bibitem[\protect\citeauthoryear{{Bordas} et~al.,}{{Bordas}
  et~al.}{2010}]{2010ATel.2919....1B}
{Bordas} P.,  et~al., 2010, The Astronomer's Telegram, \href
  {https://ui.adsabs.harvard.edu/abs/2010ATel.2919....1B} {2919, 1}

\bibitem[\protect\citeauthoryear{{Bozzo}, {Beardmore}, {Papitto}, {Ferrigno}
  \& {Gibaud}}{{Bozzo} et~al.}{2011}]{2011ATel.3558....1B}
{Bozzo} E.,  {Beardmore} A.,  {Papitto} A.,  {Ferrigno} C.,   {Gibaud} L.,
  2011, The Astronomer's Telegram, \href
  {https://ui.adsabs.harvard.edu/abs/2011ATel.3558....1B} {3558, 1}

\bibitem[\protect\citeauthoryear{{Bozzo}, {Romano}, {Falanga}, {Ferrigno},
  {Papitto}  \& {Krimm}}{{Bozzo} et~al.}{2015}]{2015A&A...579A..56B}
{Bozzo} E.,  {Romano} P.,  {Falanga} M.,  {Ferrigno} C.,  {Papitto} A.,
  {Krimm} H.~A.,  2015, \mn@doi [\aap] {10.1051/0004-6361/201526150}, \href
  {https://ui.adsabs.harvard.edu/abs/2015A&A...579A..56B} {579, A56}

\bibitem[\protect\citeauthoryear{{Bozzo} et~al.,}{{Bozzo}
  et~al.}{2018a}]{2018A&A...613A..22B}
{Bozzo} E.,  et~al., 2018a, \mn@doi [\aap] {10.1051/0004-6361/201832588}, \href
  {https://ui.adsabs.harvard.edu/abs/2018A&A...613A..22B} {613, A22}

\bibitem[\protect\citeauthoryear{{Bozzo}, {Savchenko}, {Ferrigno}, {Ducci},
  {Kuulkers}, {Ubertini}  \& {Laurent}}{{Bozzo}
  et~al.}{2018b}]{2018ATel11478....1B}
{Bozzo} E.,  {Savchenko} V.,  {Ferrigno} C.,  {Ducci} L.,  {Kuulkers} E.,
  {Ubertini} P.,   {Laurent} P.,  2018b, The Astronomer's Telegram, \href
  {https://ui.adsabs.harvard.edu/abs/2018ATel11478....1B} {11478, 1}

\bibitem[\protect\citeauthoryear{{Brandt}, {Budtz-J{\o}rgensen}  \&
  {Chenevez}}{{Brandt} et~al.}{2006a}]{2006ATel..778....1B}
{Brandt} S.,  {Budtz-J{\o}rgensen} C.,   {Chenevez} J.,  2006a, The
  Astronomer's Telegram, \href
  {https://ui.adsabs.harvard.edu/abs/2006ATel..778....1B} {778, 1}

\bibitem[\protect\citeauthoryear{{Brandt}, {Budtz-Jorgensen}, {Chenevez},
  {Lund}, {Oxborrow}  \& {Westergaard}}{{Brandt}
  et~al.}{2006b}]{2006ATel..970....1B}
{Brandt} S.,  {Budtz-Jorgensen} C.,  {Chenevez} J.,  {Lund} N.,  {Oxborrow}
  C.~A.,   {Westergaard} N.~J.,  2006b, The Astronomer's Telegram, \href
  {https://ui.adsabs.harvard.edu/abs/2006ATel..970....1B} {970, 1}

\bibitem[\protect\citeauthoryear{{Brandt}, {Budtz-J{\o}rgensen}, {Gotz},
  {Hurley}  \& {Frontera}}{{Brandt} et~al.}{2007}]{2007ATel.1054....1B}
{Brandt} S.,  {Budtz-J{\o}rgensen} C.,  {Gotz} D.,  {Hurley} K.,   {Frontera}
  F.,  2007, The Astronomer's Telegram, \href
  {https://ui.adsabs.harvard.edu/abs/2007ATel.1054....1B} {1054, 1}

\bibitem[\protect\citeauthoryear{{Brandt} et~al.,}{{Brandt}
  et~al.}{2008}]{2008ATel.1400....1B}
{Brandt} S.,  et~al., 2008, The Astronomer's Telegram, \href
  {https://ui.adsabs.harvard.edu/abs/2008ATel.1400....1B} {1400, 1}

\bibitem[\protect\citeauthoryear{{Britt} et~al.,}{{Britt}
  et~al.}{2013}]{2013ApJ...769..120B}
{Britt} C.~T.,  et~al., 2013, \mn@doi [\apj] {10.1088/0004-637X/769/2/120},
  \href {https://ui.adsabs.harvard.edu/abs/2013ApJ...769..120B} {769, 120}

\bibitem[\protect\citeauthoryear{{Burenin}, {Mescheryakov}, {Revnivtsev},
  {Bikmaev}  \& {Sunyaev}}{{Burenin} et~al.}{2006a}]{2006ATel..880....1B}
{Burenin} R.,  {Mescheryakov} A.,  {Revnivtsev} M.,  {Bikmaev} I.,   {Sunyaev}
  R.,  2006a, The Astronomer's Telegram, \href
  {https://ui.adsabs.harvard.edu/abs/2006ATel..880....1B} {880, 1}

\bibitem[\protect\citeauthoryear{{Burenin}, {Mescheryakov}, {Sazonov},
  {Revnivtsev}, {Bikmaev}  \& {Sunyaev}}{{Burenin}
  et~al.}{2006b}]{2006ATel..883....1B}
{Burenin} R.,  {Mescheryakov} A.,  {Sazonov} S.,  {Revnivtsev} M.,  {Bikmaev}
  I.,   {Sunyaev} R.,  2006b, The Astronomer's Telegram, \href
  {https://ui.adsabs.harvard.edu/abs/2006ATel..883....1B} {883, 1}

\bibitem[\protect\citeauthoryear{{Burenin}, {Revnivtsev}, {Mescheryakov},
  {Bikmaev}, {Pavlinsky}  \& {Sunyaev}}{{Burenin}
  et~al.}{2007}]{2007ATel.1270....1B}
{Burenin} R.,  {Revnivtsev} M.,  {Mescheryakov} A.,  {Bikmaev} I.,  {Pavlinsky}
  M.,   {Sunyaev} R.,  2007, The Astronomer's Telegram, \href
  {https://ui.adsabs.harvard.edu/abs/2007ATel.1270....1B} {1270, 1}

\bibitem[\protect\citeauthoryear{{Burenin}, {Mescheryakov}, {Revnivtsev},
  {Sazonov}, {Bikmaev}, {Pavlinsky}  \& {Sunyaev}}{{Burenin}
  et~al.}{2008}]{2008AstL...34..367B}
{Burenin} R.~A.,  {Mescheryakov} A.~V.,  {Revnivtsev} M.~G.,  {Sazonov} S.~Y.,
  {Bikmaev} I.~F.,  {Pavlinsky} M.~N.,   {Sunyaev} R.~A.,  2008, \mn@doi
  [Astronomy Letters] {10.1134/S1063773708060017}, \href
  {https://ui.adsabs.harvard.edu/abs/2008AstL...34..367B} {34, 367}

\bibitem[\protect\citeauthoryear{{Burenin}, {Makarov}, {Uklein}, {Revnivtsev}
  \& {Lutovinov}}{{Burenin} et~al.}{2009}]{2009ATel.2193....1R}
{Burenin} R.,  {Makarov} D.,  {Uklein} R.,  {Revnivtsev} M.,   {Lutovinov} A.,
  2009, The Astronomer's Telegram, \href
  {https://ui.adsabs.harvard.edu/abs/2009ATel.2193....1B} {2193, 1}

\bibitem[\protect\citeauthoryear{{Butler} et~al.,}{{Butler}
  et~al.}{2009}]{2009ApJ...698..502B}
{Butler} S.~C.,  et~al., 2009, \mn@doi [\apj] {10.1088/0004-637X/698/1/502},
  \href {https://ui.adsabs.harvard.edu/abs/2009ApJ...698..502B} {698, 502}

\bibitem[\protect\citeauthoryear{{Caballero} et~al.,}{{Caballero}
  et~al.}{2013}]{2013arXiv1304.1349C}
{Caballero} I.,  et~al., 2013, arXiv e-prints, \href
  {https://ui.adsabs.harvard.edu/abs/2013arXiv1304.1349C} {p. arXiv:1304.1349}

\bibitem[\protect\citeauthoryear{{Caccianiga} et~al.,}{{Caccianiga}
  et~al.}{2008}]{2008AandA...477..735C}
{Caccianiga} A.,  et~al., 2008, \mn@doi [\aap] {10.1051/0004-6361:20078568},
  \href {https://ui.adsabs.harvard.edu/abs/2008A&A...477..735C} {477, 735}

\bibitem[\protect\citeauthoryear{{Capitanio} et~al.,}{{Capitanio}
  et~al.}{2006}]{2006ApJ...643..376C}
{Capitanio} F.,  et~al., 2006, \mn@doi [\apj] {10.1086/502641}, \href
  {https://ui.adsabs.harvard.edu/abs/2006ApJ...643..376C} {643, 376}

\bibitem[\protect\citeauthoryear{{Chakrabarty}, {Jonker}  \&
  {Markwardt}}{{Chakrabarty} et~al.}{2011a}]{2011ATel.3218....1C}
{Chakrabarty} D.,  {Jonker} P.,   {Markwardt} C.~B.,  2011a, The Astronomer's
  Telegram, \href {https://ui.adsabs.harvard.edu/abs/2011ATel.3218....1C}
  {3218, 1}

\bibitem[\protect\citeauthoryear{{Chakrabarty}, {Markwardt}, {Linares}  \&
  {Jonker}}{{Chakrabarty} et~al.}{2011b}]{2011ATel.3606....1C}
{Chakrabarty} D.,  {Markwardt} C.~B.,  {Linares} M.,   {Jonker} P.~G.,  2011b,
  The Astronomer's Telegram, \href
  {https://ui.adsabs.harvard.edu/abs/2011ATel.3606....1C} {3606, 1}

\bibitem[\protect\citeauthoryear{{Chaty}, {Rahoui}, {Foellmi}, {Tomsick},
  {Rodriguez}  \& {Walter}}{{Chaty} et~al.}{2008}]{2008A&A...484..783C}
{Chaty} S.,  {Rahoui} F.,  {Foellmi} C.,  {Tomsick} J.~A.,  {Rodriguez} J.,
  {Walter} R.,  2008, \mn@doi [\aap] {10.1051/0004-6361:20078768}, \href
  {https://ui.adsabs.harvard.edu/abs/2008A&A...484..783C} {484, 783}

\bibitem[\protect\citeauthoryear{{Chelovekov} \& {Grebenev}}{{Chelovekov} \&
  {Grebenev}}{2007}]{2007AstL...33..807C}
{Chelovekov} I.~V.,  {Grebenev} S.~A.,  2007, \mn@doi [Astronomy Letters]
  {10.1134/S1063773707120043}, \href
  {https://ui.adsabs.harvard.edu/abs/2007AstL...33..807C} {33, 807}

\bibitem[\protect\citeauthoryear{{Chenevez} et~al.,}{{Chenevez}
  et~al.}{2012}]{2012ATel.4050....1C}
{Chenevez} J.,  et~al., 2012, The Astronomer's Telegram, \href
  {https://ui.adsabs.harvard.edu/abs/2012ATel.4050....1C} {4050, 1}

\bibitem[\protect\citeauthoryear{{Chernyakova}, {Lutovinov}, {Capitanio},
  {Lund}  \& {Gehrels}}{{Chernyakova} et~al.}{2003}]{2003ATel..157....1C}
{Chernyakova} M.,  {Lutovinov} A.,  {Capitanio} F.,  {Lund} N.,   {Gehrels} N.,
   2003, The Astronomer's Telegram, \href
  {https://ui.adsabs.harvard.edu/abs/2003ATel..157....1C} {157, 1}

\bibitem[\protect\citeauthoryear{{Chernyakova}, {Lutovinov}, {Rodr{\'\i}guez}
  \& {Revnivtsev}}{{Chernyakova} et~al.}{2005}]{2005MNRAS.364..455C}
{Chernyakova} M.,  {Lutovinov} A.,  {Rodr{\'\i}guez} J.,   {Revnivtsev} M.,
  2005, \mn@doi [\mnras] {10.1111/j.1365-2966.2005.09548.x}, \href
  {https://ui.adsabs.harvard.edu/abs/2005MNRAS.364..455C} {364, 455}

\bibitem[\protect\citeauthoryear{{Churazov} et~al.,}{{Churazov}
  et~al.}{2007}]{2007A&A...467..529C}
{Churazov} E.,  et~al., 2007, \mn@doi [\aap] {10.1051/0004-6361:20066230},
  \href {https://ui.adsabs.harvard.edu/abs/2007A&A...467..529C} {467, 529}

\bibitem[\protect\citeauthoryear{{Churazov} et~al.,}{{Churazov}
  et~al.}{2014}]{2014Natur.512..406C}
{Churazov} E.,  et~al., 2014, \mn@doi [\nat] {10.1038/nature13672}, \href
  {https://ui.adsabs.harvard.edu/abs/2014Natur.512..406C} {512, 406}

\bibitem[\protect\citeauthoryear{{Clavel} et~al.,}{{Clavel}
  et~al.}{2016}]{2016MNRAS.461..304C}
{Clavel} M.,  et~al., 2016, \mn@doi [\mnras] {10.1093/mnras/stw1330}, \href
  {https://ui.adsabs.harvard.edu/abs/2016MNRAS.461..304C} {461, 304}

\bibitem[\protect\citeauthoryear{{Clavel}, {Tomsick}, {Hare}, {Krivonos},
  {Mori}  \& {Stern}}{{Clavel} et~al.}{2019}]{2019ApJ...887...32C}
{Clavel} M.,  {Tomsick} J.~A.,  {Hare} J.,  {Krivonos} R.,  {Mori} K.,
  {Stern} D.,  2019, \mn@doi [\apj] {10.3847/1538-4357/ab4b55}, \href
  {https://ui.adsabs.harvard.edu/abs/2019ApJ...887...32C} {887, 32}

\bibitem[\protect\citeauthoryear{{Cocchi}, {Bazzano}, {Natalucci}, {Ubertini},
  {Heise}, {Muller}  \& {in 't Zand}}{{Cocchi}
  et~al.}{1999}]{1999A&A...346L..45C}
{Cocchi} M.,  {Bazzano} A.,  {Natalucci} L.,  {Ubertini} P.,  {Heise} J.,
  {Muller} J.~M.,   {in 't Zand} J.~J.~M.,  1999, \aap, \href
  {https://ui.adsabs.harvard.edu/abs/1999A&A...346L..45C} {346, L45}

\bibitem[\protect\citeauthoryear{{Coe}, {Bartlett}, {Bird}, {Haberl}, {Kennea},
  {McBride}, {Townsend}  \& {Udalski}}{{Coe}
  et~al.}{2015}]{2015MNRAS.447.2387C}
{Coe} M.~J.,  {Bartlett} E.~S.,  {Bird} A.~J.,  {Haberl} F.,  {Kennea} J.~A.,
  {McBride} V.~A.,  {Townsend} L.~J.,   {Udalski} A.,  2015, \mn@doi [\mnras]
  {10.1093/mnras/stu2568}, \href
  {https://ui.adsabs.harvard.edu/abs/2015MNRAS.447.2387C} {447, 2387}

\bibitem[\protect\citeauthoryear{{Coleiro}, {Chaty}, {Zurita Heras}, {Rahoui}
  \& {Tomsick}}{{Coleiro} et~al.}{2013}]{2013A&A...560A.108C}
{Coleiro} A.,  {Chaty} S.,  {Zurita Heras} J.~A.,  {Rahoui} F.,   {Tomsick}
  J.~A.,  2013, \mn@doi [\aap] {10.1051/0004-6361/201322382}, \href
  {https://ui.adsabs.harvard.edu/abs/2013A&A...560A.108C} {560, A108}

\bibitem[\protect\citeauthoryear{{Cornelisse}, {Verbunt}, {in't Zand},
  {Kuulkers}  \& {Heise}}{{Cornelisse} et~al.}{2002}]{2002A&A...392..931C}
{Cornelisse} R.,  {Verbunt} F.,  {in't Zand} J.~J.~M.,  {Kuulkers} E.,
  {Heise} J.,  2002, \mn@doi [\aap] {10.1051/0004-6361:20021183}, \href
  {https://ui.adsabs.harvard.edu/abs/2002A&A...392..931C} {392, 931}

\bibitem[\protect\citeauthoryear{{Courvoisier}, {Walter}, {Rodriguez},
  {Bouchet}  \& {Lutovinov}}{{Courvoisier} et~al.}{2003}]{2003IAUC.8063....3C}
{Courvoisier} T.~J.~L.,  {Walter} R.,  {Rodriguez} J.,  {Bouchet} L.,
  {Lutovinov} A.~A.,  2003, \iaucirc, \href
  {https://ui.adsabs.harvard.edu/abs/2003IAUC.8063....3C} {8063, 3}

\bibitem[\protect\citeauthoryear{{Crawford}, {Jauncey}  \&
  {Murdoch}}{{Crawford} et~al.}{1970}]{1970ApJ...162..405C}
{Crawford} D.~F.,  {Jauncey} D.~L.,   {Murdoch} H.~S.,  1970, \mn@doi [\apj]
  {10.1086/150672}, \href
  {https://ui.adsabs.harvard.edu/abs/1970ApJ...162..405C} {162, 405}

\bibitem[\protect\citeauthoryear{{Curran}, {Chaty}  \& {Zurita Heras}}{{Curran}
  et~al.}{2011}]{2011A&A...533A...3C}
{Curran} P.~A.,  {Chaty} S.,   {Zurita Heras} J.~A.,  2011, \mn@doi [\aap]
  {10.1051/0004-6361/201117208}, \href
  {https://ui.adsabs.harvard.edu/abs/2011A&A...533A...3C} {533, A3}

\bibitem[\protect\citeauthoryear{{Cusumano} et~al.,}{{Cusumano}
  et~al.}{2010}]{2010A&A...510A..48C}
{Cusumano} G.,  et~al., 2010, \mn@doi [\aap] {10.1051/0004-6361/200811184},
  \href {https://ui.adsabs.harvard.edu/abs/2010A&A...510A..48C} {510, A48}

\bibitem[\protect\citeauthoryear{{Cusumano}, {D'A{\`\i}}, {Segreto}, {La
  Parola}  \& {Del Santo}}{{Cusumano} et~al.}{2020}]{2020MNRAS.498.2750C}
{Cusumano} G.,  {D'A{\`\i}} A.,  {Segreto} A.,  {La Parola} V.,   {Del Santo}
  M.,  2020, \mn@doi [\mnras] {10.1093/mnras/staa2505}, \href
  {https://ui.adsabs.harvard.edu/abs/2020MNRAS.498.2750C} {498, 2750}

\bibitem[\protect\citeauthoryear{{D'A{\`\i}}, {La Parola}, {Cusumano},
  {Segreto}, {Romano}, {Vercellone}  \& {Robba}}{{D'A{\`\i}}
  et~al.}{2011}]{2011A&A...529A..30D}
{D'A{\`\i}} A.,  {La Parola} V.,  {Cusumano} G.,  {Segreto} A.,  {Romano} P.,
  {Vercellone} S.,   {Robba} N.~R.,  2011, \mn@doi [\aap]
  {10.1051/0004-6361/201016401}, \href
  {https://ui.adsabs.harvard.edu/abs/2011A&A...529A..30D} {529, A30}

\bibitem[\protect\citeauthoryear{{Degenaar} et~al.,}{{Degenaar}
  et~al.}{2007}]{2007ATel.1136....1D}
{Degenaar} N.,  et~al., 2007, The Astronomer's Telegram, \href
  {https://ui.adsabs.harvard.edu/abs/2007ATel.1136....1D} {1136, 1}

\bibitem[\protect\citeauthoryear{{Degenaar} et~al.,}{{Degenaar}
  et~al.}{2010}]{2010MNRAS.404.1591D}
{Degenaar} N.,  et~al., 2010, \mn@doi [\mnras]
  {10.1111/j.1365-2966.2010.16388.x}, \href
  {https://ui.adsabs.harvard.edu/abs/2010MNRAS.404.1591D} {404, 1591}

\bibitem[\protect\citeauthoryear{{Degenaar}, {Yang}  \& {Wijnands}}{{Degenaar}
  et~al.}{2011}]{2011ATel.3741....1D}
{Degenaar} N.,  {Yang} Y.~J.,   {Wijnands} R.,  2011, The Astronomer's
  Telegram, \href {https://ui.adsabs.harvard.edu/abs/2011ATel.3741....1D}
  {3741, 1}

\bibitem[\protect\citeauthoryear{{Degenaar}, {Altamirano}  \&
  {Wijnands}}{{Degenaar} et~al.}{2012}]{2012ATel.4219....1D}
{Degenaar} N.,  {Altamirano} D.,   {Wijnands} R.,  2012, The Astronomer's
  Telegram, \href {https://ui.adsabs.harvard.edu/abs/2012ATel.4219....1D}
  {4219, 1}

\bibitem[\protect\citeauthoryear{{Del Monte} et~al.,}{{Del Monte}
  et~al.}{2008a}]{2008int..workE.122D}
{Del Monte} E.,  et~al., 2008a, in The 7th INTEGRAL Workshop. p.~122

\bibitem[\protect\citeauthoryear{{Del Monte} et~al.,}{{Del Monte}
  et~al.}{2008b}]{2008ATel.1445....1D}
{Del Monte} E.,  et~al., 2008b, The Astronomer's Telegram, \href
  {https://ui.adsabs.harvard.edu/abs/2008ATel.1445....1D} {1445, 1}

\bibitem[\protect\citeauthoryear{{Del Santo} et~al.,}{{Del Santo}
  et~al.}{2011}]{2011ATel.3210....1D}
{Del Santo} M.,  et~al., 2011, The Astronomer's Telegram, \href
  {https://ui.adsabs.harvard.edu/abs/2011ATel.3210....1D} {3210, 1}

\bibitem[\protect\citeauthoryear{{Del Santo}, {Nucita}, {Lodato}, {Manni}, {De
  Paolis}, {Farihi}, {De Cesare}  \& {Segreto}}{{Del Santo}
  et~al.}{2014}]{2014MNRAS.444...93D}
{Del Santo} M.,  {Nucita} A.~A.,  {Lodato} G.,  {Manni} L.,  {De Paolis} F.,
  {Farihi} J.,  {De Cesare} G.,   {Segreto} A.,  2014, \mn@doi [\mnras]
  {10.1093/mnras/stu1436}, \href
  {https://ui.adsabs.harvard.edu/abs/2014MNRAS.444...93D} {444, 93}

\bibitem[\protect\citeauthoryear{{Del Santo} et~al.,}{{Del Santo}
  et~al.}{2016}]{2016MNRAS.456.3585D}
{Del Santo} M.,  et~al., 2016, \mn@doi [\mnras] {10.1093/mnras/stv2901}, \href
  {https://ui.adsabs.harvard.edu/abs/2016MNRAS.456.3585D} {456, 3585}

\bibitem[\protect\citeauthoryear{{Doroshenko} et~al.,}{{Doroshenko}
  et~al.}{2020a}]{2020MNRAS.491.1857D}
{Doroshenko} V.,  et~al., 2020a, \mn@doi [\mnras] {10.1093/mnras/stz2879},
  \href {https://ui.adsabs.harvard.edu/abs/2020MNRAS.491.1857D} {491, 1857}

\bibitem[\protect\citeauthoryear{{Doroshenko}, {Tsygankov}, {Long},
  {Santangelo}, {Molkov}, {Lutovinov}, {Kong}  \& {Zhang}}{{Doroshenko}
  et~al.}{2020b}]{2020A&A...634A..89D}
{Doroshenko} V.,  {Tsygankov} S.,  {Long} J.,  {Santangelo} A.,  {Molkov} S.,
  {Lutovinov} A.,  {Kong} L.~D.,   {Zhang} S.,  2020b, \mn@doi [\aap]
  {10.1051/0004-6361/201937036}, \href
  {https://ui.adsabs.harvard.edu/abs/2020A&A...634A..89D} {634, A89}

\bibitem[\protect\citeauthoryear{{Ducci}, {Kuulkers}, {Grinberg}, {Paizis},
  {Sidoli}, {Bozzo}, {Ferrigno}  \& {Savchenko}}{{Ducci}
  et~al.}{2018}]{2018ATel11941....1D}
{Ducci} L.,  {Kuulkers} E.,  {Grinberg} V.,  {Paizis} A.,  {Sidoli} L.,
  {Bozzo} E.,  {Ferrigno} C.,   {Savchenko} V.,  2018, The Astronomer's
  Telegram, \href {https://ui.adsabs.harvard.edu/abs/2018ATel11941....1D}
  {11941, 1}

\bibitem[\protect\citeauthoryear{{Eckert}, {Walter}, {Kretschmar}, {Mas-Hesse},
  {Palumbo}, {Roques}, {Ubertini}  \& {Winkler}}{{Eckert}
  et~al.}{2004}]{2004ATel..352....1E}
{Eckert} D.,  {Walter} R.,  {Kretschmar} P.,  {Mas-Hesse} M.,  {Palumbo}
  G.~G.~C.,  {Roques} J.~P.,  {Ubertini} P.,   {Winkler} C.,  2004, The
  Astronomer's Telegram, \href
  {https://ui.adsabs.harvard.edu/abs/2004ATel..352....1E} {352, 1}

\bibitem[\protect\citeauthoryear{{Eckert} et~al.,}{{Eckert}
  et~al.}{2013}]{2013ATel.4925....1E}
{Eckert} D.,  et~al., 2013, The Astronomer's Telegram, \href
  {https://ui.adsabs.harvard.edu/abs/2013ATel.4925....1E} {4925, 1}

\bibitem[\protect\citeauthoryear{{Edelson} \& {Malkan}}{{Edelson} \&
  {Malkan}}{2012}]{2012ApJ...751...52E}
{Edelson} R.,  {Malkan} M.,  2012, \mn@doi [\apj] {10.1088/0004-637X/751/1/52},
  \href {https://ui.adsabs.harvard.edu/abs/2012ApJ...751...52E} {751, 52}

\bibitem[\protect\citeauthoryear{{Esposito}, {Israel}, {Sidoli}, {Mason},
  {Rodr{\'\i}guez Castillo}, {Halpern}, {Moretti}  \& {G{\"o}tz}}{{Esposito}
  et~al.}{2013}]{2013MNRAS.433.2028E}
{Esposito} P.,  {Israel} G.~L.,  {Sidoli} L.,  {Mason} E.,  {Rodr{\'\i}guez
  Castillo} G.~A.,  {Halpern} J.~P.,  {Moretti} A.,   {G{\"o}tz} D.,  2013,
  \mn@doi [\mnras] {10.1093/mnras/stt870}, \href
  {https://ui.adsabs.harvard.edu/abs/2013MNRAS.433.2028E} {433, 2028}

\bibitem[\protect\citeauthoryear{{Evans} et~al.,}{{Evans}
  et~al.}{2010}]{2010ApJS..189...37E}
{Evans} I.~N.,  et~al., 2010, \mn@doi [\apjs] {10.1088/0067-0049/189/1/37},
  \href {https://ui.adsabs.harvard.edu/abs/2010ApJS..189...37E} {189, 37}

\bibitem[\protect\citeauthoryear{{Evans} et~al.,}{{Evans}
  et~al.}{2020}]{2020ApJS..247...54E}
{Evans} P.~A.,  et~al., 2020, \mn@doi [\apjs] {10.3847/1538-4365/ab7db9}, \href
  {https://ui.adsabs.harvard.edu/abs/2020ApJS..247...54E} {247, 54}

\bibitem[\protect\citeauthoryear{{Falanga}, {Bozzo}, {Walter}, {Sarty}  \&
  {Stella}}{{Falanga} et~al.}{2011}]{2011JAVSO..39..110F}
{Falanga} M.,  {Bozzo} E.,  {Walter} R.,  {Sarty} G.~E.,   {Stella} L.,  2011,
  Journal of the American Association of Variable Star Observers (JAAVSO),
  \href {https://ui.adsabs.harvard.edu/abs/2011JAVSO..39..110F} {39, 110}

\bibitem[\protect\citeauthoryear{{Ferrigno}, {Bozzo}  \& {Belloni}}{{Ferrigno}
  et~al.}{2011}]{2011ATel.3560....1F}
{Ferrigno} C.,  {Bozzo} E.,   {Belloni} L. G. A. P. T.~M.,  2011, The
  Astronomer's Telegram, \href
  {https://ui.adsabs.harvard.edu/abs/2011ATel.3560....1F} {3560, 1}

\bibitem[\protect\citeauthoryear{{Ferrigno}, {Bozzo}, {Del Santo}  \&
  {Capitanio}}{{Ferrigno} et~al.}{2012}]{2012A&A...537L...7F}
{Ferrigno} C.,  {Bozzo} E.,  {Del Santo} M.,   {Capitanio} F.,  2012, \mn@doi
  [\aap] {10.1051/0004-6361/201118474}, \href
  {https://ui.adsabs.harvard.edu/abs/2012A&A...537L...7F} {537, L7}

\bibitem[\protect\citeauthoryear{{Fiocchi}, {Bassani}, {Bazzano}, {Ubertini},
  {Landi}, {Capitanio}  \& {Bird}}{{Fiocchi}
  et~al.}{2010}]{2010ApJ...720..987F}
{Fiocchi} M.,  {Bassani} L.,  {Bazzano} A.,  {Ubertini} P.,  {Landi} R.,
  {Capitanio} F.,   {Bird} A.~J.,  2010, \mn@doi [\apj]
  {10.1088/0004-637X/720/2/987}, \href
  {https://ui.adsabs.harvard.edu/abs/2010ApJ...720..987F} {720, 987}

\bibitem[\protect\citeauthoryear{{Fornasini}, {Tomsick}, {Bachetti},
  {Krivonos}, {F{\"u}rst}, {Natalucci}, {Pottschmidt}  \& {Wilms}}{{Fornasini}
  et~al.}{2017}]{2017ApJ...841...35F}
{Fornasini} F.~M.,  {Tomsick} J.~A.,  {Bachetti} M.,  {Krivonos} R.~A.,
  {F{\"u}rst} F.,  {Natalucci} L.,  {Pottschmidt} K.,   {Wilms} J.,  2017,
  \mn@doi [\apj] {10.3847/1538-4357/aa6ff4}, \href
  {https://ui.adsabs.harvard.edu/abs/2017ApJ...841...35F} {841, 35}

\bibitem[\protect\citeauthoryear{{Fortin}, {Chaty}, {Coleiro}, {Tomsick}  \&
  {Nitschelm}}{{Fortin} et~al.}{2018}]{2018A&A...618A.150F}
{Fortin} F.,  {Chaty} S.,  {Coleiro} A.,  {Tomsick} J.~A.,   {Nitschelm}
  C.~H.~R.,  2018, \mn@doi [\aap] {10.1051/0004-6361/201731265}, \href
  {https://ui.adsabs.harvard.edu/abs/2018A&A...618A.150F} {618, A150}

\bibitem[\protect\citeauthoryear{{Fuerst} et~al.,}{{Fuerst}
  et~al.}{2018}]{2018ATel11357....1F}
{Fuerst} F.,  et~al., 2018, The Astronomer's Telegram, \href
  {https://ui.adsabs.harvard.edu/abs/2018ATel11357....1F} {11357, 1}

\bibitem[\protect\citeauthoryear{{Funk} et~al.,}{{Funk}
  et~al.}{2007}]{2007A&A...470..249F}
{Funk} S.,  et~al., 2007, \mn@doi [\aap] {10.1051/0004-6361:20066779}, \href
  {https://ui.adsabs.harvard.edu/abs/2007A&A...470..249F} {470, 249}

\bibitem[\protect\citeauthoryear{{G{\"a}nsicke} et~al.,}{{G{\"a}nsicke}
  et~al.}{2005}]{2005MNRAS.361..141G}
{G{\"a}nsicke} B.~T.,  et~al., 2005, \mn@doi [\mnras]
  {10.1111/j.1365-2966.2005.09138.x}, \href
  {https://ui.adsabs.harvard.edu/abs/2005MNRAS.361..141G} {361, 141}

\bibitem[\protect\citeauthoryear{{Gehrels} et~al.,}{{Gehrels}
  et~al.}{2004}]{2004ApJ...611.1005G}
{Gehrels} N.,  et~al., 2004, \mn@doi [\apj] {10.1086/422091}, \href
  {https://ui.adsabs.harvard.edu/abs/2004ApJ...611.1005G} {611, 1005}

\bibitem[\protect\citeauthoryear{{Georgakakis}, {Nandra}, {Laird}, {Aird}  \&
  {Trichas}}{{Georgakakis} et~al.}{2008}]{2008MNRAS.388.1205G}
{Georgakakis} A.,  {Nandra} K.,  {Laird} E.~S.,  {Aird} J.,   {Trichas} M.,
  2008, \mn@doi [\mnras] {10.1111/j.1365-2966.2008.13423.x}, \href
  {https://ui.adsabs.harvard.edu/abs/2008MNRAS.388.1205G} {388, 1205}

\bibitem[\protect\citeauthoryear{{Giann{\'\i}}, {de Rosa}, {Bassani},
  {Bazzano}, {Dean}  \& {Ubertini}}{{Giann{\'\i}}
  et~al.}{2011}]{2011MNRAS.411.2137G}
{Giann{\'\i}} S.,  {de Rosa} A.,  {Bassani} L.,  {Bazzano} A.,  {Dean} T.,
  {Ubertini} P.,  2011, \mn@doi [\mnras] {10.1111/j.1365-2966.2010.17725.x},
  \href {https://ui.adsabs.harvard.edu/abs/2011MNRAS.411.2137G} {411, 2137}

\bibitem[\protect\citeauthoryear{{Gibaud} et~al.,}{{Gibaud}
  et~al.}{2011}]{2011ATel.3565....1G}
{Gibaud} L.,  et~al., 2011, The Astronomer's Telegram, \href
  {https://ui.adsabs.harvard.edu/abs/2011ATel.3565....1G} {3565, 1}

\bibitem[\protect\citeauthoryear{{Goncalves}, {Martin}, {Halpern}, {Eracleous}
  \& {Pavlov}}{{Goncalves} et~al.}{2008}]{2008ATel.1623....1G}
{Goncalves} T.~S.,  {Martin} D.~C.,  {Halpern} J.~P.,  {Eracleous} M.,
  {Pavlov} G.~G.,  2008, The Astronomer's Telegram, \href
  {https://ui.adsabs.harvard.edu/abs/2008ATel.1623....1G} {1623, 1}

\bibitem[\protect\citeauthoryear{{Gonz{\'a}lez-Mart{\'\i}n}
  et~al.,}{{Gonz{\'a}lez-Mart{\'\i}n} et~al.}{2011}]{2011A&A...527A.142G}
{Gonz{\'a}lez-Mart{\'\i}n} O.,  et~al., 2011, \mn@doi [\aap]
  {10.1051/0004-6361/201016097}, \href
  {https://ui.adsabs.harvard.edu/abs/2011A&A...527A.142G} {527, A142}

\bibitem[\protect\citeauthoryear{{Goossens}, {Bird}, {Hill}, {Sguera}  \&
  {Drave}}{{Goossens} et~al.}{2019}]{2019MNRAS.485..286G}
{Goossens} M.~E.,  {Bird} A.~J.,  {Hill} A.~B.,  {Sguera} V.,   {Drave} S.~P.,
  2019, \mn@doi [\mnras] {10.1093/mnras/sty3236}, \href
  {https://ui.adsabs.harvard.edu/abs/2019MNRAS.485..286G} {485, 286}

\bibitem[\protect\citeauthoryear{{G{\'o}rski}, {Hivon}, {Banday}, {Wandelt},
  {Hansen}, {Reinecke}  \& {Bartelmann}}{{G{\'o}rski}
  et~al.}{2005}]{2005ApJ...622..759G}
{G{\'o}rski} K.~M.,  {Hivon} E.,  {Banday} A.~J.,  {Wandelt} B.~D.,  {Hansen}
  F.~K.,  {Reinecke} M.,   {Bartelmann} M.,  2005, \mn@doi [\apj]
  {10.1086/427976}, \href
  {https://ui.adsabs.harvard.edu/abs/2005ApJ...622..759G} {622, 759}

\bibitem[\protect\citeauthoryear{{Grebenev} \& {Sunyaev}}{{Grebenev} \&
  {Sunyaev}}{2004}]{2004ATel..342....1G}
{Grebenev} S.~A.,  {Sunyaev} R.~A.,  2004, The Astronomer's Telegram, \href
  {https://ui.adsabs.harvard.edu/abs/2004ATel..342....1G} {342, 1}

\bibitem[\protect\citeauthoryear{{Grebenev} \& {Sunyaev}}{{Grebenev} \&
  {Sunyaev}}{2007}]{2007AstL...33..149G}
{Grebenev} S.~A.,  {Sunyaev} R.~A.,  2007, \mn@doi [Astronomy Letters]
  {10.1134/S1063773707030024}, \href
  {https://ui.adsabs.harvard.edu/abs/2007AstL...33..149G} {33, 149}

\bibitem[\protect\citeauthoryear{{Grebenev} \& {Sunyaev}}{{Grebenev} \&
  {Sunyaev}}{2010}]{2010AstL...36..533G}
{Grebenev} S.~A.,  {Sunyaev} R.~A.,  2010, \mn@doi [Astronomy Letters]
  {10.1134/S1063773710080013}, \href
  {https://ui.adsabs.harvard.edu/abs/2010AstL...36..533G} {36, 533}

\bibitem[\protect\citeauthoryear{{Grebenev}, {Ubertini}, {Chenevez}, {Orr}  \&
  {Sunyaev}}{{Grebenev} et~al.}{2004}]{2004ATel..275....1G}
{Grebenev} S.~A.,  {Ubertini} P.,  {Chenevez} J.,  {Orr} A.,   {Sunyaev} R.~A.,
   2004, The Astronomer's Telegram, \href
  {https://ui.adsabs.harvard.edu/abs/2004ATel..275....1G} {275, 1}

\bibitem[\protect\citeauthoryear{{Grebenev}, {Molkov}  \& {Sunyaev}}{{Grebenev}
  et~al.}{2005a}]{2005ATel..444....1G}
{Grebenev} S.~A.,  {Molkov} S.~V.,   {Sunyaev} R.~A.,  2005a, The Astronomer's
  Telegram, \href {https://ui.adsabs.harvard.edu/abs/2005ATel..444....1G} {444,
  1}

\bibitem[\protect\citeauthoryear{{Grebenev}, {Bird}, {Molkov}, {Soldi},
  {Kretschmar}, {Diehl}, {Budz-Joergensen}  \& {McBreen}}{{Grebenev}
  et~al.}{2005b}]{2005ATel..457....1G}
{Grebenev} S.~A.,  {Bird} A.~J.,  {Molkov} S.~V.,  {Soldi} S.,  {Kretschmar}
  P.,  {Diehl} R.,  {Budz-Joergensen} C.,   {McBreen} B.,  2005b, The
  Astronomer's Telegram, \href
  {https://ui.adsabs.harvard.edu/abs/2005ATel..457....1G} {457, 1}

\bibitem[\protect\citeauthoryear{{Grebenev}, {Molkov}  \& {Sunyaev}}{{Grebenev}
  et~al.}{2005c}]{2005ATel..467....1G}
{Grebenev} S.~A.,  {Molkov} S.~V.,   {Sunyaev} R.~A.,  2005c, The Astronomer's
  Telegram, \href {https://ui.adsabs.harvard.edu/abs/2005ATel..467....1G} {467,
  1}

\bibitem[\protect\citeauthoryear{{Grebenev}, {Molkov}, {Revnivtsev}  \&
  {Sunyaev}}{{Grebenev} et~al.}{2007a}]{2007ESASP.622..373G}
{Grebenev} S.~A.,  {Molkov} S.~V.,  {Revnivtsev} M.~G.,   {Sunyaev} R.~A.,
  2007a, in The Obscured Universe. Proceedings of the VI INTEGRAL Workshop. p.
  3730376 (\mn@eprint {arXiv} {0709.2313})

\bibitem[\protect\citeauthoryear{{Grebenev}, {Revnivtsev}  \&
  {Sunyaev}}{{Grebenev} et~al.}{2007b}]{2007ATel.1319....1G}
{Grebenev} S.~A.,  {Revnivtsev} M.~G.,   {Sunyaev} R.~A.,  2007b, The
  Astronomer's Telegram, \href
  {https://ui.adsabs.harvard.edu/abs/2007ATel.1319....1G} {1319, 1}

\bibitem[\protect\citeauthoryear{{Grebenev}, {Lutovinov}, {Tsygankov}  \&
  {Mereminskiy}}{{Grebenev} et~al.}{2013}]{2013MNRAS.428...50G}
{Grebenev} S.~A.,  {Lutovinov} A.~A.,  {Tsygankov} S.~S.,   {Mereminskiy}
  I.~A.,  2013, \mn@doi [\mnras] {10.1093/mnras/sts008}, \href
  {https://ui.adsabs.harvard.edu/abs/2013MNRAS.428...50G} {428, 50}

\bibitem[\protect\citeauthoryear{{Grebenev}, {Mereminskiy}, {Prosvetov},
  {Ducci}, {Bozzo}, {Savchenko}  \& {Ferrigno}}{{Grebenev}
  et~al.}{2018}]{2018ATel11306....1G}
{Grebenev} S.~A.,  {Mereminskiy} I.~A.,  {Prosvetov} A.~V.,  {Ducci} L.,
  {Bozzo} E.,  {Savchenko} V.,   {Ferrigno} C.,  2018, The Astronomer's
  Telegram, \href {https://ui.adsabs.harvard.edu/abs/2018ATel11306....1G}
  {11306, 1}

\bibitem[\protect\citeauthoryear{{Grebenev}, {Mereminsky}, {Bozzo}, {Ferrigno},
  {Savchenko}  \& {Ducci}}{{Grebenev} et~al.}{2019}]{2019ATel13155....1G}
{Grebenev} S.~A.,  {Mereminsky} I.~A.,  {Bozzo} E.,  {Ferrigno} C.,
  {Savchenko} V.,   {Ducci} L.,  2019, The Astronomer's Telegram, \href
  {https://ui.adsabs.harvard.edu/abs/2019ATel13155....1G} {13155, 1}

\bibitem[\protect\citeauthoryear{{Greiss}, {Steeghs}, {Maccarone}, {Jonker},
  {Torres}, {Gonzalez}, {Masetti}  \& {Rojas}}{{Greiss}
  et~al.}{2011}]{2011ATel.3562....1G}
{Greiss} S.,  {Steeghs} D.,  {Maccarone} T.,  {Jonker} P.~G.,  {Torres}
  M.~A.~P.,  {Gonzalez} O.,  {Masetti} N.,   {Rojas} A.,  2011, The
  Astronomer's Telegram, \href
  {https://ui.adsabs.harvard.edu/abs/2011ATel.3562....1G} {3562, 1}

\bibitem[\protect\citeauthoryear{{Gros}, {Goldwurm}, {Cadolle-Bel}, {Goldoni},
  {Rodriguez}, {Foschini}, {Del Santo}  \& {Blay}}{{Gros}
  et~al.}{2003}]{2003A&A...411L.179G}
{Gros} A.,  {Goldwurm} A.,  {Cadolle-Bel} M.,  {Goldoni} P.,  {Rodriguez} J.,
  {Foschini} L.,  {Del Santo} M.,   {Blay} P.,  2003, \mn@doi [\aap]
  {10.1051/0004-6361:20031584}, \href
  {https://ui.adsabs.harvard.edu/abs/2003A&A...411L.179G} {411, L179}

\bibitem[\protect\citeauthoryear{{Halpern}}{{Halpern}}{2006}]{2006ATel..847....1H}
{Halpern} J.~P.,  2006, The Astronomer's Telegram, \href
  {https://ui.adsabs.harvard.edu/abs/2006ATel..847....1H} {847, 1}

\bibitem[\protect\citeauthoryear{{Halpern} \& {Gotthelf}}{{Halpern} \&
  {Gotthelf}}{2008}]{2008ATel.1457....1H}
{Halpern} J.~P.,  {Gotthelf} E.~V.,  2008, The Astronomer's Telegram, \href
  {https://ui.adsabs.harvard.edu/abs/2008ATel.1457....1H} {1457, 1}

\bibitem[\protect\citeauthoryear{{Halpern} \& {Thorstensen}}{{Halpern} \&
  {Thorstensen}}{2018}]{2018ATel11787....1H}
{Halpern} J.~P.,  {Thorstensen} J.~R.,  2018, The Astronomer's Telegram, \href
  {https://ui.adsabs.harvard.edu/abs/2018ATel11787....1H} {11787, 1}

\bibitem[\protect\citeauthoryear{{Halpern} \& {Tyagi}}{{Halpern} \&
  {Tyagi}}{2005}]{2005ATel..681....1H}
{Halpern} J.~P.,  {Tyagi} S.,  2005, The Astronomer's Telegram, \href
  {https://ui.adsabs.harvard.edu/abs/2005ATel..681....1H} {681, 1}

\bibitem[\protect\citeauthoryear{{Halpern}, {Thorstensen}, {Cho}, {Collver},
  {Motsoaledi}, {Breytenbach}, {Buckley}  \& {Woudt}}{{Halpern}
  et~al.}{2018}]{2018AJ....155..247H}
{Halpern} J.~P.,  {Thorstensen} J.~R.,  {Cho} P.,  {Collver} G.,  {Motsoaledi}
  M.,  {Breytenbach} H.,  {Buckley} D. A.~H.,   {Woudt} P.~A.,  2018, \mn@doi
  [\aj] {10.3847/1538-3881/aabfd0}, \href
  {https://ui.adsabs.harvard.edu/abs/2018AJ....155..247H} {155, 247}

\bibitem[\protect\citeauthoryear{{Hannikainen}, {Rodriguez}  \&
  {Pottschmidt}}{{Hannikainen} et~al.}{2003}]{2003IAUC.8088....4H}
{Hannikainen} D.~C.,  {Rodriguez} J.,   {Pottschmidt} K.,  2003, \iaucirc,
  \href {https://ui.adsabs.harvard.edu/abs/2003IAUC.8088....4H} {8088, 4}

\bibitem[\protect\citeauthoryear{{Hare}, {Halpern}, {Clavel}, {Grindlay},
  {Rahoui}  \& {Tomsick}}{{Hare} et~al.}{2019}]{2019ApJ...878...15H}
{Hare} J.,  {Halpern} J.~P.,  {Clavel} M.,  {Grindlay} J.~E.,  {Rahoui} F.,
  {Tomsick} J.~A.,  2019, \mn@doi [\apj] {10.3847/1538-4357/ab1cbe}, \href
  {https://ui.adsabs.harvard.edu/abs/2019ApJ...878...15H} {878, 15}

\bibitem[\protect\citeauthoryear{{Hare}, {Halpern}, {Tomsick}, {Thorstensen},
  {Bodaghee}, {Clavel}, {Krivonos}  \& {Mori}}{{Hare}
  et~al.}{2021}]{2021ApJ...914...85H}
{Hare} J.,  {Halpern} J.~P.,  {Tomsick} J.~A.,  {Thorstensen} J.~R.,
  {Bodaghee} A.,  {Clavel} M.,  {Krivonos} R.,   {Mori} K.,  2021, \mn@doi
  [\apj] {10.3847/1538-4357/abfa96}, \href
  {https://ui.adsabs.harvard.edu/abs/2021ApJ...914...85H} {914, 85}

\bibitem[\protect\citeauthoryear{{Harrison} et~al.,}{{Harrison}
  et~al.}{2016}]{2016ApJ...831..185H}
{Harrison} F.~A.,  et~al., 2016, \mn@doi [\apj] {10.3847/0004-637X/831/2/185},
  \href {https://ui.adsabs.harvard.edu/abs/2016ApJ...831..185H} {831, 185}

\bibitem[\protect\citeauthoryear{{Heinke}, {Tomsick}, {Yusef-Zadeh}  \&
  {Grindlay}}{{Heinke} et~al.}{2009}]{2009ApJ...701.1627H}
{Heinke} C.~O.,  {Tomsick} J.~A.,  {Yusef-Zadeh} F.,   {Grindlay} J.~E.,  2009,
  \mn@doi [\apj] {10.1088/0004-637X/701/2/1627}, \href
  {https://ui.adsabs.harvard.edu/abs/2009ApJ...701.1627H} {701, 1627}

\bibitem[\protect\citeauthoryear{{Heinke} et~al.,}{{Heinke}
  et~al.}{2019}]{2019ATel12843....1H}
{Heinke} C.~O.,  et~al., 2019, The Astronomer's Telegram, \href
  {https://ui.adsabs.harvard.edu/abs/2019ATel12843....1H} {12843, 1}

\bibitem[\protect\citeauthoryear{{Hiemstra}, {M{\'e}ndez}, {Done}, {D{\'\i}az
  Trigo}, {Altamirano}  \& {Casella}}{{Hiemstra}
  et~al.}{2011}]{2011MNRAS.411..137H}
{Hiemstra} B.,  {M{\'e}ndez} M.,  {Done} C.,  {D{\'\i}az Trigo} M.,
  {Altamirano} D.,   {Casella} P.,  2011, \mn@doi [\mnras]
  {10.1111/j.1365-2966.2010.17661.x}, \href
  {https://ui.adsabs.harvard.edu/abs/2011MNRAS.411..137H} {411, 137}

\bibitem[\protect\citeauthoryear{{Huchra} et~al.,}{{Huchra}
  et~al.}{2012}]{2012ApJS..199...26H}
{Huchra} J.~P.,  et~al., 2012, \mn@doi [\apjs] {10.1088/0067-0049/199/2/26},
  \href {https://ui.adsabs.harvard.edu/abs/2012ApJS..199...26H} {199, 26}

\bibitem[\protect\citeauthoryear{{Jauncey}}{{Jauncey}}{1967}]{1967Natur.216..877J}
{Jauncey} D.~L.,  1967, \mn@doi [\nat] {10.1038/216877a0}, \href
  {https://ui.adsabs.harvard.edu/abs/1967Natur.216..877J} {216, 877}

\bibitem[\protect\citeauthoryear{{Kaaret}, {Morgan}, {Vanderspek}  \&
  {Tomsick}}{{Kaaret} et~al.}{2006}]{2006ApJ...638..963K}
{Kaaret} P.,  {Morgan} E.~H.,  {Vanderspek} R.,   {Tomsick} J.~A.,  2006,
  \mn@doi [\apj] {10.1086/498886}, \href
  {https://ui.adsabs.harvard.edu/abs/2006ApJ...638..963K} {638, 963}

\bibitem[\protect\citeauthoryear{{Kalamkar}, {Homan}, {Altamirano}, {van der
  Klis}, {Casella}  \& {Linares}}{{Kalamkar}
  et~al.}{2011}]{2011ApJ...731L...2K}
{Kalamkar} M.,  {Homan} J.,  {Altamirano} D.,  {van der Klis} M.,  {Casella}
  P.,   {Linares} M.,  2011, \mn@doi [\apjl] {10.1088/2041-8205/731/1/L2},
  \href {https://ui.adsabs.harvard.edu/abs/2011ApJ...731L...2K} {731, L2}

\bibitem[\protect\citeauthoryear{{Kalemci}, {Tomsick}, {Rothschild},
  {Pottschmidt}, {Corbel}  \& {Kaaret}}{{Kalemci}
  et~al.}{2006}]{2006ApJ...639..340K}
{Kalemci} E.,  {Tomsick} J.~A.,  {Rothschild} R.~E.,  {Pottschmidt} K.,
  {Corbel} S.,   {Kaaret} P.,  2006, \mn@doi [\apj] {10.1086/499222}, \href
  {https://ui.adsabs.harvard.edu/abs/2006ApJ...639..340K} {639, 340}

\bibitem[\protect\citeauthoryear{{Karasev}, {Lutovinov}  \&
  {Grebenev}}{{Karasev} et~al.}{2007}]{2007AstL...33..159K}
{Karasev} D.~I.,  {Lutovinov} A.~A.,   {Grebenev} S.~A.,  2007, \mn@doi
  [Astronomy Letters] {10.1134/S1063773707030036}, \href
  {https://ui.adsabs.harvard.edu/abs/2007AstL...33..159K} {33, 159}

\bibitem[\protect\citeauthoryear{{Karasev}, {Lutovinov}  \&
  {Burenin}}{{Karasev} et~al.}{2008}]{2008AstL...34..753K}
{Karasev} D.~I.,  {Lutovinov} A.~A.,   {Burenin} R.~A.,  2008, \mn@doi
  [Astronomy Letters] {10.1134/S1063773708110042}, \href
  {https://ui.adsabs.harvard.edu/abs/2008AstL...34..753K} {34, 753}

\bibitem[\protect\citeauthoryear{{Karasev}, {Lutovinov}  \&
  {Burenin}}{{Karasev} et~al.}{2010}]{2010MNRAS.409L..69K}
{Karasev} D.~I.,  {Lutovinov} A.~A.,   {Burenin} R.~A.,  2010, \mn@doi [\mnras]
  {10.1111/j.1745-3933.2010.00949.x}, \href
  {https://ui.adsabs.harvard.edu/abs/2010MNRAS.409L..69K} {409, L69}

\bibitem[\protect\citeauthoryear{{Karasev}, {Lutovinov}, {Revnivtsev}  \&
  {Krivonos}}{{Karasev} et~al.}{2012}]{2012AstL...38..629K}
{Karasev} D.~I.,  {Lutovinov} A.~A.,  {Revnivtsev} M.~G.,   {Krivonos} R.~A.,
  2012, \mn@doi [Astronomy Letters] {10.1134/S1063773712100039}, \href
  {https://ui.adsabs.harvard.edu/abs/2012AstL...38..629K} {38, 629}

\bibitem[\protect\citeauthoryear{{Karasev} et~al.,}{{Karasev}
  et~al.}{2018}]{2018AstL...44..522K}
{Karasev} D.~I.,  et~al., 2018, \mn@doi [Astronomy Letters]
  {10.1134/S1063773718090037}, \href
  {https://ui.adsabs.harvard.edu/abs/2018AstL...44..522K} {44, 522}

\bibitem[\protect\citeauthoryear{{Karasev} et~al.,}{{Karasev}
  et~al.}{2020}]{2020AstL...45..836K}
{Karasev} D.~I.,  et~al., 2020, \mn@doi [Astronomy Letters]
  {10.1134/S1063773719120028}, \href
  {https://ui.adsabs.harvard.edu/abs/2020AstL...45..836K} {45, 836}

\bibitem[\protect\citeauthoryear{{Kaur}, {Wijnands}, {Paul}, {Patruno}  \&
  {Degenaar}}{{Kaur} et~al.}{2010}]{2010MNRAS.402.2388K}
{Kaur} R.,  {Wijnands} R.,  {Paul} B.,  {Patruno} A.,   {Degenaar} N.,  2010,
  \mn@doi [\mnras] {10.1111/j.1365-2966.2009.15919.x}, \href
  {https://ui.adsabs.harvard.edu/abs/2010MNRAS.402.2388K} {402, 2388}

\bibitem[\protect\citeauthoryear{{Kaur}, {Kotulla}, {Degenaar}, {Wijnands}  \&
  {Kaplan}}{{Kaur} et~al.}{2011}]{2011ATel.3268....1K}
{Kaur} R.,  {Kotulla} R.,  {Degenaar} N.,  {Wijnands} R.,   {Kaplan} D.,  2011,
  The Astronomer's Telegram, \href
  {https://ui.adsabs.harvard.edu/abs/2011ATel.3268....1K} {3268, 1}

\bibitem[\protect\citeauthoryear{{Kaur}, {Wijnands}, {Kamble}, {Cackett},
  {Kutulla}, {Kaplan}  \& {Degenaar}}{{Kaur}
  et~al.}{2017}]{2017MNRAS.464..170K}
{Kaur} R.,  {Wijnands} R.,  {Kamble} A.,  {Cackett} E.~M.,  {Kutulla} R.,
  {Kaplan} D.,   {Degenaar} N.,  2017, \mn@doi [\mnras]
  {10.1093/mnras/stw2319}, \href
  {https://ui.adsabs.harvard.edu/abs/2017MNRAS.464..170K} {464, 170}

\bibitem[\protect\citeauthoryear{{Keek}, {Kuiper}  \& {Hermsen}}{{Keek}
  et~al.}{2006}]{2006ATel..810....1K}
{Keek} S.,  {Kuiper} L.,   {Hermsen} W.,  2006, The Astronomer's Telegram,
  \href {https://ui.adsabs.harvard.edu/abs/2006ATel..810....1K} {810, 1}

\bibitem[\protect\citeauthoryear{{Kennea} et~al.,}{{Kennea}
  et~al.}{2019}]{2019ATel13195....1K}
{Kennea} J.~A.,  et~al., 2019, The Astronomer's Telegram, \href
  {https://ui.adsabs.harvard.edu/abs/2019ATel13195....1K} {13195, 1}

\bibitem[\protect\citeauthoryear{{Kniazev}, {Revnivtsev}, {Burenin}  \&
  {Tkachenko}}{{Kniazev} et~al.}{2010}]{2010ATel.2457....1K}
{Kniazev} A.,  {Revnivtsev} M.,  {Burenin} R.,   {Tkachenko} A.,  2010, The
  Astronomer's Telegram, \href
  {https://ui.adsabs.harvard.edu/abs/2010ATel.2457....1K} {2457, 1}

\bibitem[\protect\citeauthoryear{{Koss}, {Mushotzky}, {Veilleux}, {Winter},
  {Baumgartner}, {Tueller}, {Gehrels}  \& {Valencic}}{{Koss}
  et~al.}{2011}]{2011ApJ...739...57K}
{Koss} M.,  {Mushotzky} R.,  {Veilleux} S.,  {Winter} L.~M.,  {Baumgartner} W.,
   {Tueller} J.,  {Gehrels} N.,   {Valencic} L.,  2011, \mn@doi [\apj]
  {10.1088/0004-637X/739/2/57}, \href
  {https://ui.adsabs.harvard.edu/abs/2011ApJ...739...57K} {739, 57}

\bibitem[\protect\citeauthoryear{{Koss} et~al.,}{{Koss}
  et~al.}{2017}]{2017ApJ...850...74K}
{Koss} M.,  et~al., 2017, \mn@doi [\apj] {10.3847/1538-4357/aa8ec9}, \href
  {https://ui.adsabs.harvard.edu/abs/2017ApJ...850...74K} {850, 74}

\bibitem[\protect\citeauthoryear{{Kretschmar}, {Mereghetti}, {Hermsen},
  {Ubertini}, {Winkler}, {Brandt}  \& {Diehl}}{{Kretschmar}
  et~al.}{2004}]{2004ATel..345....1K}
{Kretschmar} P.,  {Mereghetti} S.,  {Hermsen} W.,  {Ubertini} P.,  {Winkler}
  C.,  {Brandt} S.,   {Diehl} R.,  2004, The Astronomer's Telegram, \href
  {https://ui.adsabs.harvard.edu/abs/2004ATel..345....1K} {345, 1}

\bibitem[\protect\citeauthoryear{{Kretschmar} et~al.,}{{Kretschmar}
  et~al.}{2019}]{2019NewAR..8601546K}
{Kretschmar} P.,  et~al., 2019, \mn@doi [\nar] {10.1016/j.newar.2020.101546},
  \href {https://ui.adsabs.harvard.edu/abs/2019NewAR..8601546K} {86, 101546}

\bibitem[\protect\citeauthoryear{{Krimm}, {Tomsick}, {Markwardt}, {Brocksopp},
  {Gris{\'e}}, {Kaaret}  \& {Romano}}{{Krimm}
  et~al.}{2011}]{2011ApJ...735..104K}
{Krimm} H.~A.,  {Tomsick} J.~A.,  {Markwardt} C.~B.,  {Brocksopp} C.,
  {Gris{\'e}} F.,  {Kaaret} P.,   {Romano} P.,  2011, \mn@doi [\apj]
  {10.1088/0004-637X/735/2/104}, \href
  {https://ui.adsabs.harvard.edu/abs/2011ApJ...735..104K} {735, 104}

\bibitem[\protect\citeauthoryear{{Krimm} et~al.,}{{Krimm}
  et~al.}{2012}]{2012ATel.4130....1K}
{Krimm} H.~A.,  et~al., 2012, The Astronomer's Telegram, \href
  {https://ui.adsabs.harvard.edu/abs/2012ATel.4130....1K} {4130, 1}

\bibitem[\protect\citeauthoryear{{Krimm} et~al.,}{{Krimm}
  et~al.}{2013a}]{2013ApJS..209...14K}
{Krimm} H.~A.,  et~al., 2013a, \mn@doi [\apjs] {10.1088/0067-0049/209/1/14},
  \href {https://ui.adsabs.harvard.edu/abs/2013ApJS..209...14K} {209, 14}

\bibitem[\protect\citeauthoryear{{Krimm} et~al.,}{{Krimm}
  et~al.}{2013b}]{2013ATel.4769....1K}
{Krimm} H.~A.,  et~al., 2013b, The Astronomer's Telegram, \href
  {https://ui.adsabs.harvard.edu/abs/2013ATel.4769....1K} {4769, 1}

\bibitem[\protect\citeauthoryear{{Krivonos}, {Vikhlinin}, {Churazov},
  {Lutovinov}, {Molkov}  \& {Sunyaev}}{{Krivonos}
  et~al.}{2005}]{2005ApJ...625...89K}
{Krivonos} R.,  {Vikhlinin} A.,  {Churazov} E.,  {Lutovinov} A.,  {Molkov} S.,
   {Sunyaev} R.,  2005, \mn@doi [\apj] {10.1086/429657}, \href
  {https://ui.adsabs.harvard.edu/abs/2005ApJ...625...89K} {625, 89}

\bibitem[\protect\citeauthoryear{{Krivonos}, {Revnivtsev}, {Churazov},
  {Sazonov}, {Grebenev}  \& {Sunyaev}}{{Krivonos}
  et~al.}{2007a}]{2007A&A...463..957K}
{Krivonos} R.,  {Revnivtsev} M.,  {Churazov} E.,  {Sazonov} S.,  {Grebenev} S.,
    {Sunyaev} R.,  2007a, \mn@doi [\aap] {10.1051/0004-6361:20065626}, \href
  {https://ui.adsabs.harvard.edu/abs/2007A&A...463..957K} {463, 957}

\bibitem[\protect\citeauthoryear{{Krivonos}, {Revnivtsev}, {Lutovinov},
  {Sazonov}, {Churazov}  \& {Sunyaev}}{{Krivonos}
  et~al.}{2007b}]{2007AandA...475..775K}
{Krivonos} R.,  {Revnivtsev} M.,  {Lutovinov} A.,  {Sazonov} S.,  {Churazov}
  E.,   {Sunyaev} R.,  2007b, \mn@doi [\aap] {10.1051/0004-6361:20077191},
  \href {https://ui.adsabs.harvard.edu/abs/2007A&A...475..775K} {475, 775}

\bibitem[\protect\citeauthoryear{{Krivonos}, {Revnivtsev}, {Lutovinov},
  {Sazonov}, {Churazov}  \& {Sunyaev}}{{Krivonos}
  et~al.}{2007c}]{2007AA...475..775K}
{Krivonos} R.,  {Revnivtsev} M.,  {Lutovinov} A.,  {Sazonov} S.,  {Churazov}
  E.,   {Sunyaev} R.,  2007c, \mn@doi [\aap] {10.1051/0004-6361:20077191},
  \href {https://ui.adsabs.harvard.edu/abs/2007A&A...475..775K} {475, 775}

\bibitem[\protect\citeauthoryear{{Krivonos}, {Tsygankov}, {Sunyaev},
  {Melnikov}, {Bikmaev}, {Pavlinsky}  \& {Burenin}}{{Krivonos}
  et~al.}{2009}]{2009ATel.2170....1K}
{Krivonos} R.,  {Tsygankov} S.,  {Sunyaev} R.,  {Melnikov} S.,  {Bikmaev} I.,
  {Pavlinsky} M.,   {Burenin} R.,  2009, The Astronomer's Telegram, \href
  {https://ui.adsabs.harvard.edu/abs/2009ATel.2170....1K} {2170, 1}

\bibitem[\protect\citeauthoryear{{Krivonos}, {Revnivtsev}, {Tsygankov},
  {Sazonov}, {Vikhlinin}, {Pavlinsky}, {Churazov}  \& {Sunyaev}}{{Krivonos}
  et~al.}{2010a}]{2010A&A...519A.107K}
{Krivonos} R.,  {Revnivtsev} M.,  {Tsygankov} S.,  {Sazonov} S.,  {Vikhlinin}
  A.,  {Pavlinsky} M.,  {Churazov} E.,   {Sunyaev} R.,  2010a, \mn@doi [\aap]
  {10.1051/0004-6361/200913814}, \href
  {https://ui.adsabs.harvard.edu/abs/2010A&A...519A.107K} {519, A107}

\bibitem[\protect\citeauthoryear{{Krivonos}, {Tsygankov}, {Revnivtsev},
  {Grebenev}, {Churazov}  \& {Sunyaev}}{{Krivonos}
  et~al.}{2010b}]{2010AandA...523A..61K}
{Krivonos} R.,  {Tsygankov} S.,  {Revnivtsev} M.,  {Grebenev} S.,  {Churazov}
  E.,   {Sunyaev} R.,  2010b, \mn@doi [\aap] {10.1051/0004-6361/201014935},
  \href {https://ui.adsabs.harvard.edu/abs/2010A&A...523A..61K} {523, A61}

\bibitem[\protect\citeauthoryear{{Krivonos}, {Tsygankov}, {Revnivtsev},
  {Grebenev}, {Churazov}  \& {Sunyaev}}{{Krivonos}
  et~al.}{2010c}]{2010AA...523A..61K}
{Krivonos} R.,  {Tsygankov} S.,  {Revnivtsev} M.,  {Grebenev} S.,  {Churazov}
  E.,   {Sunyaev} R.,  2010c, \mn@doi [\aap] {10.1051/0004-6361/201014935},
  \href {https://ui.adsabs.harvard.edu/abs/2010A&A...523A..61K} {523, A61}

\bibitem[\protect\citeauthoryear{{Krivonos}, {Tsygankov}, {Burenin},
  {Revnivtsev}  \& {Lutovinov}}{{Krivonos} et~al.}{2011}]{2011ATel.3382....1K}
{Krivonos} R.,  {Tsygankov} S.,  {Burenin} R.,  {Revnivtsev} M.,   {Lutovinov}
  A.,  2011, The Astronomer's Telegram, \href
  {https://ui.adsabs.harvard.edu/abs/2011ATel.3382....1K} {3382, 1}

\bibitem[\protect\citeauthoryear{{Krivonos}, {Tsygankov}, {Lutovinov},
  {Revnivtsev}, {Churazov}  \& {Sunyaev}}{{Krivonos}
  et~al.}{2012}]{2012AandA...545A..27K}
{Krivonos} R.,  {Tsygankov} S.,  {Lutovinov} A.,  {Revnivtsev} M.,  {Churazov}
  E.,   {Sunyaev} R.,  2012, \mn@doi [\aap] {10.1051/0004-6361/201219617},
  \href {https://ui.adsabs.harvard.edu/abs/2012A&A...545A..27K} {545, A27}

\bibitem[\protect\citeauthoryear{{Krivonos}, {Lutovinov}, {Molkov},
  {Revnivtsev}, {Tsygankov}  \& {Sunyaev}}{{Krivonos}
  et~al.}{2013}]{2013ATel.4924....1K}
{Krivonos} R.,  {Lutovinov} A.,  {Molkov} S.,  {Revnivtsev} M.,  {Tsygankov}
  S.,   {Sunyaev} R.,  2013, The Astronomer's Telegram, \href
  {https://ui.adsabs.harvard.edu/abs/2013ATel.4924....1K} {4924, 1}

\bibitem[\protect\citeauthoryear{{Krivonos}, {Tsygankov}, {Lutovinov},
  {Revnivtsev}, {Churazov}  \& {Sunyaev}}{{Krivonos}
  et~al.}{2015}]{2015MNRAS.448.3766K}
{Krivonos} R.,  {Tsygankov} S.,  {Lutovinov} A.,  {Revnivtsev} M.,  {Churazov}
  E.,   {Sunyaev} R.,  2015, \mn@doi [\mnras] {10.1093/mnras/stv150}, \href
  {https://ui.adsabs.harvard.edu/abs/2015MNRAS.448.3766K} {448, 3766}

\bibitem[\protect\citeauthoryear{{Krivonos}, {Tsygankov}, {Mereminskiy},
  {Lutovinov}, {Sazonov}  \& {Sunyaev}}{{Krivonos}
  et~al.}{2017}]{2017MNRAS.470..512K}
{Krivonos} R.~A.,  {Tsygankov} S.~S.,  {Mereminskiy} I.~A.,  {Lutovinov} A.~A.,
   {Sazonov} S.~Y.,   {Sunyaev} R.~A.,  2017, \mn@doi [\mnras]
  {10.1093/mnras/stx1276}, \href
  {https://ui.adsabs.harvard.edu/abs/2017MNRAS.470..512K} {470, 512}

\bibitem[\protect\citeauthoryear{{Krivonos}, {Sazonov}, {Tsygankov}  \&
  {Poutanen}}{{Krivonos} et~al.}{2018}]{2018MNRAS.480.2357K}
{Krivonos} R.,  {Sazonov} S.,  {Tsygankov} S.~S.,   {Poutanen} J.,  2018,
  \mn@doi [\mnras] {10.1093/mnras/sty1995}, \href
  {https://ui.adsabs.harvard.edu/abs/2018MNRAS.480.2357K} {480, 2357}

\bibitem[\protect\citeauthoryear{{Krivonos} et~al.,}{{Krivonos}
  et~al.}{2021}]{2021NewAR..9201612K}
{Krivonos} R.~A.,  et~al., 2021, \mn@doi [\nar] {10.1016/j.newar.2021.101612},
  \href {https://ui.adsabs.harvard.edu/abs/2021NewAR..9201612K} {92, 101612}

\bibitem[\protect\citeauthoryear{{Kuiper} \& {Hermsen}}{{Kuiper} \&
  {Hermsen}}{2009}]{2009A&A...501.1031K}
{Kuiper} L.,  {Hermsen} W.,  2009, \mn@doi [\aap]
  {10.1051/0004-6361/200811580}, \href
  {https://ui.adsabs.harvard.edu/abs/2009A&A...501.1031K} {501, 1031}

\bibitem[\protect\citeauthoryear{{Kuiper}, {Hermsen}, {in't Zand}  \& {den
  Hartog}}{{Kuiper} et~al.}{2005}]{2005ATel..662....1K}
{Kuiper} L.,  {Hermsen} W.,  {in't Zand} J.,   {den Hartog} P.~R.,  2005, The
  Astronomer's Telegram, \href
  {https://ui.adsabs.harvard.edu/abs/2005ATel..662....1K} {662, 1}

\bibitem[\protect\citeauthoryear{{Kuiper}, {Keek}, {Hermsen}, {Jonker}  \&
  {Steeghs}}{{Kuiper} et~al.}{2006a}]{2006ATel..684....1K}
{Kuiper} L.,  {Keek} S.,  {Hermsen} W.,  {Jonker} P.~G.,   {Steeghs} D.,
  2006a, The Astronomer's Telegram, \href
  {https://ui.adsabs.harvard.edu/abs/2006ATel..684....1K} {684, 1}

\bibitem[\protect\citeauthoryear{{Kuiper}, {den Hartog}  \& {Hermsen}}{{Kuiper}
  et~al.}{2006b}]{2006ATel..939....1K}
{Kuiper} L.,  {den Hartog} P.~R.,   {Hermsen} W.,  2006b, The Astronomer's
  Telegram, \href {https://ui.adsabs.harvard.edu/abs/2006ATel..939....1K} {939,
  1}

\bibitem[\protect\citeauthoryear{{Kuiper}, {Jonker}, {Torres}, {Rest}  \&
  {Keek}}{{Kuiper} et~al.}{2008}]{2008ATel.1774....1K}
{Kuiper} L.,  {Jonker} P.~G.,  {Torres} M.~A.~P.,  {Rest} A.,   {Keek} S.,
  2008, The Astronomer's Telegram, \href
  {https://ui.adsabs.harvard.edu/abs/2008ATel.1774....1K} {1774, 1}

\bibitem[\protect\citeauthoryear{{Kuulkers}, {Lutovinov}, {Parmar},
  {Capitanio}, {Mowlavi}  \& {Hermsen}}{{Kuulkers}
  et~al.}{2003}]{2003ATel..149....1K}
{Kuulkers} E.,  {Lutovinov} A.,  {Parmar} A.,  {Capitanio} F.,  {Mowlavi} N.,
  {Hermsen} W.,  2003, The Astronomer's Telegram, \href
  {https://ui.adsabs.harvard.edu/abs/2003ATel..149....1K} {149, 1}

\bibitem[\protect\citeauthoryear{{Kuulkers} et~al.,}{{Kuulkers}
  et~al.}{2006}]{2006ATel..874....1K}
{Kuulkers} E.,  et~al., 2006, The Astronomer's Telegram, \href
  {https://ui.adsabs.harvard.edu/abs/2006ATel..874....1K} {874, 1}

\bibitem[\protect\citeauthoryear{{Kuulkers} et~al.,}{{Kuulkers}
  et~al.}{2013}]{2013ATel.4804....1K}
{Kuulkers} E.,  et~al., 2013, The Astronomer's Telegram, \href
  {https://ui.adsabs.harvard.edu/abs/2013ATel.4804....1K} {4804, 1}

\bibitem[\protect\citeauthoryear{{Kuznetsova}, {Krivonos}, {Churazov},
  {Lyskova}  \& {Lutovinov}}{{Kuznetsova} et~al.}{2019}]{2019MNRAS.489.1828K}
{Kuznetsova} E.,  {Krivonos} R.,  {Churazov} E.,  {Lyskova} N.,   {Lutovinov}
  A.,  2019, \mn@doi [\mnras] {10.1093/mnras/stz2261}, \href
  {https://ui.adsabs.harvard.edu/abs/2019MNRAS.489.1828K} {489, 1828}

\bibitem[\protect\citeauthoryear{{Kuznetsova}, {Krivonos}, {Lutovinov}  \&
  {Clavel}}{{Kuznetsova} et~al.}{2022}]{2022MNRAS.509.1605K}
{Kuznetsova} E.,  {Krivonos} R.,  {Lutovinov} A.,   {Clavel} M.,  2022, \mn@doi
  [\mnras] {10.1093/mnras/stab3004}, \href
  {https://ui.adsabs.harvard.edu/abs/2022MNRAS.509.1605K} {509, 1605}

\bibitem[\protect\citeauthoryear{{La Parola}, {Cusumano}, {Segreto},
  {D'A{\`\i}}, {Masetti}  \& {D'Elia}}{{La Parola}
  et~al.}{2013}]{2013ApJ...775L..24L}
{La Parola} V.,  {Cusumano} G.,  {Segreto} A.,  {D'A{\`\i}} A.,  {Masetti} N.,
   {D'Elia} V.,  2013, \mn@doi [\apjl] {10.1088/2041-8205/775/1/L24}, \href
  {https://ui.adsabs.harvard.edu/abs/2013ApJ...775L..24L} {775, L24}

\bibitem[\protect\citeauthoryear{{Lamperti} et~al.,}{{Lamperti}
  et~al.}{2017}]{2017MNRAS.467..540L}
{Lamperti} I.,  et~al., 2017, \mn@doi [\mnras] {10.1093/mnras/stx055}, \href
  {https://ui.adsabs.harvard.edu/abs/2017MNRAS.467..540L} {467, 540}

\bibitem[\protect\citeauthoryear{{Landi} et~al.,}{{Landi}
  et~al.}{2007}]{2007ATel.1288....1L}
{Landi} R.,  et~al., 2007, The Astronomer's Telegram, \href
  {https://ui.adsabs.harvard.edu/abs/2007ATel.1288....1L} {1288, 1}

\bibitem[\protect\citeauthoryear{{Landi} et~al.,}{{Landi}
  et~al.}{2009}]{2009A&A...493..893L}
{Landi} R.,  et~al., 2009, \mn@doi [\aap] {10.1051/0004-6361:200810503}, \href
  {https://ui.adsabs.harvard.edu/abs/2009A&A...493..893L} {493, 893}

\bibitem[\protect\citeauthoryear{{Landi}, {Bassani}, {Malizia}, {Stephen},
  {Bazzano}, {Fiocchi}  \& {Bird}}{{Landi} et~al.}{2010a}]{2010MNRAS.403..945L}
{Landi} R.,  {Bassani} L.,  {Malizia} A.,  {Stephen} J.~B.,  {Bazzano} A.,
  {Fiocchi} M.,   {Bird} A.~J.,  2010a, \mn@doi [\mnras]
  {10.1111/j.1365-2966.2010.16183.x}, \href
  {https://ui.adsabs.harvard.edu/abs/2010MNRAS.403..945L} {403, 945}

\bibitem[\protect\citeauthoryear{{Landi} et~al.,}{{Landi}
  et~al.}{2010b}]{2010ATel.2853....1L}
{Landi} R.,  et~al., 2010b, The Astronomer's Telegram, \href
  {https://ui.adsabs.harvard.edu/abs/2010ATel.2853....1L} {2853, 1}

\bibitem[\protect\citeauthoryear{{Landi}, {Malizia}, {Bazzano}, {Fiocchi},
  {Bird}  \& {Gehrels}}{{Landi} et~al.}{2011a}]{2011ATel.3184....1L}
{Landi} R.,  {Malizia} A.,  {Bazzano} A.,  {Fiocchi} M.,  {Bird} A.~J.,
  {Gehrels} N.,  2011a, The Astronomer's Telegram, \href
  {https://ui.adsabs.harvard.edu/abs/2011ATel.3184....1L} {3184, 1}

\bibitem[\protect\citeauthoryear{{Landi}, {Bassani}, {Masetti}, {Bazzano}  \&
  {Bird}}{{Landi} et~al.}{2011b}]{2011ATel.3271....1L}
{Landi} R.,  {Bassani} L.,  {Masetti} N.,  {Bazzano} A.,   {Bird} A.~J.,
  2011b, The Astronomer's Telegram, \href
  {https://ui.adsabs.harvard.edu/abs/2011ATel.3271....1L} {3271, 1}

\bibitem[\protect\citeauthoryear{{Landi}, {Bassani}, {Masetti}, {Bazzano}  \&
  {Bird}}{{Landi} et~al.}{2011c}]{2011ATel.3272....1L}
{Landi} R.,  {Bassani} L.,  {Masetti} N.,  {Bazzano} A.,   {Bird} A.~J.,
  2011c, The Astronomer's Telegram, \href
  {https://ui.adsabs.harvard.edu/abs/2011ATel.3272....1L} {3272, 1}

\bibitem[\protect\citeauthoryear{{Landi}, {Bassani}, {Masetti}, {Bazzano},
  {Tarana}  \& {Bird}}{{Landi} et~al.}{2012}]{2012ATel.4233....1L}
{Landi} R.,  {Bassani} L.,  {Masetti} N.,  {Bazzano} A.,  {Tarana} A.,   {Bird}
  A.~J.,  2012, The Astronomer's Telegram, \href
  {https://ui.adsabs.harvard.edu/abs/2012ATel.4233....1L} {4233, 1}

\bibitem[\protect\citeauthoryear{{Landi} et~al.,}{{Landi}
  et~al.}{2017}]{2017MNRAS.470.1107L}
{Landi} R.,  et~al., 2017, \mn@doi [\mnras] {10.1093/mnras/stx908}, \href
  {https://ui.adsabs.harvard.edu/abs/2017MNRAS.470.1107L} {470, 1107}

\bibitem[\protect\citeauthoryear{{Lebrun} et~al.,}{{Lebrun}
  et~al.}{2003}]{2003A&A...411L.141L}
{Lebrun} F.,  et~al., 2003, \mn@doi [\aap] {10.1051/0004-6361:20031367}, \href
  {https://ui.adsabs.harvard.edu/abs/2003A&A...411L.141L} {411, L141}

\bibitem[\protect\citeauthoryear{{Leyder}, {Walter}  \& {Rauw}}{{Leyder}
  et~al.}{2008}]{2008A&A...477L..29L}
{Leyder} J.~C.,  {Walter} R.,   {Rauw} G.,  2008, \mn@doi [\aap]
  {10.1051/0004-6361:20078981}, \href
  {https://ui.adsabs.harvard.edu/abs/2008A&A...477L..29L} {477, L29}

\bibitem[\protect\citeauthoryear{{Lubi{\'n}ski}, {Cadolle Bel}, {von Kienlin},
  {Budtz-Jorgensen}, {McBreen}, {Kretschmar}, {Hermsen}  \&
  {Shtykovsky}}{{Lubi{\'n}ski} et~al.}{2005}]{2005ATel..469....1L}
{Lubi{\'n}ski} P.,  {Cadolle Bel} M.,  {von Kienlin} A.,  {Budtz-Jorgensen} C.,
   {McBreen} B.,  {Kretschmar} P.,  {Hermsen} W.,   {Shtykovsky} P.,  2005, The
  Astronomer's Telegram, \href
  {https://ui.adsabs.harvard.edu/abs/2005ATel..469....1L} {469, 1}

\bibitem[\protect\citeauthoryear{{Lutovinov} \& {Revnivtsev}}{{Lutovinov} \&
  {Revnivtsev}}{2003}]{2003AstL...29..719L}
{Lutovinov} A.~A.,  {Revnivtsev} M.~G.,  2003, \mn@doi [Astronomy Letters]
  {10.1134/1.1624457}, \href
  {https://ui.adsabs.harvard.edu/abs/2003AstL...29..719L} {29, 719}

\bibitem[\protect\citeauthoryear{{Lutovinov}, {Shaw}, {Foschini}  \&
  {Paul}}{{Lutovinov} et~al.}{2003a}]{2003ATel..154....1L}
{Lutovinov} A.,  {Shaw} S.,  {Foschini} L.,   {Paul} J.,  2003a, The
  Astronomer's Telegram, \href
  {https://ui.adsabs.harvard.edu/abs/2003ATel..154....1L} {154, 1}

\bibitem[\protect\citeauthoryear{{Lutovinov}, {Walter}, {Belanger}, {Lund},
  {Grebenev}  \& {Winkler}}{{Lutovinov} et~al.}{2003b}]{2003ATel..155....1L}
{Lutovinov} A.,  {Walter} R.,  {Belanger} G.,  {Lund} N.,  {Grebenev} S.,
  {Winkler} C.,  2003b, The Astronomer's Telegram, \href
  {https://ui.adsabs.harvard.edu/abs/2003ATel..155....1L} {155, 1}

\bibitem[\protect\citeauthoryear{{Lutovinov}, {Cadolle Bel}, {Belanger},
  {Goldwurm}, {Budtz-Jorgensen}, {Mowlavi}, {Paul}  \& {Orr}}{{Lutovinov}
  et~al.}{2004a}]{2004ATel..328....1L}
{Lutovinov} A.,  {Cadolle Bel} M.,  {Belanger} G.,  {Goldwurm} A.,
  {Budtz-Jorgensen} C.,  {Mowlavi} N.,  {Paul} J.,   {Orr} A.,  2004a, The
  Astronomer's Telegram, \href
  {https://ui.adsabs.harvard.edu/abs/2004ATel..328....1L} {328, 1}

\bibitem[\protect\citeauthoryear{{Lutovinov}, {Rodrigues}, {Budtz-Jorgensen},
  {Grebenev}  \& {Winkler}}{{Lutovinov} et~al.}{2004b}]{2004ATel..329....1L}
{Lutovinov} A.,  {Rodrigues} J.,  {Budtz-Jorgensen} C.,  {Grebenev} S.,
  {Winkler} C.,  2004b, The Astronomer's Telegram, \href
  {https://ui.adsabs.harvard.edu/abs/2004ATel..329....1L} {329, 1}

\bibitem[\protect\citeauthoryear{{Lutovinov}, {Rodriguez}, {Revnivtsev}  \&
  {Shtykovskiy}}{{Lutovinov} et~al.}{2005a}]{2005A&A...433L..41L}
{Lutovinov} A.,  {Rodriguez} J.,  {Revnivtsev} M.,   {Shtykovskiy} P.,  2005a,
  \mn@doi [\aap] {10.1051/0004-6361:200500092}, \href
  {https://ui.adsabs.harvard.edu/abs/2005A&A...433L..41L} {433, L41}

\bibitem[\protect\citeauthoryear{{Lutovinov}, {Revnivtsev}, {Gilfanov},
  {Shtykovskiy}, {Molkov}  \& {Sunyaev}}{{Lutovinov}
  et~al.}{2005b}]{2005A&A...444..821L}
{Lutovinov} A.,  {Revnivtsev} M.,  {Gilfanov} M.,  {Shtykovskiy} P.,  {Molkov}
  S.,   {Sunyaev} R.,  2005b, \mn@doi [\aap] {10.1051/0004-6361:20042392},
  \href {https://ui.adsabs.harvard.edu/abs/2005A&A...444..821L} {444, 821}

\bibitem[\protect\citeauthoryear{{Lutovinov}, {Burenin}, {Revnivtsev},
  {Suleimanov}  \& {Tkachenko}}{{Lutovinov}
  et~al.}{2010a}]{2010AstL...36..904L}
{Lutovinov} A.~A.,  {Burenin} R.~A.,  {Revnivtsev} M.~G.,  {Suleimanov} V.~F.,
   {Tkachenko} A.~Y.,  2010a, \mn@doi [Astronomy Letters]
  {10.1134/S1063773710120042}, \href
  {https://ui.adsabs.harvard.edu/abs/2010AstL...36..904L} {36, 904}

\bibitem[\protect\citeauthoryear{{Lutovinov}, {Burenin}, {Sazonov},
  {Revnivtsev}, {Moiseev}  \& {Dodonov}}{{Lutovinov}
  et~al.}{2010b}]{2010ATel.2759....1L}
{Lutovinov} A.,  {Burenin} R.,  {Sazonov} S.,  {Revnivtsev} M.,  {Moiseev} A.,
   {Dodonov} S.,  2010b, The Astronomer's Telegram, \href
  {https://ui.adsabs.harvard.edu/abs/2010ATel.2759....1L} {2759, 1}

\bibitem[\protect\citeauthoryear{{Lutovinov}, {Burenin}, {Revnivtsev}  \&
  {Bikmaev}}{{Lutovinov} et~al.}{2012}]{2012AstL...38....1L}
{Lutovinov} A.~A.,  {Burenin} R.~A.,  {Revnivtsev} M.~G.,   {Bikmaev} I.~F.,
  2012, \mn@doi [Astronomy Letters] {10.1134/S1063773712010045}, \href
  {https://ui.adsabs.harvard.edu/abs/2012AstL...38....1L} {38, 1}

\bibitem[\protect\citeauthoryear{{Lutovinov}, {Mironov}, {Burenin},
  {Revnivtsev}, {Tsygankov}, {Pavlinsky}, {Korobtsev}  \&
  {Eselevich}}{{Lutovinov} et~al.}{2013}]{2013AstL...39..513L}
{Lutovinov} A.~A.,  {Mironov} A.~I.,  {Burenin} R.~A.,  {Revnivtsev} M.~G.,
  {Tsygankov} S.~S.,  {Pavlinsky} M.~N.,  {Korobtsev} I.~V.,   {Eselevich}
  M.~V.,  2013, \mn@doi [Astronomy Letters] {10.1134/S1063773713080069}, \href
  {https://ui.adsabs.harvard.edu/abs/2013AstL...39..513L} {39, 513}

\bibitem[\protect\citeauthoryear{{Lutovinov}, {Suleimanov}, {Manuel Luna},
  {Sazonov}, {de Martino}, {Ducci}, {Doroshenko}  \& {Falanga}}{{Lutovinov}
  et~al.}{2020}]{2020NewAR..9101547L}
{Lutovinov} A.,  {Suleimanov} V.,  {Manuel Luna} G.~J.,  {Sazonov} S.,  {de
  Martino} D.,  {Ducci} L.,  {Doroshenko} V.,   {Falanga} M.,  2020, \mn@doi
  [\nar] {10.1016/j.newar.2020.101547}, \href
  {https://ui.adsabs.harvard.edu/abs/2020NewAR..9101547L} {91, 101547}

\bibitem[\protect\citeauthoryear{{Malizia} et~al.,}{{Malizia}
  et~al.}{2005}]{2005ApJ...630L.157M}
{Malizia} A.,  et~al., 2005, \mn@doi [\apjl] {10.1086/491653}, \href
  {https://ui.adsabs.harvard.edu/abs/2005ApJ...630L.157M} {630, L157}

\bibitem[\protect\citeauthoryear{{Malizia} et~al.,}{{Malizia}
  et~al.}{2007}]{2007ApJ...668...81M}
{Malizia} A.,  et~al., 2007, \mn@doi [\apj] {10.1086/520874}, \href
  {https://ui.adsabs.harvard.edu/abs/2007ApJ...668...81M} {668, 81}

\bibitem[\protect\citeauthoryear{{Malizia}, {Bassani}, {Sguera}, {Stephen},
  {Bazzano}, {Fiocchi}  \& {Bird}}{{Malizia}
  et~al.}{2010}]{2010MNRAS.408..975M}
{Malizia} A.,  {Bassani} L.,  {Sguera} V.,  {Stephen} J.~B.,  {Bazzano} A.,
  {Fiocchi} M.,   {Bird} A.~J.,  2010, \mn@doi [\mnras]
  {10.1111/j.1365-2966.2010.17157.x}, \href
  {https://ui.adsabs.harvard.edu/abs/2010MNRAS.408..975M} {408, 975}

\bibitem[\protect\citeauthoryear{{Malizia}, {Bassani}, {Bazzano}, {Bird},
  {Masetti}, {Panessa}, {Stephen}  \& {Ubertini}}{{Malizia}
  et~al.}{2012}]{2012MNRAS.426.1750M}
{Malizia} A.,  {Bassani} L.,  {Bazzano} A.,  {Bird} A.~J.,  {Masetti} N.,
  {Panessa} F.,  {Stephen} J.~B.,   {Ubertini} P.,  2012, \mn@doi [\mnras]
  {10.1111/j.1365-2966.2012.21755.x}, \href
  {https://ui.adsabs.harvard.edu/abs/2012MNRAS.426.1750M} {426, 1750}

\bibitem[\protect\citeauthoryear{{Malizia}, {Landi}, {Molina}, {Bassani},
  {Bazzano}, {Bird}  \& {Ubertini}}{{Malizia}
  et~al.}{2016}]{2016MNRAS.460...19M}
{Malizia} A.,  {Landi} R.,  {Molina} M.,  {Bassani} L.,  {Bazzano} A.,  {Bird}
  A.~J.,   {Ubertini} P.,  2016, \mn@doi [\mnras] {10.1093/mnras/stw972}, \href
  {https://ui.adsabs.harvard.edu/abs/2016MNRAS.460...19M} {460, 19}

\bibitem[\protect\citeauthoryear{{Malizia}, {Bassani}, {Sguera}, {Bazzano},
  {Fiocchi}, {Ubertini}  \& {Bird}}{{Malizia}
  et~al.}{2017}]{2017ATel10411....1M}
{Malizia} A.,  {Bassani} L.,  {Sguera} V.,  {Bazzano} A.,  {Fiocchi} M.~T.,
  {Ubertini} P.,   {Bird} A.~J.,  2017, The Astronomer's Telegram, \href
  {https://ui.adsabs.harvard.edu/abs/2017ATel10411....1M} {10411, 1}

\bibitem[\protect\citeauthoryear{{Malizia}, {Sazonov}, {Bassani}, {Pian},
  {Beckmann}, {Molina}, {Mereminskiy}  \& {Belanger}}{{Malizia}
  et~al.}{2020}]{2020NewAR..9001545M}
{Malizia} A.,  {Sazonov} S.,  {Bassani} L.,  {Pian} E.,  {Beckmann} V.,
  {Molina} M.,  {Mereminskiy} I.,   {Belanger} G.,  2020, \mn@doi [\nar]
  {10.1016/j.newar.2020.101545}, \href
  {https://ui.adsabs.harvard.edu/abs/2020NewAR..9001545M} {90, 101545}

\bibitem[\protect\citeauthoryear{{Margutti} et~al.,}{{Margutti}
  et~al.}{2019}]{2019ApJ...872...18M}
{Margutti} R.,  et~al., 2019, \mn@doi [\apj] {10.3847/1538-4357/aafa01}, \href
  {https://ui.adsabs.harvard.edu/abs/2019ApJ...872...18M} {872, 18}

\bibitem[\protect\citeauthoryear{{Markwardt}}{{Markwardt}}{2008}]{2008ATel.1686....1M}
{Markwardt} C.~B.,  2008, The Astronomer's Telegram, \href
  {https://ui.adsabs.harvard.edu/abs/2008ATel.1686....1M} {1686, 1}

\bibitem[\protect\citeauthoryear{{Markwardt}, {Swank}  \&
  {Strohmayer}}{{Markwardt} et~al.}{2004}]{2004ATel..353....1M}
{Markwardt} C.~B.,  {Swank} J.~H.,   {Strohmayer} T.~E.,  2004, The
  Astronomer's Telegram, \href
  {https://ui.adsabs.harvard.edu/abs/2004ATel..353....1M} {353, 1}

\bibitem[\protect\citeauthoryear{{Mart{\'\i}}, {Paredes}, {Bloom}, {Casares},
  {Rib{\'o}}  \& {Falco}}{{Mart{\'\i}} et~al.}{2004}]{2004A&A...413..309M}
{Mart{\'\i}} J.,  {Paredes} J.~M.,  {Bloom} J.~S.,  {Casares} J.,  {Rib{\'o}}
  M.,   {Falco} E.~E.,  2004, \mn@doi [\aap] {10.1051/0004-6361:20031511},
  \href {https://ui.adsabs.harvard.edu/abs/2004A&A...413..309M} {413, 309}

\bibitem[\protect\citeauthoryear{{Maselli} et~al.,}{{Maselli}
  et~al.}{2013}]{2013ApJS..206...17M}
{Maselli} A.,  et~al., 2013, \mn@doi [\apjs] {10.1088/0067-0049/206/2/17},
  \href {https://ui.adsabs.harvard.edu/abs/2013ApJS..206...17M} {206, 17}

\bibitem[\protect\citeauthoryear{{Masetti}, {Palazzi}, {Bassani}, {Malizia}  \&
  {Stephen}}{{Masetti} et~al.}{2004}]{2004A&A...426L..41M}
{Masetti} N.,  {Palazzi} E.,  {Bassani} L.,  {Malizia} A.,   {Stephen} J.~B.,
  2004, \mn@doi [\aap] {10.1051/0004-6361:200400078}, \href
  {https://ui.adsabs.harvard.edu/abs/2004A&A...426L..41M} {426, L41}

\bibitem[\protect\citeauthoryear{{Masetti} et~al.,}{{Masetti}
  et~al.}{2006a}]{2006A&A...449.1139M}
{Masetti} N.,  et~al., 2006a, \mn@doi [\aap] {10.1051/0004-6361:20054332},
  \href {https://ui.adsabs.harvard.edu/abs/2006A&A...449.1139M} {449, 1139}

\bibitem[\protect\citeauthoryear{{Masetti} et~al.,}{{Masetti}
  et~al.}{2006b}]{2006A&A...455...11M}
{Masetti} N.,  et~al., 2006b, \mn@doi [\aap] {10.1051/0004-6361:20065111},
  \href {https://ui.adsabs.harvard.edu/abs/2006A&A...455...11M} {455, 11}

\bibitem[\protect\citeauthoryear{{Masetti} et~al.,}{{Masetti}
  et~al.}{2006c}]{2006A&A...459...21M}
{Masetti} N.,  et~al., 2006c, \mn@doi [\aap] {10.1051/0004-6361:20066055},
  \href {https://ui.adsabs.harvard.edu/abs/2006A&A...459...21M} {459, 21}

\bibitem[\protect\citeauthoryear{{Masetti}, {Bassani}, {Dean}, {Ubertini}  \&
  {Walter}}{{Masetti} et~al.}{2006d}]{2006ATel..715....1M}
{Masetti} N.,  {Bassani} L.,  {Dean} A.~J.,  {Ubertini} P.,   {Walter} R.,
  2006d, The Astronomer's Telegram, \href
  {https://ui.adsabs.harvard.edu/abs/2006ATel..715....1M} {715, 1}

\bibitem[\protect\citeauthoryear{{Masetti}, {Morelli}, {Palazzi}, {Stephen},
  {Bazzano}, {Dean}, {Walter}  \& {Minniti}}{{Masetti}
  et~al.}{2006e}]{2006ATel..783....1M}
{Masetti} N.,  {Morelli} L.,  {Palazzi} E.,  {Stephen} J.,  {Bazzano} A.,
  {Dean} A.~J.,  {Walter} R.,   {Minniti} D.,  2006e, The Astronomer's
  Telegram, \href {https://ui.adsabs.harvard.edu/abs/2006ATel..783....1M} {783,
  1}

\bibitem[\protect\citeauthoryear{{Masetti}, {Rigon}, {Maiorano}, {Cusumano},
  {Palazzi}, {Orlandini}, {Amati}  \& {Frontera}}{{Masetti}
  et~al.}{2007a}]{2007A&A...464..277M}
{Masetti} N.,  {Rigon} E.,  {Maiorano} E.,  {Cusumano} G.,  {Palazzi} E.,
  {Orlandini} M.,  {Amati} L.,   {Frontera} F.,  2007a, \mn@doi [\aap]
  {10.1051/0004-6361:20066517}, \href
  {https://ui.adsabs.harvard.edu/abs/2007A&A...464..277M} {464, 277}

\bibitem[\protect\citeauthoryear{{Masetti} et~al.,}{{Masetti}
  et~al.}{2007b}]{2007A&A...470..331M}
{Masetti} N.,  et~al., 2007b, \mn@doi [\aap] {10.1051/0004-6361:20077509},
  \href {https://ui.adsabs.harvard.edu/abs/2007A&A...470..331M} {470, 331}

\bibitem[\protect\citeauthoryear{{Masetti} et~al.,}{{Masetti}
  et~al.}{2007c}]{2007ATel.1034....1M}
{Masetti} N.,  et~al., 2007c, The Astronomer's Telegram, \href
  {https://ui.adsabs.harvard.edu/abs/2007ATel.1034....1M} {1034, 1}

\bibitem[\protect\citeauthoryear{{Masetti} et~al.,}{{Masetti}
  et~al.}{2008}]{2008A&A...482..113M}
{Masetti} N.,  et~al., 2008, \mn@doi [\aap] {10.1051/0004-6361:20079332}, \href
  {https://ui.adsabs.harvard.edu/abs/2008A&A...482..113M} {482, 113}

\bibitem[\protect\citeauthoryear{{Masetti} et~al.,}{{Masetti}
  et~al.}{2009}]{2009A&A...495..121M}
{Masetti} N.,  et~al., 2009, \mn@doi [\aap] {10.1051/0004-6361:200811322},
  \href {https://ui.adsabs.harvard.edu/abs/2009A&A...495..121M} {495, 121}

\bibitem[\protect\citeauthoryear{{Masetti} et~al.,}{{Masetti}
  et~al.}{2010a}]{2010A&A...511A..48M}
{Masetti} N.,  et~al., 2010a, \mn@doi [\aap] {10.1051/0004-6361/200913404},
  \href {https://ui.adsabs.harvard.edu/abs/2010A&A...511A..48M} {511, A48}

\bibitem[\protect\citeauthoryear{{Masetti} et~al.,}{{Masetti}
  et~al.}{2010b}]{2010A&A...519A..96M}
{Masetti} N.,  et~al., 2010b, \mn@doi [\aap] {10.1051/0004-6361/201014852},
  \href {https://ui.adsabs.harvard.edu/abs/2010A&A...519A..96M} {519, A96}

\bibitem[\protect\citeauthoryear{{Masetti} et~al.,}{{Masetti}
  et~al.}{2012a}]{2012A&A...538A.123M}
{Masetti} N.,  et~al., 2012a, \mn@doi [\aap] {10.1051/0004-6361/201118559},
  \href {https://ui.adsabs.harvard.edu/abs/2012A&A...538A.123M} {538, A123}

\bibitem[\protect\citeauthoryear{{Masetti}, {Nucita}  \& {Parisi}}{{Masetti}
  et~al.}{2012b}]{2012A&A...544A.114M}
{Masetti} N.,  {Nucita} A.~A.,   {Parisi} P.,  2012b, \mn@doi [\aap]
  {10.1051/0004-6361/201219334}, \href
  {https://ui.adsabs.harvard.edu/abs/2012A&A...544A.114M} {544, A114}

\bibitem[\protect\citeauthoryear{{Masetti}, {Jimenez-Bailon}, {Chavushyan},
  {Parisi}, {Bazzano}, {Landi}  \& {Bird}}{{Masetti}
  et~al.}{2012c}]{2012ATel.4248....1M}
{Masetti} N.,  {Jimenez-Bailon} E.,  {Chavushyan} V.,  {Parisi} P.,  {Bazzano}
  A.,  {Landi} R.,   {Bird} A.~J.,  2012c, The Astronomer's Telegram, \href
  {https://ui.adsabs.harvard.edu/abs/2012ATel.4248....1M} {4248, 1}

\bibitem[\protect\citeauthoryear{{Masetti} et~al.,}{{Masetti}
  et~al.}{2013}]{2013A&A...556A.120M}
{Masetti} N.,  et~al., 2013, \mn@doi [\aap] {10.1051/0004-6361/201322026},
  \href {https://ui.adsabs.harvard.edu/abs/2013A&A...556A.120M} {556, A120}

\bibitem[\protect\citeauthoryear{{McCollum} \& {Laine}}{{McCollum} \&
  {Laine}}{2019}]{2019ATel13211....1M}
{McCollum} B.,  {Laine} S.,  2019, The Astronomer's Telegram, \href
  {https://ui.adsabs.harvard.edu/abs/2019ATel13211....1M} {13211, 1}

\bibitem[\protect\citeauthoryear{{Mereminskiy}, {Krivonos}, {Lutovinov},
  {Sazonov}, {Revnivtsev}  \& {Sunyaev}}{{Mereminskiy}
  et~al.}{2016}]{2016MNRAS.459..140M}
{Mereminskiy} I.~A.,  {Krivonos} R.~A.,  {Lutovinov} A.~A.,  {Sazonov} S.~Y.,
  {Revnivtsev} M.~G.,   {Sunyaev} R.~A.,  2016, \mn@doi [\mnras]
  {10.1093/mnras/stw613}, \href
  {https://ui.adsabs.harvard.edu/abs/2016MNRAS.459..140M} {459, 140}

\bibitem[\protect\citeauthoryear{{Mescheryakov}, {Burenin}, {Sazonov},
  {Revnivtsev}, {Bikmaev}  \& {Sunyaev}}{{Mescheryakov}
  et~al.}{2006}]{2006ATel..948....1M}
{Mescheryakov} A.,  {Burenin} R.,  {Sazonov} S.,  {Revnivtsev} M.,  {Bikmaev}
  I.,   {Sunyaev} R.,  2006, The Astronomer's Telegram, \href
  {https://ui.adsabs.harvard.edu/abs/2006ATel..948....1M} {948, 1}

\bibitem[\protect\citeauthoryear{{Mescheryakov}, {Burenin}, {Sazonov},
  {Revnivtsev}, {Bikmaev}, {Pavlinsky}  \& {Sunyaev}}{{Mescheryakov}
  et~al.}{2009}]{2009ATel.2132....1M}
{Mescheryakov} A.,  {Burenin} R.,  {Sazonov} S.,  {Revnivtsev} M.,  {Bikmaev}
  I.,  {Pavlinsky} M.,   {Sunyaev} R.,  2009, The Astronomer's Telegram, \href
  {https://ui.adsabs.harvard.edu/abs/2009ATel.2132....1M} {2132, 1}

\bibitem[\protect\citeauthoryear{{Milisavljevic}, {Fesen}, {Parrent}  \&
  {Thorstensen}}{{Milisavljevic} et~al.}{2011}]{2011ATel.3146....1M}
{Milisavljevic} D.,  {Fesen} R.~A.,  {Parrent} J.~T.,   {Thorstensen} J.~R.,
  2011, The Astronomer's Telegram, \href
  {https://ui.adsabs.harvard.edu/abs/2011ATel.3146....1M} {3146, 1}

\bibitem[\protect\citeauthoryear{{Miyasaka}, {Tomsick}, {Xu}  \&
  {Harrison}}{{Miyasaka} et~al.}{2018}]{2018ATel12340....1M}
{Miyasaka} H.,  {Tomsick} J.~A.,  {Xu} Y.,   {Harrison} F.~A.,  2018, The
  Astronomer's Telegram, \href
  {https://ui.adsabs.harvard.edu/abs/2018ATel12340....1M} {12340, 1}

\bibitem[\protect\citeauthoryear{{Molina}, {Venturi}, {Malizia}, {Bassani},
  {Dallacasa}, {Lal}, {Bird}  \& {Ubertini}}{{Molina}
  et~al.}{2015}]{2015MNRAS.451.2370M}
{Molina} M.,  {Venturi} T.,  {Malizia} A.,  {Bassani} L.,  {Dallacasa} D.,
  {Lal} D.~V.,  {Bird} A.~J.,   {Ubertini} P.,  2015, \mn@doi [\mnras]
  {10.1093/mnras/stv1116}, \href
  {https://ui.adsabs.harvard.edu/abs/2015MNRAS.451.2370M} {451, 2370}

\bibitem[\protect\citeauthoryear{{Molkov}, {Mowlavi}, {Goldwurm}, {Strong},
  {Lund}, {Paul}  \& {Oosterbroek}}{{Molkov}
  et~al.}{2003}]{2003ATel..176....1M}
{Molkov} S.,  {Mowlavi} N.,  {Goldwurm} A.,  {Strong} A.,  {Lund} N.,  {Paul}
  J.,   {Oosterbroek} T.,  2003, The Astronomer's Telegram, \href
  {https://ui.adsabs.harvard.edu/abs/2003ATel..176....1M} {176, 1}

\bibitem[\protect\citeauthoryear{{Molkov}, {Cherepashchuk}, {Lutovinov},
  {Revnivtsev}, {Postnov}  \& {Sunyaev}}{{Molkov}
  et~al.}{2004}]{2004AstL...30..534M}
{Molkov} S.~V.,  {Cherepashchuk} A.~M.,  {Lutovinov} A.~A.,  {Revnivtsev}
  M.~G.,  {Postnov} K.~A.,   {Sunyaev} R.~A.,  2004, \mn@doi [Astronomy
  Letters] {10.1134/1.1784495}, \href
  {https://ui.adsabs.harvard.edu/abs/2004AstL...30..534M} {30, 534}

\bibitem[\protect\citeauthoryear{{Morelli}, {Masetti}, {Bassani}, {Landi},
  {Malizia}, {Bird}, {Ubertini}  \& {Galaz}}{{Morelli}
  et~al.}{2006}]{2006ATel..785....1M}
{Morelli} L.,  {Masetti} N.,  {Bassani} L.,  {Landi} R.,  {Malizia} A.,  {Bird}
  A.~J.,  {Ubertini} P.,   {Galaz} G.,  2006, The Astronomer's Telegram, \href
  {https://ui.adsabs.harvard.edu/abs/2006ATel..785....1M} {785, 1}

\bibitem[\protect\citeauthoryear{{Morgan}, {Swank}, {Markwardt}  \&
  {Gehrels}}{{Morgan} et~al.}{2005}]{2005ATel..550....1M}
{Morgan} E.,  {Swank} J.,  {Markwardt} C.,   {Gehrels} N.,  2005, The
  Astronomer's Telegram, \href
  {https://ui.adsabs.harvard.edu/abs/2005ATel..550....1M} {550, 1}

\bibitem[\protect\citeauthoryear{{Mori} et~al.,}{{Mori}
  et~al.}{2015}]{2015ApJ...814...94M}
{Mori} K.,  et~al., 2015, \mn@doi [\apj] {10.1088/0004-637X/814/2/94}, \href
  {https://ui.adsabs.harvard.edu/abs/2015ApJ...814...94M} {814, 94}

\bibitem[\protect\citeauthoryear{{Nabizadeh}, {Tsygankov}, {Karasev},
  {M{\"o}nkk{\"o}nen}, {Lutovinov}, {Nagirner}  \& {Poutanen}}{{Nabizadeh}
  et~al.}{2019}]{2019A&A...622A.198N}
{Nabizadeh} A.,  {Tsygankov} S.~S.,  {Karasev} D.~I.,  {M{\"o}nkk{\"o}nen} J.,
  {Lutovinov} A.~A.,  {Nagirner} D.~I.,   {Poutanen} J.,  2019, \mn@doi [\aap]
  {10.1051/0004-6361/201834635}, \href
  {https://ui.adsabs.harvard.edu/abs/2019A&A...622A.198N} {622, A198}

\bibitem[\protect\citeauthoryear{{Nakahira} et~al.,}{{Nakahira}
  et~al.}{2013}]{2013ATel.5474....1N}
{Nakahira} S.,  et~al., 2013, The Astronomer's Telegram, \href
  {https://ui.adsabs.harvard.edu/abs/2013ATel.5474....1N} {5474, 1}

\bibitem[\protect\citeauthoryear{{Natalucci}, {Fiocchi}, {Bazzano}, {Kuulkers}
  \& {Sanchez}}{{Natalucci} et~al.}{2011}]{2011ATel.3181....1N}
{Natalucci} L.,  {Fiocchi} M.,  {Bazzano} A.,  {Kuulkers} E.,   {Sanchez} C.,
  2011, The Astronomer's Telegram, \href
  {https://ui.adsabs.harvard.edu/abs/2011ATel.3181....1N} {3181, 1}

\bibitem[\protect\citeauthoryear{{Nebot G{\'o}mez-Mor{\'a}n}, {Motch},
  {Pineau}, {Carrera}, {Pakull}  \& {Riddick}}{{Nebot G{\'o}mez-Mor{\'a}n}
  et~al.}{2015}]{2015MNRAS.452..884N}
{Nebot G{\'o}mez-Mor{\'a}n} A.,  {Motch} C.,  {Pineau} F.~X.,  {Carrera} F.~J.,
   {Pakull} M.~W.,   {Riddick} F.,  2015, \mn@doi [\mnras]
  {10.1093/mnras/stv1020}, \href
  {https://ui.adsabs.harvard.edu/abs/2015MNRAS.452..884N} {452, 884}

\bibitem[\protect\citeauthoryear{{Negoro} et~al.,}{{Negoro}
  et~al.}{2018}]{2018ATel12254....1N}
{Negoro} H.,  et~al., 2018, The Astronomer's Telegram, \href
  {https://ui.adsabs.harvard.edu/abs/2018ATel12254....1N} {12254, 1}

\bibitem[\protect\citeauthoryear{{Negueruela} \& {Schurch}}{{Negueruela} \&
  {Schurch}}{2007}]{2007A&A...461..631N}
{Negueruela} I.,  {Schurch} M.~P.~E.,  2007, \mn@doi [\aap]
  {10.1051/0004-6361:20066054}, \href
  {https://ui.adsabs.harvard.edu/abs/2007A&A...461..631N} {461, 631}

\bibitem[\protect\citeauthoryear{{Negueruela}, {Smith}  \&
  {Chaty}}{{Negueruela} et~al.}{2005}]{2005ATel..470....1N}
{Negueruela} I.,  {Smith} D.~M.,   {Chaty} S.,  2005, The Astronomer's
  Telegram, \href {https://ui.adsabs.harvard.edu/abs/2005ATel..470....1N} {470,
  1}

\bibitem[\protect\citeauthoryear{{Negueruela}, {Torrej{\'o}n}  \&
  {McBride}}{{Negueruela} et~al.}{2007}]{2007ATel.1239....1N}
{Negueruela} I.,  {Torrej{\'o}n} J.~M.,   {McBride} V.,  2007, The Astronomer's
  Telegram, \href {https://ui.adsabs.harvard.edu/abs/2007ATel.1239....1N}
  {1239, 1}

\bibitem[\protect\citeauthoryear{{Nespoli}, {Fabregat}  \&
  {Mennickent}}{{Nespoli} et~al.}{2008a}]{2008A&A...486..911N}
{Nespoli} E.,  {Fabregat} J.,   {Mennickent} R.~E.,  2008a, \mn@doi [\aap]
  {10.1051/0004-6361:200809645}, \href
  {https://ui.adsabs.harvard.edu/abs/2008A&A...486..911N} {486, 911}

\bibitem[\protect\citeauthoryear{{Nespoli}, {Fabregat}  \&
  {Mennickent}}{{Nespoli} et~al.}{2008b}]{2008ATel.1396....1N}
{Nespoli} E.,  {Fabregat} J.,   {Mennickent} R.~E.,  2008b, The Astronomer's
  Telegram, \href {https://ui.adsabs.harvard.edu/abs/2008ATel.1396....1N}
  {1396, 1}

\bibitem[\protect\citeauthoryear{{Nespoli}, {Fabregat}  \&
  {Mennickent}}{{Nespoli} et~al.}{2010}]{2010A&A...516A..94N}
{Nespoli} E.,  {Fabregat} J.,   {Mennickent} R.~E.,  2010, \mn@doi [\aap]
  {10.1051/0004-6361/200913410}, \href
  {https://ui.adsabs.harvard.edu/abs/2010A&A...516A..94N} {516, A94}

\bibitem[\protect\citeauthoryear{{Neustroev}, {Veledina}, {Poutanen},
  {Zharikov}, {Tsygankov}, {Sjoberg}  \& {Kajava}}{{Neustroev}
  et~al.}{2014}]{2014MNRAS.445.2424N}
{Neustroev} V.~V.,  {Veledina} A.,  {Poutanen} J.,  {Zharikov} S.~V.,
  {Tsygankov} S.~S.,  {Sjoberg} G.,   {Kajava} J. J.~E.,  2014, \mn@doi
  [\mnras] {10.1093/mnras/stu1924}, \href
  {https://ui.adsabs.harvard.edu/abs/2014MNRAS.445.2424N} {445, 2424}

\bibitem[\protect\citeauthoryear{{Nucita}, {Carpano}  \& {Guainazzi}}{{Nucita}
  et~al.}{2007}]{2007A&A...474L...1N}
{Nucita} A.~A.,  {Carpano} S.,   {Guainazzi} M.,  2007, \mn@doi [\aap]
  {10.1051/0004-6361:20078005}, \href
  {https://ui.adsabs.harvard.edu/abs/2007A&A...474L...1N} {474, L1}

\bibitem[\protect\citeauthoryear{{Nucita}, {De Paolis}, {Saxton}  \&
  {Read}}{{Nucita} et~al.}{2012}]{2012NewA...17..589N}
{Nucita} A.~A.,  {De Paolis} F.,  {Saxton} R.,   {Read} A.~M.,  2012, \mn@doi
  [\na] {10.1016/j.newast.2012.02.001}, \href
  {https://ui.adsabs.harvard.edu/abs/2012NewA...17..589N} {17, 589}

\bibitem[\protect\citeauthoryear{{Ochsenbein}, {Bauer}  \&
  {Marcout}}{{Ochsenbein} et~al.}{2000}]{vizier}
{Ochsenbein} F.,  {Bauer} P.,   {Marcout} J.,  2000, \mn@doi [\aaps]
  {10.1051/aas:2000169}, \href
  {https://ui.adsabs.harvard.edu/abs/2000A&AS..143...23O} {143, 23}

\bibitem[\protect\citeauthoryear{{Oda} et~al.,}{{Oda}
  et~al.}{2019}]{2019PASJ...71..108O}
{Oda} S.,  et~al., 2019, \mn@doi [\pasj] {10.1093/pasj/psz091}, \href
  {https://ui.adsabs.harvard.edu/abs/2019PASJ...71..108O} {71, 108}

\bibitem[\protect\citeauthoryear{{Oh}, {Yi}, {Schawinski}, {Koss},
  {Trakhtenbrot}  \& {Soto}}{{Oh} et~al.}{2015}]{2015ApJS..219....1O}
{Oh} K.,  {Yi} S.~K.,  {Schawinski} K.,  {Koss} M.,  {Trakhtenbrot} B.,
  {Soto} K.,  2015, \mn@doi [\apjs] {10.1088/0067-0049/219/1/1}, \href
  {https://ui.adsabs.harvard.edu/abs/2015ApJS..219....1O} {219, 1}

\bibitem[\protect\citeauthoryear{{Oh} et~al.,}{{Oh}
  et~al.}{2018}]{2018ApJS..235....4O}
{Oh} K.,  et~al., 2018, \mn@doi [\apjs] {10.3847/1538-4365/aaa7fd}, \href
  {https://ui.adsabs.harvard.edu/abs/2018ApJS..235....4O} {235, 4}

\bibitem[\protect\citeauthoryear{{Paizis} et~al.,}{{Paizis}
  et~al.}{2007}]{2007ApJ...657L.109P}
{Paizis} A.,  et~al., 2007, \mn@doi [\apjl] {10.1086/513313}, \href
  {https://ui.adsabs.harvard.edu/abs/2007ApJ...657L.109P} {657, L109}

\bibitem[\protect\citeauthoryear{{Papitto}, {Ferrigno}, {Bozzo}, {Gibaud},
  {Burderi}, {di Salvo}  \& {Riggio}}{{Papitto}
  et~al.}{2011}]{2011ATel.3556....1P}
{Papitto} A.,  {Ferrigno} C.,  {Bozzo} E.,  {Gibaud} L.,  {Burderi} L.,  {di
  Salvo} T.,   {Riggio} A.,  2011, The Astronomer's Telegram, \href
  {https://ui.adsabs.harvard.edu/abs/2011ATel.3556....1P} {3556, 1}

\bibitem[\protect\citeauthoryear{{Parikh}, {Russell}, {Wijnands},
  {Miller-Jones}, {Sivakoff}  \& {Tetarenko}}{{Parikh}
  et~al.}{2019}]{2019ApJ...878L..28P}
{Parikh} A.~S.,  {Russell} T.~D.,  {Wijnands} R.,  {Miller-Jones} J.~C.~A.,
  {Sivakoff} G.~R.,   {Tetarenko} A.~J.,  2019, \mn@doi [\apjl]
  {10.3847/2041-8213/ab2636}, \href
  {https://ui.adsabs.harvard.edu/abs/2019ApJ...878L..28P} {878, L28}

\bibitem[\protect\citeauthoryear{{Parisi} et~al.,}{{Parisi}
  et~al.}{2008}]{2008ATel.1540....1P}
{Parisi} P.,  et~al., 2008, The Astronomer's Telegram, \href
  {https://ui.adsabs.harvard.edu/abs/2008ATel.1540....1P} {1540, 1}

\bibitem[\protect\citeauthoryear{{Parisi} et~al.,}{{Parisi}
  et~al.}{2012}]{2012ATel.4151....1P}
{Parisi} P.,  et~al., 2012, The Astronomer's Telegram, \href
  {https://ui.adsabs.harvard.edu/abs/2012ATel.4151....1P} {4151, 1}

\bibitem[\protect\citeauthoryear{{Parisi} et~al.,}{{Parisi}
  et~al.}{2014}]{2014A&A...561A..67P}
{Parisi} P.,  et~al., 2014, \mn@doi [\aap] {10.1051/0004-6361/201322409}, \href
  {https://ui.adsabs.harvard.edu/abs/2014A&A...561A..67P} {561, A67}

\bibitem[\protect\citeauthoryear{{Pavan}, {Bozzo}, {Ferrigno}, {Ricci},
  {Manousakis}, {Walter}  \& {Stella}}{{Pavan}
  et~al.}{2011}]{2011A&A...526A.122P}
{Pavan} L.,  {Bozzo} E.,  {Ferrigno} C.,  {Ricci} C.,  {Manousakis} A.,
  {Walter} R.,   {Stella} L.,  2011, \mn@doi [\aap]
  {10.1051/0004-6361/201015561}, \href
  {https://ui.adsabs.harvard.edu/abs/2011A&A...526A.122P} {526, A122}

\bibitem[\protect\citeauthoryear{{Pavlinsky}, {Grebenev}  \&
  {Sunyaev}}{{Pavlinsky} et~al.}{1994}]{1994ApJ...425..110P}
{Pavlinsky} M.~N.,  {Grebenev} S.~A.,   {Sunyaev} R.~A.,  1994, \mn@doi [\apj]
  {10.1086/173967}, \href
  {https://ui.adsabs.harvard.edu/abs/1994ApJ...425..110P} {425, 110}

\bibitem[\protect\citeauthoryear{{Pavlinsky} et~al.,}{{Pavlinsky}
  et~al.}{2021}]{2021arXiv210705879P}
{Pavlinsky} M.,  et~al., 2021, arXiv e-prints, \href
  {https://ui.adsabs.harvard.edu/abs/2021arXiv210705879P} {p. arXiv:2107.05879}

\bibitem[\protect\citeauthoryear{{Pearlman}, {Coley}, {Corbet}  \&
  {Pottschmidt}}{{Pearlman} et~al.}{2019}]{2019ApJ...873...86P}
{Pearlman} A.~B.,  {Coley} J.~B.,  {Corbet} R. H.~D.,   {Pottschmidt} K.,
  2019, \mn@doi [\apj] {10.3847/1538-4357/aaf001}, \href
  {https://ui.adsabs.harvard.edu/abs/2019ApJ...873...86P} {873, 86}

\bibitem[\protect\citeauthoryear{{Produit}, {Ballet}  \& {Mowlavi}}{{Produit}
  et~al.}{2004}]{2004ATel..278....1P}
{Produit} N.,  {Ballet} J.,   {Mowlavi} N.,  2004, The Astronomer's Telegram,
  \href {https://ui.adsabs.harvard.edu/abs/2004ATel..278....1P} {278, 1}

\bibitem[\protect\citeauthoryear{{Rahoui}, {Tomsick}  \& {Krivonos}}{{Rahoui}
  et~al.}{2017}]{2017MNRAS.465.1563R}
{Rahoui} F.,  {Tomsick} J.~A.,   {Krivonos} R.,  2017, \mn@doi [\mnras]
  {10.1093/mnras/stw2830}, \href
  {https://ui.adsabs.harvard.edu/abs/2017MNRAS.465.1563R} {465, 1563}

\bibitem[\protect\citeauthoryear{{Ratti}, {Bassa}, {Torres}, {Kuiper},
  {Miller-Jones}  \& {Jonker}}{{Ratti} et~al.}{2010}]{2010MNRAS.408.1866R}
{Ratti} E.~M.,  {Bassa} C.~G.,  {Torres} M.~A.~P.,  {Kuiper} L.,
  {Miller-Jones} J.~C.~A.,   {Jonker} P.~G.,  2010, \mn@doi [\mnras]
  {10.1111/j.1365-2966.2010.17252.x}, \href
  {https://ui.adsabs.harvard.edu/abs/2010MNRAS.408.1866R} {408, 1866}

\bibitem[\protect\citeauthoryear{{Rau}}{{Rau}}{2018}]{2018ATel11332....1R}
{Rau} A.,  2018, The Astronomer's Telegram, \href
  {https://ui.adsabs.harvard.edu/abs/2018ATel11332....1R} {11332, 1}

\bibitem[\protect\citeauthoryear{{Renaud} et~al.,}{{Renaud}
  et~al.}{2010}]{2010ApJ...716..663R}
{Renaud} M.,  et~al., 2010, \mn@doi [\apj] {10.1088/0004-637X/716/1/663}, \href
  {https://ui.adsabs.harvard.edu/abs/2010ApJ...716..663R} {716, 663}

\bibitem[\protect\citeauthoryear{{Revnivtsev}, {Sazonov}, {Gilfanov}  \&
  {Sunyaev}}{{Revnivtsev} et~al.}{2003a}]{2003AstL...29..587R}
{Revnivtsev} M.~G.,  {Sazonov} S.~Y.,  {Gilfanov} M.~R.,   {Sunyaev} R.~A.,
  2003a, \mn@doi [Astronomy Letters] {10.1134/1.1607496}, \href
  {https://ui.adsabs.harvard.edu/abs/2003AstL...29..587R} {29, 587}

\bibitem[\protect\citeauthoryear{{Revnivtsev}, {Chernyakova}, {Capitanio},
  {Westergaard}, {Shoenfelder}, {Gehrels}  \& {Winkler}}{{Revnivtsev}
  et~al.}{2003b}]{2003ATel..132....1R}
{Revnivtsev} M.,  {Chernyakova} M.,  {Capitanio} F.,  {Westergaard} N.~J.,
  {Shoenfelder} V.,  {Gehrels} N.,   {Winkler} C.,  2003b, The Astronomer's
  Telegram, \href {https://ui.adsabs.harvard.edu/abs/2003ATel..132....1R} {132,
  1}

\bibitem[\protect\citeauthoryear{{Revnivtsev}, {Tuerler}, {Del Santo},
  {Westergaard}, {Gehrels}  \& {Winkler}}{{Revnivtsev}
  et~al.}{2003c}]{2003IAUC.8097....2R}
{Revnivtsev} M.,  {Tuerler} M.,  {Del Santo} M.,  {Westergaard} N.~J.,
  {Gehrels} N.,   {Winkler} C.,  2003c, \iaucirc, \href
  {https://ui.adsabs.harvard.edu/abs/2003IAUC.8097....2R} {8097, 2}

\bibitem[\protect\citeauthoryear{{Revnivtsev} et~al.,}{{Revnivtsev}
  et~al.}{2004a}]{2004AstL...30..382R}
{Revnivtsev} M.~G.,  et~al., 2004a, \mn@doi [Astronomy Letters]
  {10.1134/1.1764884}, \href
  {https://ui.adsabs.harvard.edu/abs/2004AstL...30..382R} {30, 382}

\bibitem[\protect\citeauthoryear{{Revnivtsev} et~al.,}{{Revnivtsev}
  et~al.}{2004b}]{2004A&A...425L..49R}
{Revnivtsev} M.~G.,  et~al., 2004b, \mn@doi [\aap]
  {10.1051/0004-6361:200400064}, \href
  {https://ui.adsabs.harvard.edu/abs/2004A&A...425L..49R} {425, L49}

\bibitem[\protect\citeauthoryear{{Revnivtsev}, {Sazonov}, {Molkov},
  {Lutovinov}, {Churazov}  \& {Sunyaev}}{{Revnivtsev}
  et~al.}{2006}]{2006AstL...32..145R}
{Revnivtsev} M.~G.,  {Sazonov} S.~Y.,  {Molkov} S.~V.,  {Lutovinov} A.~A.,
  {Churazov} E.~M.,   {Sunyaev} R.~A.,  2006, \mn@doi [Astronomy Letters]
  {10.1134/S1063773706030017}, \href
  {https://ui.adsabs.harvard.edu/abs/2006AstL...32..145R} {32, 145}

\bibitem[\protect\citeauthoryear{{Revnivtsev}, {Sunyaev}, {Lutovinov}  \&
  {Sazonov}}{{Revnivtsev} et~al.}{2007}]{2007ATel.1253....1R}
{Revnivtsev} M.,  {Sunyaev} R.,  {Lutovinov} A.,   {Sazonov} S.,  2007, The
  Astronomer's Telegram, \href
  {https://ui.adsabs.harvard.edu/abs/2007ATel.1253....1R} {1253, 1}

\bibitem[\protect\citeauthoryear{{Revnivtsev}, {Lutovinov}, {Churazov},
  {Sazonov}, {Gilfanov}, {Grebenev}  \& {Sunyaev}}{{Revnivtsev}
  et~al.}{2008}]{2008A&A...491..209R}
{Revnivtsev} M.,  {Lutovinov} A.,  {Churazov} E.,  {Sazonov} S.,  {Gilfanov}
  M.,  {Grebenev} S.,   {Sunyaev} R.,  2008, \mn@doi [\aap]
  {10.1051/0004-6361:200810115}, \href
  {https://ui.adsabs.harvard.edu/abs/2008A&A...491..209R} {491, 209}

\bibitem[\protect\citeauthoryear{{Revnivtsev} et~al.,}{{Revnivtsev}
  et~al.}{2009}]{2009AstL...35...33R}
{Revnivtsev} M.~G.,  et~al., 2009, \mn@doi [Astronomy Letters]
  {10.1134/S1063773709010046}, \href
  {https://ui.adsabs.harvard.edu/abs/2009AstL...35...33R} {35, 33}

\bibitem[\protect\citeauthoryear{{Reynolds} et~al.,}{{Reynolds}
  et~al.}{2012}]{2012ATel.3951....1R}
{Reynolds} M.~T.,  et~al., 2012, The Astronomer's Telegram, \href
  {https://ui.adsabs.harvard.edu/abs/2012ATel.3951....1R} {3951, 1}

\bibitem[\protect\citeauthoryear{{Ricci} et~al.,}{{Ricci}
  et~al.}{2017}]{2017ApJS..233...17R}
{Ricci} C.,  et~al., 2017, \mn@doi [\apjs] {10.3847/1538-4365/aa96ad}, \href
  {https://ui.adsabs.harvard.edu/abs/2017ApJS..233...17R} {233, 17}

\bibitem[\protect\citeauthoryear{{Rodes-Roca}, {Bernabeu}, {Magazz{\`u}},
  {Torrej{\'o}n}  \& {Solano}}{{Rodes-Roca} et~al.}{2018}]{2018MNRAS.476.2110R}
{Rodes-Roca} J.~J.,  {Bernabeu} G.,  {Magazz{\`u}} A.,  {Torrej{\'o}n} J.~M.,
  {Solano} E.,  2018, \mn@doi [\mnras] {10.1093/mnras/sty333}, \href
  {https://ui.adsabs.harvard.edu/abs/2018MNRAS.476.2110R} {476, 2110}

\bibitem[\protect\citeauthoryear{{Rodriguez} et~al.,}{{Rodriguez}
  et~al.}{2004}]{2004ATel..340....1R}
{Rodriguez} J.,  et~al., 2004, The Astronomer's Telegram, \href
  {https://ui.adsabs.harvard.edu/abs/2004ATel..340....1R} {340, 1}

\bibitem[\protect\citeauthoryear{{Rodriguez}, {Cadolle Bel}, {Tomsick},
  {Corbel}, {Brocksopp}, {Paizis}, {Shaw}  \& {Bodaghee}}{{Rodriguez}
  et~al.}{2007}]{2007ApJ...655L..97R}
{Rodriguez} J.,  {Cadolle Bel} M.,  {Tomsick} J.~A.,  {Corbel} S.,  {Brocksopp}
  C.,  {Paizis} A.,  {Shaw} S.~E.,   {Bodaghee} A.,  2007, \mn@doi [\apjl]
  {10.1086/511819}, \href
  {https://ui.adsabs.harvard.edu/abs/2007ApJ...655L..97R} {655, L97}

\bibitem[\protect\citeauthoryear{{Rodriguez}, {Tomsick}  \&
  {Chaty}}{{Rodriguez} et~al.}{2008}]{2008A&A...482..731R}
{Rodriguez} J.,  {Tomsick} J.~A.,   {Chaty} S.,  2008, \mn@doi [\aap]
  {10.1051/0004-6361:20079208}, \href
  {https://ui.adsabs.harvard.edu/abs/2008A&A...482..731R} {482, 731}

\bibitem[\protect\citeauthoryear{{Rodriguez}, {Tomsick}  \&
  {Chaty}}{{Rodriguez} et~al.}{2009}]{2009A&A...494..417R}
{Rodriguez} J.,  {Tomsick} J.~A.,   {Chaty} S.,  2009, \mn@doi [\aap]
  {10.1051/0004-6361:200810773}, \href
  {https://ui.adsabs.harvard.edu/abs/2009A&A...494..417R} {494, 417}

\bibitem[\protect\citeauthoryear{{Rodriguez}, {Tomsick}  \&
  {Bodaghee}}{{Rodriguez} et~al.}{2010}]{2010A&A...517A..14R}
{Rodriguez} J.,  {Tomsick} J.~A.,   {Bodaghee} A.,  2010, \mn@doi [\aap]
  {10.1051/0004-6361/200913967}, \href
  {https://ui.adsabs.harvard.edu/abs/2010A&A...517A..14R} {517, A14}

\bibitem[\protect\citeauthoryear{{Rojas} et~al.,}{{Rojas}
  et~al.}{2017}]{2017A&A...602A.124R}
{Rojas} A.~F.,  et~al., 2017, \mn@doi [\aap] {10.1051/0004-6361/201629463},
  \href {https://ui.adsabs.harvard.edu/abs/2017A&A...602A.124R} {602, A124}

\bibitem[\protect\citeauthoryear{{Russell}, {Lewis}, {Altamirano}  \&
  {Roche}}{{Russell} et~al.}{2011}]{2011ATel.3622....1R}
{Russell} D.~M.,  {Lewis} F.,  {Altamirano} D.,   {Roche} P.,  2011, The
  Astronomer's Telegram, \href
  {https://ui.adsabs.harvard.edu/abs/2011ATel.3622....1R} {3622, 1}

\bibitem[\protect\citeauthoryear{{Russell} et~al.,}{{Russell}
  et~al.}{2019}]{2019ApJ...883..198R}
{Russell} T.~D.,  et~al., 2019, \mn@doi [\apj] {10.3847/1538-4357/ab3d36},
  \href {https://ui.adsabs.harvard.edu/abs/2019ApJ...883..198R} {883, 198}

\bibitem[\protect\citeauthoryear{{Sanchez-Fernandez}, {Eckert}, {Bozzo},
  {Kajava}, {Kuulkers}  \& {Chenevez}}{{Sanchez-Fernandez}
  et~al.}{2015}]{2015ATel.7946....1S}
{Sanchez-Fernandez} C.,  {Eckert} D.,  {Bozzo} E.,  {Kajava} J.,  {Kuulkers}
  E.,   {Chenevez} J.,  2015, The Astronomer's Telegram, \href
  {https://ui.adsabs.harvard.edu/abs/2015ATel.7946....1S} {7946, 1}

\bibitem[\protect\citeauthoryear{{Sanna} et~al.,}{{Sanna}
  et~al.}{2017}]{2017A&A...598A..34S}
{Sanna} A.,  et~al., 2017, \mn@doi [\aap] {10.1051/0004-6361/201629406}, \href
  {https://ui.adsabs.harvard.edu/abs/2017A&A...598A..34S} {598, A34}

\bibitem[\protect\citeauthoryear{{Sanna} et~al.,}{{Sanna}
  et~al.}{2018}]{2018A&A...617L...8S}
{Sanna} A.,  et~al., 2018, \mn@doi [\aap] {10.1051/0004-6361/201834160}, \href
  {https://ui.adsabs.harvard.edu/abs/2018A&A...617L...8S} {617, L8}

\bibitem[\protect\citeauthoryear{{Saxton}, {Read}, {Esquej}, {Freyberg},
  {Altieri}  \& {Bermejo}}{{Saxton} et~al.}{2008}]{2008A&A...480..611S}
{Saxton} R.~D.,  {Read} A.~M.,  {Esquej} P.,  {Freyberg} M.~J.,  {Altieri} B.,
   {Bermejo} D.,  2008, \mn@doi [\aap] {10.1051/0004-6361:20079193}, \href
  {https://ui.adsabs.harvard.edu/abs/2008A&A...480..611S} {480, 611}

\bibitem[\protect\citeauthoryear{{Sazonov}, {Churazov}, {Revnivtsev},
  {Vikhlinin}  \& {Sunyaev}}{{Sazonov} et~al.}{2005}]{2005A&A...444L..37S}
{Sazonov} S.,  {Churazov} E.,  {Revnivtsev} M.,  {Vikhlinin} A.,   {Sunyaev}
  R.,  2005, \mn@doi [\aap] {10.1051/0004-6361:200500205}, \href
  {https://ui.adsabs.harvard.edu/abs/2005A&A...444L..37S} {444, L37}

\bibitem[\protect\citeauthoryear{{Sazonov}, {Revnivtsev}, {Krivonos},
  {Churazov}  \& {Sunyaev}}{{Sazonov} et~al.}{2007}]{2007A&A...462...57S}
{Sazonov} S.,  {Revnivtsev} M.,  {Krivonos} R.,  {Churazov} E.,   {Sunyaev} R.,
   2007, \mn@doi [\aap] {10.1051/0004-6361:20066277}, \href
  {https://ui.adsabs.harvard.edu/abs/2007A&A...462...57S} {462, 57}

\bibitem[\protect\citeauthoryear{{Sazonov}, {Revnivtsev}, {Burenin},
  {Churazov}, {Sunyaev}, {Forman}  \& {Murray}}{{Sazonov}
  et~al.}{2008}]{2008A&A...487..509S}
{Sazonov} S.,  {Revnivtsev} M.,  {Burenin} R.,  {Churazov} E.,  {Sunyaev} R.,
  {Forman} W.~R.,   {Murray} S.~S.,  2008, \mn@doi [\aap]
  {10.1051/0004-6361:200809528}, \href
  {https://ui.adsabs.harvard.edu/abs/2008A&A...487..509S} {487, 509}

\bibitem[\protect\citeauthoryear{{Sazonov} et~al.,}{{Sazonov}
  et~al.}{2020}]{2020NewAR..8801536S}
{Sazonov} S.,  et~al., 2020, \mn@doi [\nar] {10.1016/j.newar.2020.101536},
  \href {https://ui.adsabs.harvard.edu/abs/2020NewAR..8801536S} {88, 101536}

\bibitem[\protect\citeauthoryear{{Scott}, {Finger}, {Wilson}, {Koh}, {Prince},
  {Vaughan}  \& {Chakrabarty}}{{Scott} et~al.}{1997}]{1997ApJ...488..831S}
{Scott} D.~M.,  {Finger} M.~H.,  {Wilson} R.~B.,  {Koh} D.~T.,  {Prince} T.~A.,
   {Vaughan} B.~A.,   {Chakrabarty} D.,  1997, \mn@doi [\apj] {10.1086/304740},
  \href {https://ui.adsabs.harvard.edu/abs/1997ApJ...488..831S} {488, 831}

\bibitem[\protect\citeauthoryear{{Segreto}, {Cusumano}, {La Parola},
  {D'A{\`\i}}, {Masetti}  \& {D'Avanzo}}{{Segreto}
  et~al.}{2013}]{2013A&A...557A.113S}
{Segreto} A.,  {Cusumano} G.,  {La Parola} V.,  {D'A{\`\i}} A.,  {Masetti} N.,
   {D'Avanzo} P.,  2013, \mn@doi [\aap] {10.1051/0004-6361/201321897}, \href
  {https://ui.adsabs.harvard.edu/abs/2013A&A...557A.113S} {557, A113}

\bibitem[\protect\citeauthoryear{{Sguera}, {Drave}, {Sidoli}, {Masetti},
  {Landi}, {Bird}  \& {Bazzano}}{{Sguera} et~al.}{2013}]{2013A&A...556A..27S}
{Sguera} V.,  {Drave} S.~P.,  {Sidoli} L.,  {Masetti} N.,  {Landi} R.,  {Bird}
  A.~J.,   {Bazzano} A.,  2013, \mn@doi [\aap] {10.1051/0004-6361/201220785},
  \href {https://ui.adsabs.harvard.edu/abs/2013A&A...556A..27S} {556, A27}

\bibitem[\protect\citeauthoryear{{Sguera}, {Sidoli}, {Bird}, {Paizis}  \&
  {Bazzano}}{{Sguera} et~al.}{2020}]{2020MNRAS.491.4543S}
{Sguera} V.,  {Sidoli} L.,  {Bird} A.~J.,  {Paizis} A.,   {Bazzano} A.,  2020,
  \mn@doi [\mnras] {10.1093/mnras/stz3330}, \href
  {https://ui.adsabs.harvard.edu/abs/2020MNRAS.491.4543S} {491, 4543}

\bibitem[\protect\citeauthoryear{{Shaw}, {Heinke}, {Degenaar}, {Wijnands},
  {Kaur}  \& {Forestell}}{{Shaw} et~al.}{2017}]{2017MNRAS.471.2508S}
{Shaw} A.~W.,  {Heinke} C.~O.,  {Degenaar} N.,  {Wijnands} R.,  {Kaur} R.,
  {Forestell} L.~M.,  2017, \mn@doi [\mnras] {10.1093/mnras/stx1732}, \href
  {https://ui.adsabs.harvard.edu/abs/2017MNRAS.471.2508S} {471, 2508}

\bibitem[\protect\citeauthoryear{{Shidatsu} et~al.,}{{Shidatsu}
  et~al.}{2018}]{2018ApJ...868...54S}
{Shidatsu} M.,  et~al., 2018, \mn@doi [\apj] {10.3847/1538-4357/aae929}, \href
  {https://ui.adsabs.harvard.edu/abs/2018ApJ...868...54S} {868, 54}

\bibitem[\protect\citeauthoryear{{Sidoli}, {Paizis}  \& {Mereghetti}}{{Sidoli}
  et~al.}{2006}]{2006A&A...450L...9S}
{Sidoli} L.,  {Paizis} A.,   {Mereghetti} S.,  2006, \mn@doi [\aap]
  {10.1051/0004-6361:20064940}, \href
  {https://ui.adsabs.harvard.edu/abs/2006A&A...450L...9S} {450, L9}

\bibitem[\protect\citeauthoryear{{Sidoli}, {Paizis}, {Mereghetti}, {G{\"o}tz}
  \& {Del Santo}}{{Sidoli} et~al.}{2011}]{2011MNRAS.415.2373S}
{Sidoli} L.,  {Paizis} A.,  {Mereghetti} S.,  {G{\"o}tz} D.,   {Del Santo} M.,
  2011, \mn@doi [\mnras] {10.1111/j.1365-2966.2011.18865.x}, \href
  {https://ui.adsabs.harvard.edu/abs/2011MNRAS.415.2373S} {415, 2373}

\bibitem[\protect\citeauthoryear{{Smith} \& {Hartigan}}{{Smith} \&
  {Hartigan}}{2006}]{2006ApJ...638.1045S}
{Smith} N.,  {Hartigan} P.,  2006, \mn@doi [\apj] {10.1086/498860}, \href
  {https://ui.adsabs.harvard.edu/abs/2006ApJ...638.1045S} {638, 1045}

\bibitem[\protect\citeauthoryear{{Soldi}, {Brandt}, {Garau}, {Grebenev},
  {Kuulkers}, {Palumbo}  \& {Tarana}}{{Soldi}
  et~al.}{2005}]{2005ATel..456....1S}
{Soldi} S.,  {Brandt} S.,  {Garau} A.~D.,  {Grebenev} S.~A.,  {Kuulkers} E.,
  {Palumbo} G.~G.~C.,   {Tarana} A.,  2005, The Astronomer's Telegram, \href
  {https://ui.adsabs.harvard.edu/abs/2005ATel..456....1S} {456, 1}

\bibitem[\protect\citeauthoryear{{Soldi} et~al.,}{{Soldi}
  et~al.}{2006}]{2006ATel..885....1S}
{Soldi} S.,  et~al., 2006, The Astronomer's Telegram, \href
  {https://ui.adsabs.harvard.edu/abs/2006ATel..885....1S} {885, 1}

\bibitem[\protect\citeauthoryear{{Spiro} et~al.,}{{Spiro}
  et~al.}{2013}]{2013ATel.5537....1S}
{Spiro} S.,  et~al., 2013, The Astronomer's Telegram, \href
  {https://ui.adsabs.harvard.edu/abs/2013ATel.5537....1S} {5537, 1}

\bibitem[\protect\citeauthoryear{{Steeghs}, {Knigge}, {Drew}, {Unruh}  \&
  {Greimel}}{{Steeghs} et~al.}{2008}]{2008ATel.1653....1S}
{Steeghs} D.,  {Knigge} C.,  {Drew} J.,  {Unruh} Y.,   {Greimel} R.,  2008, The
  Astronomer's Telegram, \href
  {https://ui.adsabs.harvard.edu/abs/2008ATel.1653....1S} {1653, 1}

\bibitem[\protect\citeauthoryear{{Stephen}, {Bassani}, {Malizia}, {Masetti}  \&
  {Ubertini}}{{Stephen} et~al.}{2018}]{2018ATel11341....1S}
{Stephen} J.~B.,  {Bassani} L.,  {Malizia} A.,  {Masetti} N.,   {Ubertini} P.,
  2018, The Astronomer's Telegram, \href
  {https://ui.adsabs.harvard.edu/abs/2018ATel11341....1S} {11341, 1}

\bibitem[\protect\citeauthoryear{{Sunyaev}, {Lutovinov}, {Molkov}  \&
  {Deluit}}{{Sunyaev} et~al.}{2003a}]{2003ATel..181....1S}
{Sunyaev} R.,  {Lutovinov} A.,  {Molkov} S.,   {Deluit} S.,  2003a, The
  Astronomer's Telegram, \href
  {https://ui.adsabs.harvard.edu/abs/2003ATel..181....1S} {181, 1}

\bibitem[\protect\citeauthoryear{{Sunyaev}, {Grebenev}, {Lutovinov},
  {Rodriguez}, {Mereghetti}, {Gotz}  \& {Courvoisier}}{{Sunyaev}
  et~al.}{2003b}]{2003ATel..190....1S}
{Sunyaev} R.~A.,  {Grebenev} S.~A.,  {Lutovinov} A.~A.,  {Rodriguez} J.,
  {Mereghetti} S.,  {Gotz} D.,   {Courvoisier} T.,  2003b, The Astronomer's
  Telegram, \href {https://ui.adsabs.harvard.edu/abs/2003ATel..190....1S} {190,
  1}

\bibitem[\protect\citeauthoryear{{Terrier}, {Mattana}, {Djannati-Atai},
  {Marandon}, {Renaud}  \& {Dubois}}{{Terrier}
  et~al.}{2008}]{2008AIPC.1085..312T}
{Terrier} R.,  {Mattana} F.,  {Djannati-Atai} A.,  {Marandon} V.,  {Renaud} M.,
    {Dubois} F.,  2008, in {Aharonian} F.~A.,  {Hofmann} W.,   {Rieger} F.,
  eds,  American Institute of Physics Conference Series Vol. 1085, American
  Institute of Physics Conference Series. pp 312--315,
  \mn@doi{10.1063/1.3076669}

\bibitem[\protect\citeauthoryear{{Terrier} et~al.,}{{Terrier}
  et~al.}{2010}]{2010ApJ...719..143T}
{Terrier} R.,  et~al., 2010, \mn@doi [\apj] {10.1088/0004-637X/719/1/143},
  \href {https://ui.adsabs.harvard.edu/abs/2010ApJ...719..143T} {719, 143}

\bibitem[\protect\citeauthoryear{{Tomsick}, {Lingenfelter}, {Walter},
  {Rodriguez}, {Goldwurm}, {Corbel}  \& {Kaaret}}{{Tomsick}
  et~al.}{2003}]{2003IAUC.8076....1T}
{Tomsick} J.~A.,  {Lingenfelter} R.,  {Walter} R.,  {Rodriguez} J.,  {Goldwurm}
  A.,  {Corbel} S.,   {Kaaret} P.,  2003, \iaucirc, \href
  {https://ui.adsabs.harvard.edu/abs/2003IAUC.8076....1T} {8076, 1}

\bibitem[\protect\citeauthoryear{{Tomsick}, {Lingenfelter}, {Corbel},
  {Goldwurm}  \& {Kaaret}}{{Tomsick} et~al.}{2004}]{2004ATel..224....1T}
{Tomsick} J.~A.,  {Lingenfelter} R.,  {Corbel} S.,  {Goldwurm} A.,   {Kaaret}
  P.,  2004, The Astronomer's Telegram, \href
  {https://ui.adsabs.harvard.edu/abs/2004ATel..224....1T} {224, 1}

\bibitem[\protect\citeauthoryear{{Tomsick}, {Chaty}, {Rodriguez}, {Foschini},
  {Walter}  \& {Kaaret}}{{Tomsick} et~al.}{2006}]{2006ApJ...647.1309T}
{Tomsick} J.~A.,  {Chaty} S.,  {Rodriguez} J.,  {Foschini} L.,  {Walter} R.,
  {Kaaret} P.,  2006, \mn@doi [\apj] {10.1086/505595}, \href
  {https://ui.adsabs.harvard.edu/abs/2006ApJ...647.1309T} {647, 1309}

\bibitem[\protect\citeauthoryear{{Tomsick}, {Chaty}, {Rodriguez}, {Walter}  \&
  {Kaaret}}{{Tomsick} et~al.}{2008}]{2008ApJ...685.1143T}
{Tomsick} J.~A.,  {Chaty} S.,  {Rodriguez} J.,  {Walter} R.,   {Kaaret} P.,
  2008, \mn@doi [\apj] {10.1086/591040}, \href
  {https://ui.adsabs.harvard.edu/abs/2008ApJ...685.1143T} {685, 1143}

\bibitem[\protect\citeauthoryear{{Tomsick}, {Chaty}, {Rodriguez}, {Walter}  \&
  {Kaaret}}{{Tomsick} et~al.}{2009}]{2009ApJ...701..811T}
{Tomsick} J.~A.,  {Chaty} S.,  {Rodriguez} J.,  {Walter} R.,   {Kaaret} P.,
  2009, \mn@doi [\apj] {10.1088/0004-637X/701/1/811}, \href
  {https://ui.adsabs.harvard.edu/abs/2009ApJ...701..811T} {701, 811}

\bibitem[\protect\citeauthoryear{{Tomsick}, {Bodaghee}, {Chaty}, {Rodriguez},
  {Rahoui}, {Halpern}, {Kalemci}  \& {{\"O}zbey Arabaci}}{{Tomsick}
  et~al.}{2012}]{2012ApJ...754..145T}
{Tomsick} J.~A.,  {Bodaghee} A.,  {Chaty} S.,  {Rodriguez} J.,  {Rahoui} F.,
  {Halpern} J.,  {Kalemci} E.,   {{\"O}zbey Arabaci} M.,  2012, \mn@doi [\apj]
  {10.1088/0004-637X/754/2/145}, \href
  {https://ui.adsabs.harvard.edu/abs/2012ApJ...754..145T} {754, 145}

\bibitem[\protect\citeauthoryear{{Tomsick}, {Krivonos}, {Rahoui}, {Ajello},
  {Rodriguez}, {Barri{\`e}re}, {Bodaghee}  \& {Chaty}}{{Tomsick}
  et~al.}{2015}]{2015MNRAS.449..597T}
{Tomsick} J.~A.,  {Krivonos} R.,  {Rahoui} F.,  {Ajello} M.,  {Rodriguez} J.,
  {Barri{\`e}re} N.,  {Bodaghee} A.,   {Chaty} S.,  2015, \mn@doi [\mnras]
  {10.1093/mnras/stv325}, \href
  {https://ui.adsabs.harvard.edu/abs/2015MNRAS.449..597T} {449, 597}

\bibitem[\protect\citeauthoryear{{Tomsick}, {Rahoui}, {Krivonos}, {Clavel},
  {Strader}  \& {Chomiuk}}{{Tomsick} et~al.}{2016a}]{2016MNRAS.460..513T}
{Tomsick} J.~A.,  {Rahoui} F.,  {Krivonos} R.,  {Clavel} M.,  {Strader} J.,
  {Chomiuk} L.,  2016a, \mn@doi [\mnras] {10.1093/mnras/stw871}, \href
  {https://ui.adsabs.harvard.edu/abs/2016MNRAS.460..513T} {460, 513}

\bibitem[\protect\citeauthoryear{{Tomsick}, {Krivonos}, {Wang}, {Bodaghee},
  {Chaty}, {Rahoui}, {Rodriguez}  \& {Fornasini}}{{Tomsick}
  et~al.}{2016b}]{2016ApJ...816...38T}
{Tomsick} J.~A.,  {Krivonos} R.,  {Wang} Q.,  {Bodaghee} A.,  {Chaty} S.,
  {Rahoui} F.,  {Rodriguez} J.,   {Fornasini} F.~M.,  2016b, \mn@doi [\apj]
  {10.3847/0004-637X/816/1/38}, \href
  {https://ui.adsabs.harvard.edu/abs/2016ApJ...816...38T} {816, 38}

\bibitem[\protect\citeauthoryear{{Tomsick} et~al.,}{{Tomsick}
  et~al.}{2020}]{2020ApJ...889...53T}
{Tomsick} J.~A.,  et~al., 2020, \mn@doi [\apj] {10.3847/1538-4357/ab5fd2},
  \href {https://ui.adsabs.harvard.edu/abs/2020ApJ...889...53T} {889, 53}

\bibitem[\protect\citeauthoryear{{Tomsick} et~al.,}{{Tomsick}
  et~al.}{2021}]{2021ApJ...914...48T}
{Tomsick} J.~A.,  et~al., 2021, \mn@doi [\apj] {10.3847/1538-4357/abfa1a},
  \href {https://ui.adsabs.harvard.edu/abs/2021ApJ...914...48T} {914, 48}

\bibitem[\protect\citeauthoryear{{Torrej{\'o}n}, {Negueruela}, {Smith}  \&
  {Harrison}}{{Torrej{\'o}n} et~al.}{2010}]{2010A&A...510A..61T}
{Torrej{\'o}n} J.~M.,  {Negueruela} I.,  {Smith} D.~M.,   {Harrison} T.~E.,
  2010, \mn@doi [\aap] {10.1051/0004-6361/200912619}, \href
  {https://ui.adsabs.harvard.edu/abs/2010A&A...510A..61T} {510, A61}

\bibitem[\protect\citeauthoryear{{Torres}, {Garcia}, {McClintock}, {Steeghs},
  {Miller}, {Callanan}, {Zhao}  \& {Berlind}}{{Torres}
  et~al.}{2004}]{2004ATel..264....1T}
{Torres} M.~A.~P.,  {Garcia} M.~R.,  {McClintock} J.~E.,  {Steeghs} D.,
  {Miller} J.,  {Callanan} P.~J.,  {Zhao} P.,   {Berlind} P.,  2004, The
  Astronomer's Telegram, \href
  {https://ui.adsabs.harvard.edu/abs/2004ATel..264....1T} {264, 1}

\bibitem[\protect\citeauthoryear{{Torres}, {Steeghs}, {Garcia}, {McClintock},
  {Berlind}, {Zhao}, {Jonker}  \& {Callanan}}{{Torres}
  et~al.}{2006}]{2006ATel..862....1T}
{Torres} M.~A.~P.,  {Steeghs} D.,  {Garcia} M.~R.,  {McClintock} J.~E.,
  {Berlind} P.,  {Zhao} P.,  {Jonker} P.~G.,   {Callanan} P.~J.,  2006, The
  Astronomer's Telegram, \href
  {https://ui.adsabs.harvard.edu/abs/2006ATel..862....1T} {862, 1}

\bibitem[\protect\citeauthoryear{{Tovmassian} et~al.,}{{Tovmassian}
  et~al.}{2017}]{2017A&A...608A..36T}
{Tovmassian} G.,  et~al., 2017, \mn@doi [\aap] {10.1051/0004-6361/201731323},
  \href {https://ui.adsabs.harvard.edu/abs/2017A&A...608A..36T} {608, A36}

\bibitem[\protect\citeauthoryear{{Townsend} et~al.,}{{Townsend}
  et~al.}{2011}]{2011MNRAS.410.1813T}
{Townsend} L.~J.,  et~al., 2011, \mn@doi [\mnras]
  {10.1111/j.1365-2966.2010.17563.x}, \href
  {https://ui.adsabs.harvard.edu/abs/2011MNRAS.410.1813T} {410, 1813}

\bibitem[\protect\citeauthoryear{{Tueller} et~al.,}{{Tueller}
  et~al.}{2005}]{2005ATel..669....1T}
{Tueller} J.,  et~al., 2005, The Astronomer's Telegram, \href
  {https://ui.adsabs.harvard.edu/abs/2005ATel..669....1T} {669, 1}

\bibitem[\protect\citeauthoryear{{Tueller}, {Mushotzky}, {Barthelmy},
  {Cannizzo}, {Gehrels}, {Markwardt}, {Skinner}  \& {Winter}}{{Tueller}
  et~al.}{2008}]{2008ApJ...681..113T}
{Tueller} J.,  {Mushotzky} R.~F.,  {Barthelmy} S.,  {Cannizzo} J.~K.,
  {Gehrels} N.,  {Markwardt} C.~B.,  {Skinner} G.~K.,   {Winter} L.~M.,  2008,
  \mn@doi [\apj] {10.1086/588458}, \href
  {https://ui.adsabs.harvard.edu/abs/2008ApJ...681..113T} {681, 113}

\bibitem[\protect\citeauthoryear{{Tueller} et~al.,}{{Tueller}
  et~al.}{2010}]{2010ApJS..186..378T}
{Tueller} J.,  et~al., 2010, \mn@doi [\apjs] {10.1088/0067-0049/186/2/378},
  \href {https://ui.adsabs.harvard.edu/abs/2010ApJS..186..378T} {186, 378}

\bibitem[\protect\citeauthoryear{{Tuerler}, {Walter}  \& {Ferrigno}}{{Tuerler}
  et~al.}{2012}]{2012ATel.4183....1T}
{Tuerler} M.,  {Walter} R.,   {Ferrigno} C.,  2012, The Astronomer's Telegram,
  \href {https://ui.adsabs.harvard.edu/abs/2012ATel.4183....1T} {4183, 1}

\bibitem[\protect\citeauthoryear{{Ubertini} et~al.,}{{Ubertini}
  et~al.}{2003}]{2003A&A...411L.131U}
{Ubertini} P.,  et~al., 2003, \mn@doi [\aap] {10.1051/0004-6361:20031224},
  \href {https://ui.adsabs.harvard.edu/abs/2003A&A...411L.131U} {411, L131}

\bibitem[\protect\citeauthoryear{{Ubertini} et~al.,}{{Ubertini}
  et~al.}{2005}]{2005ApJ...629L.109U}
{Ubertini} P.,  et~al., 2005, \mn@doi [\apjl] {10.1086/447766}, \href
  {https://ui.adsabs.harvard.edu/abs/2005ApJ...629L.109U} {629, L109}

\bibitem[\protect\citeauthoryear{{Vasilopoulos}, {Maggi}, {Haberl}, {Sturm},
  {Pietsch}, {Bartlett}  \& {Coe}}{{Vasilopoulos}
  et~al.}{2013}]{2013A&A...558A..74V}
{Vasilopoulos} G.,  {Maggi} P.,  {Haberl} F.,  {Sturm} R.,  {Pietsch} W.,
  {Bartlett} E.~S.,   {Coe} M.~J.,  2013, \mn@doi [\aap]
  {10.1051/0004-6361/201322335}, \href
  {https://ui.adsabs.harvard.edu/abs/2013A&A...558A..74V} {558, A74}

\bibitem[\protect\citeauthoryear{{Vasilopoulos}, {Haberl}, {Sturm}, {Maggi}  \&
  {Udalski}}{{Vasilopoulos} et~al.}{2014}]{2014A&A...567A.129V}
{Vasilopoulos} G.,  {Haberl} F.,  {Sturm} R.,  {Maggi} P.,   {Udalski} A.,
  2014, \mn@doi [\aap] {10.1051/0004-6361/201423934}, \href
  {https://ui.adsabs.harvard.edu/abs/2014A&A...567A.129V} {567, A129}

\bibitem[\protect\citeauthoryear{{Voges} et~al.,}{{Voges}
  et~al.}{1999}]{voges99}
{Voges} W.,  et~al., 1999, \aap, \href
  {https://ui.adsabs.harvard.edu/abs/1999A&A...349..389V} {349, 389}

\bibitem[\protect\citeauthoryear{{Vovk} et~al.,}{{Vovk}
  et~al.}{2012}]{2012ATel.4381....1V}
{Vovk} I.,  et~al., 2012, The Astronomer's Telegram, \href
  {https://ui.adsabs.harvard.edu/abs/2012ATel.4381....1V} {4381, 1}

\bibitem[\protect\citeauthoryear{{Walter} et~al.,}{{Walter}
  et~al.}{2004}]{2004ATel..229....1W}
{Walter} R.,  et~al., 2004, The Astronomer's Telegram, \href
  {https://ui.adsabs.harvard.edu/abs/2004ATel..229....1W} {229, 1}

\bibitem[\protect\citeauthoryear{{Walter} et~al.,}{{Walter}
  et~al.}{2006}]{2006A&A...453..133W}
{Walter} R.,  et~al., 2006, \mn@doi [\aap] {10.1051/0004-6361:20053719}, \href
  {https://ui.adsabs.harvard.edu/abs/2006A&A...453..133W} {453, 133}

\bibitem[\protect\citeauthoryear{{Wang}}{{Wang}}{2010}]{2010A&A...516A..15W}
{Wang} W.,  2010, \mn@doi [\aap] {10.1051/0004-6361/200913196}, \href
  {https://ui.adsabs.harvard.edu/abs/2010A&A...516A..15W} {516, A15}

\bibitem[\protect\citeauthoryear{{Watson} et~al.,}{{Watson}
  et~al.}{2009}]{2009A&A...493..339W}
{Watson} M.~G.,  et~al., 2009, \mn@doi [\aap] {10.1051/0004-6361:200810534},
  \href {https://ui.adsabs.harvard.edu/abs/2009A&A...493..339W} {493, 339}

\bibitem[\protect\citeauthoryear{{Webb} et~al.,}{{Webb}
  et~al.}{2020}]{2020A&A...641A.136W}
{Webb} N.~A.,  et~al., 2020, \mn@doi [\aap] {10.1051/0004-6361/201937353},
  \href {https://ui.adsabs.harvard.edu/abs/2020A&A...641A.136W} {641, A136}

\bibitem[\protect\citeauthoryear{{Westergaard}, {Budtz-Jorgensen}, {Chenevez},
  {Lund}, {Brandt}  \& {Oxborrow}}{{Westergaard}
  et~al.}{2006}]{2006ATel..967....1W}
{Westergaard} N.~J.,  {Budtz-Jorgensen} C.,  {Chenevez} J.,  {Lund} N.,
  {Brandt} S.,   {Oxborrow} C.~A.,  2006, The Astronomer's Telegram, \href
  {https://ui.adsabs.harvard.edu/abs/2006ATel..967....1W} {967, 1}

\bibitem[\protect\citeauthoryear{{Wijnands}}{{Wijnands}}{2006}]{2006ATel..972....1W}
{Wijnands} R.,  2006, The Astronomer's Telegram, \href
  {https://ui.adsabs.harvard.edu/abs/2006ATel..972....1W} {972, 1}

\bibitem[\protect\citeauthoryear{{Winkler} et~al.,}{{Winkler}
  et~al.}{2003}]{2003A&A...411L...1W}
{Winkler} C.,  et~al., 2003, \mn@doi [\aap] {10.1051/0004-6361:20031288}, \href
  {https://ui.adsabs.harvard.edu/abs/2003A&A...411L...1W} {411, L1}

\bibitem[\protect\citeauthoryear{{Winter}, {Mushotzky}, {Tueller}  \&
  {Markwardt}}{{Winter} et~al.}{2008}]{2008ApJ...674..686W}
{Winter} L.~M.,  {Mushotzky} R.~F.,  {Tueller} J.,   {Markwardt} C.,  2008,
  \mn@doi [\apj] {10.1086/525274}, \href
  {https://ui.adsabs.harvard.edu/abs/2008ApJ...674..686W} {674, 686}

\bibitem[\protect\citeauthoryear{{Winter}, {Mushotzky}, {Terashima}  \&
  {Ueda}}{{Winter} et~al.}{2009}]{2009ApJ...701.1644W}
{Winter} L.~M.,  {Mushotzky} R.~F.,  {Terashima} Y.,   {Ueda} Y.,  2009,
  \mn@doi [\apj] {10.1088/0004-637X/701/2/1644}, \href
  {https://ui.adsabs.harvard.edu/abs/2009ApJ...701.1644W} {701, 1644}

\bibitem[\protect\citeauthoryear{{Worpel}, {Schwope}, {Traulsen}, {Mukai}  \&
  {Ok}}{{Worpel} et~al.}{2020}]{2020A&A...639A..17W}
{Worpel} H.,  {Schwope} A.~D.,  {Traulsen} I.,  {Mukai} K.,   {Ok} S.,  2020,
  \mn@doi [\aap] {10.1051/0004-6361/202038038}, \href
  {https://ui.adsabs.harvard.edu/abs/2020A&A...639A..17W} {639, A17}

\bibitem[\protect\citeauthoryear{{Xiao} et~al.,}{{Xiao}
  et~al.}{2019}]{2019JHEAp..24...30X}
{Xiao} G.~C.,  et~al., 2019, \mn@doi [Journal of High Energy Astrophysics]
  {10.1016/j.jheap.2019.09.005}, \href
  {https://ui.adsabs.harvard.edu/abs/2019JHEAp..24...30X} {24, 30}

\bibitem[\protect\citeauthoryear{{Xu}, {Shao}  \& {Li}}{{Xu}
  et~al.}{2019}]{2019MNRAS.489.3031X}
{Xu} X.,  {Shao} Y.,   {Li} X.-D.,  2019, \mn@doi [\mnras]
  {10.1093/mnras/stz2388}, \href
  {https://ui.adsabs.harvard.edu/abs/2019MNRAS.489.3031X} {489, 3031}

\bibitem[\protect\citeauthoryear{{Yatabe} et~al.,}{{Yatabe}
  et~al.}{2019}]{2019ATel12425....1Y}
{Yatabe} F.,  et~al., 2019, The Astronomer's Telegram, \href
  {https://ui.adsabs.harvard.edu/abs/2019ATel12425....1Y} {12425, 1}

\bibitem[\protect\citeauthoryear{{Zhang}, {Chen}, {Wang}, {Torres}  \&
  {Li}}{{Zhang} et~al.}{2009}]{2009A&A...502..231Z}
{Zhang} S.,  {Chen} Y.~P.,  {Wang} J.~M.,  {Torres} D.~F.,   {Li} T.~P.,  2009,
  \mn@doi [\aap] {10.1051/0004-6361/200911636}, \href
  {https://ui.adsabs.harvard.edu/abs/2009A&A...502..231Z} {502, 231}

\bibitem[\protect\citeauthoryear{{Zhang} et~al.,}{{Zhang}
  et~al.}{2015}]{2015ApJ...815..132Z}
{Zhang} S.,  et~al., 2015, \mn@doi [\apj] {10.1088/0004-637X/815/2/132}, \href
  {https://ui.adsabs.harvard.edu/abs/2015ApJ...815..132Z} {815, 132}

\bibitem[\protect\citeauthoryear{{Zolotukhin} \& {Revnivtsev}}{{Zolotukhin} \&
  {Revnivtsev}}{2011}]{2011MNRAS.411..620Z}
{Zolotukhin} I.~Y.,  {Revnivtsev} M.~G.,  2011, \mn@doi [\mnras]
  {10.1111/j.1365-2966.2010.17706.x}, \href
  {https://ui.adsabs.harvard.edu/abs/2011MNRAS.411..620Z} {411, 620}

\bibitem[\protect\citeauthoryear{{Zolotukhin} \& {Revnivtsev}}{{Zolotukhin} \&
  {Revnivtsev}}{2015}]{2015MNRAS.446.2418Z}
{Zolotukhin} I.~Y.,  {Revnivtsev} M.~G.,  2015, \mn@doi [\mnras]
  {10.1093/mnras/stu2212}, \href
  {https://ui.adsabs.harvard.edu/abs/2015MNRAS.446.2418Z} {446, 2418}

\bibitem[\protect\citeauthoryear{{Zurita Heras} \& {Chaty}}{{Zurita Heras} \&
  {Chaty}}{2008}]{2008A&A...489..657Z}
{Zurita Heras} J.~A.,  {Chaty} S.,  2008, \mn@doi [\aap]
  {10.1051/0004-6361:20079097}, \href
  {https://ui.adsabs.harvard.edu/abs/2008A&A...489..657Z} {489, 657}

\bibitem[\protect\citeauthoryear{{Zurita Heras}, {Chaty}  \& {Tomsick}}{{Zurita
  Heras} et~al.}{2009}]{2009A&A...502..787Z}
{Zurita Heras} J.~A.,  {Chaty} S.,   {Tomsick} J.~A.,  2009, \mn@doi [\aap]
  {10.1051/0004-6361/200912359}, \href
  {https://ui.adsabs.harvard.edu/abs/2009A&A...502..787Z} {502, 787}

\bibitem[\protect\citeauthoryear{{de Rosa} et~al.,}{{de Rosa}
  et~al.}{2012}]{2012MNRAS.420.2087D}
{de Rosa} A.,  et~al., 2012, \mn@doi [\mnras]
  {10.1111/j.1365-2966.2011.20167.x}, \href
  {https://ui.adsabs.harvard.edu/abs/2012MNRAS.420.2087D} {420, 2087}

\bibitem[\protect\citeauthoryear{{den Hartog}, {Hermsen}, {Kuiper}, {in't
  Zand}, {Winkler}  \& {Domingo}}{{den Hartog}
  et~al.}{2004a}]{2004ATel..261....1D}
{den Hartog} P.~R.,  {Hermsen} W.,  {Kuiper} L.~M.,  {in't Zand} J.~J.~M.,
  {Winkler} C.,   {Domingo} A.,  2004a, The Astronomer's Telegram, \href
  {https://ui.adsabs.harvard.edu/abs/2004ATel..261....1D} {261, 1}

\bibitem[\protect\citeauthoryear{{den Hartog}, {Kuiper}, {Corbet}, {in't Zand},
  {Hermsen}, {Vink}, {Remillard}  \& {van der Klis}}{{den Hartog}
  et~al.}{2004b}]{2004ATel..281....1D}
{den Hartog} P.~R.,  {Kuiper} L.~M.,  {Corbet} R.~H.~D.,  {in't Zand} J.~J.~M.,
   {Hermsen} W.,  {Vink} J.,  {Remillard} R.,   {van der Klis} M.,  2004b, The
  Astronomer's Telegram, \href
  {https://ui.adsabs.harvard.edu/abs/2004ATel..281....1D} {281, 1}

\bibitem[\protect\citeauthoryear{{den Hartog}, {Hermsen}, {Kuiper}, {Vink},
  {in't Zand}  \& {Collmar}}{{den Hartog} et~al.}{2006}]{2006A&A...451..587D}
{den Hartog} P.~R.,  {Hermsen} W.,  {Kuiper} L.,  {Vink} J.,  {in't Zand}
  J.~J.~M.,   {Collmar} W.,  2006, \mn@doi [\aap] {10.1051/0004-6361:20054711},
  \href {https://ui.adsabs.harvard.edu/abs/2006A&A...451..587D} {451, 587}

\bibitem[\protect\citeauthoryear{{in 't Zand}, {Bazzano}, {Cocchi}, {Ubertini},
  {Muller}  \& {Torroni}}{{in 't Zand} et~al.}{1998}]{1998IAUC.6846....2I}
{in 't Zand} J.,  {Bazzano} A.,  {Cocchi} M.,  {Ubertini} P.,  {Muller} J.~M.,
   {Torroni} V.,  1998, \iaucirc, \href
  {https://ui.adsabs.harvard.edu/abs/1998IAUC.6846....2I} {6846, 2}

\bibitem[\protect\citeauthoryear{{in't Zand}}{{in't
  Zand}}{2005}]{2005A&A...441L...1I}
{in't Zand} J.~J.~M.,  2005, \mn@doi [\aap] {10.1051/0004-6361:200500162},
  \href {https://ui.adsabs.harvard.edu/abs/2005A&A...441L...1I} {441, L1}

\makeatother
\end{thebibliography}




\appendix

\section{Catalog of sources in 17--60~keV band}
\label{sec:catalog}

\onecolumn
\begin{minipage}{0.9\textwidth}
{\normalsize {\bf Table A1.} The full list of hard X-ray sources detected during the \integral\ all-sky survey based on 17 years of observations. This catalog is only available in the online version of the paper, at the CDS via anonymous ftp to cdsarc.u-strasbg.fr (130.79.128.5) or via \url{http://cdsarc.u-strasbg.fr/viz-bin/qcat?}, and at website \url{http://integral.cosmos.ru}.}
\end{minipage}
\begin{longtable}{lrrrcclll}
\hline 
\multicolumn{1}{|c|}{Id} &
\multicolumn{1}{|c|}{Name\footnotemark[1]} & 
\multicolumn{1}{c|}{RA\footnotemark[2]} & 
\multicolumn{1}{c|}{Dec\footnotemark[2]} &
\multicolumn{1}{c|}{Flux\footnotemark[3]} &
\multicolumn{1}{c|}{Type\footnotemark[4]} &
\multicolumn{1}{c|}{Notes\footnotemark[5]} \\

\multicolumn{1}{|c|}{} &
\multicolumn{1}{|c|}{} &
\multicolumn{1}{c|}{(deg)} &
\multicolumn{1}{c|}{(deg)} &
\multicolumn{1}{c|}{(17--60~keV)} &
\multicolumn{1}{c|}{} &
\multicolumn{1}{c|}{} \\ \hline 
\endfirsthead

\multicolumn{7}{c}{{\bfseries Table A1  -- continued from previous page}} \\
\hline 
\multicolumn{1}{|c|}{Id} &
\multicolumn{1}{c|}{Name\footnotemark[1]} &          
\multicolumn{1}{c|}{RA\footnotemark[2]} &                     
\multicolumn{1}{c|}{Dec\footnotemark[2]} &
\multicolumn{1}{c|}{Flux\footnotemark[3]} &
\multicolumn{1}{c|}{Type\footnotemark[4]} &
\multicolumn{1}{c|}{Notes}  \\

\multicolumn{1}{|c|}{} &
\multicolumn{1}{c|}{} &
\multicolumn{1}{c|}{(deg)} &
\multicolumn{1}{c|}{(deg)} &
\multicolumn{1}{c|}{(17--60~keV)} &
\multicolumn{1}{c|}{} &
\multicolumn{1}{c|}{} \\ \hline 
\endhead

\hline \multicolumn{7}{|r|}{{Continued on next page}} \\ \hline
\endfoot

\hline \hline
\endlastfoot

N001	&  SWIFT J0001.6-7701	& 0.370	& -77.020	&  0.58 $\pm$ 0.12	&  SEYFERT	&  z=0.058; !B16; \\
N002	&  IGR J00040+7020	& 1.029	& 70.315	&  0.86 $\pm$ 0.08	&  SEYFERT	&  R 1; z=0.096; \\
N003	&  {\bf IGR J00115+2645}	& 2.880	& 26.760	&  1.90 $\pm$ 0.40	&  UNIDENT	&  !B16; \\
N004	&  {\bf IGR J00197+6224}	& 4.930	& 62.410	&  0.29 $\pm$ 0.06	&  UNIDENT	&  !B16; \\
N005	&  IGR J00234+6141	& 5.742	& 61.686	&  0.78 $\pm$ 0.06	&  CV	&  R 2,3; \\
N006	&  TYCHO SNR	& 6.325	& 64.150	&  0.76 $\pm$ 0.06	&  SNR	&  \\
N007	&  IGR J00255+6821	& 6.407	& 68.362	&  0.53 $\pm$ 0.07	&  SEYFERT	&  R 4,1; z=0.012; \\
N008	&  V709 Cas	& 7.201	& 59.296	&  5.52 $\pm$ 0.06	&  CV	&  \\
\end{longtable}

\section{References in the catalog}
\label{sec:refs}
(1) \cite{2008A&A...482..113M}, (2) \cite{2006A&A...451..587D}, (3) \cite{2008ApJ...685.1143T}, (4) \cite{2006ATel..939....1K}, (5) \cite{2004ATel..352....1E}, (6) \cite{2004ATel..353....1M}, (7) \cite{2009A&A...495..121M}, (8) \cite{2004ATel..281....1D}, (9) \cite{2015MNRAS.447.2387C}, (10) \cite{2017A&A...602A.124R}, (11) \cite{2018AstL...44..522K}, (12) \cite{2011MNRAS.410.1813T}, (13) \cite{2006ATel..883....1B}, (14) \cite{2010ApJS..186....1B}, (15) \cite{2010A&A...519A..96M}, (16) \cite{2012ApJ...754..145T}, (17) \cite{2013A&A...557A.113S}, (18) \cite{2005ATel..681....1H}, (19) \cite{2010A&A...516A..15W}, (20) \cite{2010A&A...517A..14R}, (21) \cite{2005ATel..662....1K}, (22) \cite{2013ApJS..207...19B}, (23) \cite{2009A&A...493..339W}, (24) \cite{2008AstL...34..367B}, (25) \cite{2008A&A...482..731R}, (26) \cite{2017ATel10968....1B}, (27) \cite{2020MNRAS.500..565B}, (28) \cite{2020MNRAS.491.1857D}, (29) \cite{1978Natur.273..450A}, (30) \cite{2010HEAD...11.1305B}, (31) \cite{2012ATel.4151....1P}, (32) \cite{2008Atel.1363....1B}, (33) \cite{2014A&A...561A..67P}, (34) \cite{2010ATel.2759....1L}, (35) \cite{2012AstL...38....1L}, (36) \cite{2006ATel..880....1B}, (37) \cite{2016MNRAS.459..140M}, (38) \cite{2018ApJS..235....4O}, (39) \cite{2015MNRAS.449..597T}, (40) \cite{2013ATel.5537....1S}, (41) \cite{2011ATel.3382....1K}, (42) \cite{2004A&A...413..309M}, (43) \cite{2013MNRAS.428...50G}, (44) \cite{2015MNRAS.453.3100B}, (45) \cite{2005A&A...444L..37S}, (46) \cite{2006A&A...459...21M}, (47) \cite{2016MNRAS.460...19M}, (48) \cite{2014A&A...567A.129V}, (49) \cite{2013A&A...558A..74V}, (50) \cite{2007ATel.1253....1R}, (51) \cite{2008ApJ...674..686W}, (52) \cite{2016ApJS..223...15B}, (53) \cite{2017MNRAS.470..512K}, (54) \cite{2006ApJ...636..765B}, (55) \cite{2004ATel..261....1D}, (56) \cite{2017MNRAS.470.1107L}, (57) \cite{2012MNRAS.426.1750M}, (58) \cite{2008A&A...487..509S}, (59) \cite{2009AstL...35...33R}, (60) \cite{2012A&A...542A..22B}, (61) \cite{2009A&A...502..787Z}, (62) \cite{2017A&A...598A..34S}, (63) \cite{2010ATel.2853....1L}, (64) \cite{2008ATel.1540....1P}, (65) \cite{2012ApJS..199...26H}, (66) \cite{2008ApJ...681..113T}, (67) \cite{2013A&A...556A.120M}, (68) \cite{2006ATel..684....1K}, (69) \cite{2013A&A...560A.108C}, (70) \cite{2006AstL...32..145R}, (71) \cite{2006ATel..715....1M}, (72) \cite{2006ATel..785....1M}, (73) \cite{2006A&A...449.1139M}, (74) \cite{2008A&A...477L..29L}, (75) \cite{2010ApJ...720..987F}, (76) \cite{2018A&A...618A.150F}, (77) \cite{2019ApJ...887...32C}, (78) \cite{2020AstL...45..836K}, (79) \cite{2007ApJS..170..175B}, (80) \cite{2005ATel..469....1L}, (81) \cite{2005ATel..470....1N}, (82) \cite{2006A&A...450L...9S}, (83) \cite{2004ATel..278....1P}, (84) \cite{2004ATel..275....1G}, (85) \cite{2007ATel.1239....1N}, (86) \cite{2011A&A...527A.142G}, (87) \cite{2009ATel.2132....1M}, (88) \cite{2010A&A...511A..48M}, (89) \cite{2012AstL...38..629K}, (90) \cite{2018MNRAS.480.2357K}, (91) \cite{2015ApJS..219....1O}, (92) \cite{2017ATel10411....1M}, (93) \cite{2021ApJ...914...48T}, (94) \cite{2005MNRAS.364..455C}, (95) \cite{2005A&A...442....1A}, (96) \cite{2007A&A...462...57S}, (97) \cite{2010MNRAS.403..945L}, (98) \cite{2007AandA...475..775K}, (99) \cite{2010ATel.2457....1K}, (100) \cite{2009ApJ...701..811T}, (101) \cite{2011A&A...529A..30D}, (102) \cite{2011ApJ...739...57K}, (103) \cite{2002A&A...392..931C}, (104) \cite{2006ATel..810....1K}, (105) \cite{2010ApJ...716..663R}, (106) \cite{2011ATel.3184....1L}, (107) \cite{2016MNRAS.460..513T}, (108) \cite{2018MNRAS.478.1185B}, (109) \cite{2015MNRAS.451.2370M}, (110) \cite{2006ATel..783....1M}, (111) \cite{2020ApJ...889...53T}, (112) \cite{2012ATel.4183....1T}, (113) \cite{2010ApJS..186..378T}, (114) \cite{2016ApJ...816...38T}, (115) \cite{2010MNRAS.408..975M}, (116) \cite{2011ATel.3146....1M}, (117) \cite{2011ApJ...735..104K}, (118) \cite{2007ATel.1034....1M}, (119) \cite{2012ApJ...751...52E}, (120) \cite{2011ATel.3271....1L}, (121) \cite{2019ApJ...872...18M}, (122) \cite{2004ATel..229....1W}, (123) \cite{2006MNRAS.372..224B}, (124) \cite{2007A&A...470..331M}, (125) \cite{2006ApJ...647.1309T}, (126) \cite{2005ATel..456....1S}, (127) \cite{2005ApJ...631..506B}, (128) \cite{2008ATel.1774....1K}, (129) \cite{2010MNRAS.408.1866R}, (130) \cite{2018ATel12340....1M}, (131) \cite{2003AstL...29..587R}, (132) \cite{2003IAUC.8063....3C}, (133) \cite{2003IAUC.8076....1T}, (134) \cite{2005A&A...433L..41L}, (135) \cite{2006A&A...453..133W}, (136) \cite{2007A&A...461..631N}, (137) \cite{2015ApJS..218...23A}, (138) \cite{2003IAUC.8097....2R}, (139) \cite{2010A&A...516A..94N}, (140) \cite{2006A&A...447.1027B}, (141) \cite{2004ATel..224....1T}, (142) \cite{2004ATel..329....1L}, (143) \cite{2005A&A...444..821L}, (144) \cite{2003ATel..176....1M}, (145) \cite{2008A&A...484..783C}, (146) \cite{2005ATel..457....1G}, (147) \cite{2008ATel.1396....1N}, (148) \cite{2019ApJ...873...86P}, (149) \cite{2017ApJ...850...74K}, (150) \cite{2011MNRAS.411..137H}, (151) \cite{2010AstL...36..904L}, (152) \cite{2007ATel.1270....1B}, (153) \cite{2019JHEAp..24...30X}, (154) \cite{2018ATel11306....1G}, (155) \cite{2011ApJ...731L...2K}, (156) \cite{2009ATel.2170....1K}, (157) \cite{2015ATel.7946....1S}, (158) \cite{2007A&A...467..529C}, (159) \cite{2012ATel.4219....1D}, (160) \cite{2003ATel..149....1K}, (161) \cite{2003AstL...29..719L}, (162) \cite{2006ApJ...643..376C}, (163) \cite{2005ATel..444....1G}, (164) \cite{2007ESASP.622..373G}, (165) \cite{2019MNRAS.489.1828K}, (166) \cite{2012ATel.4233....1L}, (167) \cite{2017MNRAS.465.1563R}, (168) \cite{2012AandA...545A..27K}, (169) \cite{2012A&A...538A.123M}, (170) \cite{2005ApJ...634L..21B}, (171) \cite{2008AandA...477..735C}, (172) \cite{2006ATel..778....1B}, (173) \cite{2005A&A...437L..27B}, (174) \cite{2011MNRAS.415.2373S}, (175) \cite{2019ATel12436....1B}, (176) \cite{2005MNRAS.361..141G}, (177) \cite{2018A&A...613A..22B}, (178) \cite{2004ATel..328....1L}, (179) \cite{2015A&A...579A..56B}, (180) \cite{2010MNRAS.404.1591D}, (181) \cite{2006ATel..874....1K}, (182) \cite{2019MNRAS.485..286G}, (183) \cite{2011ATel.3565....1G}, (184) \cite{2012NewA...17..589N}, (185) \cite{2014MNRAS.444...93D}, (186) \cite{2003ATel..181....1S}, (187) \cite{2006ApJ...638.1045S}, (188) \cite{2020MNRAS.491.4543S}, (189) \cite{2013ApJ...769..120B}, (190) \cite{2004ATel..345....1K}, (191) \cite{2009ApJ...701.1627H}, (192) \cite{2004ATel..232....1B}, (193) \cite{2004ATel..264....1T}, (194) \cite{2011JAVSO..39..110F}, (195) \cite{2013ATel.4924....1K}, (196) \cite{2014MNRAS.441..640B}, (197) \cite{2019ATel12843....1H}, (198) \cite{1999A&A...346L..45C}, (199) \cite{2012ATel.4381....1V}, (200) \cite{2006ApJ...636..275B}, (201) \cite{2007A&A...463..957K}, (202) \cite{2015ApJ...814...94M}, (203) \cite{2006ATel..948....1M}, (204) \cite{1994ApJ...425..110P}, (205) \cite{2003ATel..132....1R}, (206) \cite{2006ApJ...639..340K}, (207) \cite{1998IAUC.6846....2I}, (208) \cite{2015MNRAS.446.2418Z}, (209) \cite{2006ATel..970....1B}, (210) \cite{2006ATel..972....1W}, (211) \cite{2007ATel.1136....1D}, (212) \cite{2017MNRAS.464..170K}, (213) \cite{2005ATel..467....1G}, (214) \cite{2008ATel.1445....1D}, (215) \cite{2008ATel.1651....1A}, (216) \cite{2008int..workE.122D}, (217) \cite{2009A&A...502..231Z}, (218) \cite{2004A&A...425L..49R}, (219) \cite{2010ApJ...719..143T}, (220) \cite{2015ApJ...815..132Z}, (221) \cite{2009A&A...494..417R}, (222) \cite{2007ApJ...668...81M}, (223) \cite{2007AstL...33..149G}, (224) \cite{2008A&A...489..657Z}, (225) \cite{2010MNRAS.409L..69K}, (226) \cite{2010MNRAS.402.2388K}, (227) \cite{2015MNRAS.452..884N}, (228) \cite{1997ApJ...488..831S}, (229) \cite{2008ATel.1400....1B}, (230) \cite{2006ATel..885....1S}, (231) \cite{2007ApJ...657L.109P}, (232) \cite{2007ApJ...655L..97R}, (233) \cite{2011ATel.3556....1P}, (234) \cite{2011ATel.3560....1F}, (235) \cite{2011ATel.3558....1B}, (236) \cite{2011ATel.3562....1G}, (237) \cite{2011ATel.3622....1R}, (238) \cite{2011ATel.3606....1C}, (239) \cite{2011MNRAS.411..620Z}, (240) \cite{2008A&A...491..209R}, (241) \cite{2004ATel..342....1G}, (242) \cite{2011ATel.3181....1N}, (243) \cite{2019ATel13155....1G}, (244) \cite{2021ApJ...914...85H}, (245) \cite{2005ATel..550....1M}, (246) \cite{2014MNRAS.445.2424N}, (247) \cite{2017MNRAS.471.2508S}, (248) \cite{2013ATel.4804....1K}, (249) \cite{2013ATel.4769....1K}, (250) \cite{2013ApJS..209...14K}, (251) \cite{2007AstL...33..807C}, (252) \cite{2003ATel..190....1S}, (253) \cite{2005A&A...441L...1I}, (254) \cite{2011A&A...533A...3C}, (255) \cite{2018ATel11941....1D}, (256) \cite{2018A&A...617L...8S}, (257) \cite{2003ATel..155....1L}, (258) \cite{2007ATel.1054....1B}, (259) Coughenour et al., (in prep.), (260) \cite{2004AstL...30..382R}, (261) \cite{2010A&A...510A..61T}, (262) \cite{2004A&A...426L..41M}, (263) \cite{2012ATel.4050....1C}, (264) \cite{2007A&A...474L...1N}, (265) \cite{2012A&A...544A.114M}, (266) \cite{2011ATel.3210....1D}, (267) \cite{2011ATel.3218....1C}, (268) \cite{2011ATel.3268....1K}, (269) \cite{2018ATel12254....1N}, (270) \cite{2018ATel11357....1F}, (271) \cite{2018ATel11332....1R}, (272) \cite{2019MNRAS.485.5235A}, (273) \cite{2005ApJ...629L.109U}, (274) \cite{2007A&A...470..249F}, (275) \cite{2009ATel.2193....1R}, (276) \cite{2020A&A...639A..17W}, (277) \cite{2019A&A...622A.198N}, (278) \cite{2008ATel.1457....1H}, (279) \cite{2018ATel11478....1B}, (280) \cite{2009ApJ...698..502B}, (281) \cite{2017ApJ...841...35F}, (282) \cite{2020MNRAS.498.2750C}, (283) \cite{2013ApJ...775L..24L}, (284) \cite{2008ATel.1686....1M}, (285) \cite{2013ATel.4925....1E}, (286) \cite{2006ApJ...636L..65B}, (287) \cite{2009A&A...493..893L}, (288) \cite{2013ATel.5474....1N}, (289) \cite{2019PASJ...71..108O}, (290) \cite{2016MNRAS.461..304C}, (291) \cite{2009MNRAS.395L...1B}, (292) \cite{2003ATel..154....1L}, (293) \cite{2012A&A...537L...7F}, (294) \cite{2005ApJ...630L.157M}, (295) \cite{2021arXiv210705879P}, (296) \cite{2004ATel..340....1R}, (297) \cite{2008A&A...486..911N}, (298) \cite{2020A&A...634A..89D}, (299) \cite{2012ATel.4130....1K}, (300) \cite{2019ATel13211....1M}, (301) \cite{2019ATel13195....1K}, (302) \cite{2007ATel.1319....1G}, (303) \cite{2010AstL...36..533G}, (304) \cite{2012ApJ...753....3B}, (305) \cite{2013A&A...556A..27S}, (306) \cite{2009A&A...501.1031K}, (307) \cite{2003ATel..157....1C}, (308) \cite{2004AstL...30..534M}, (309) \cite{2008AIPC.1085..312T}, (310) \cite{2012ATel.3951....1R}, (311) \cite{2006AstL...32..221B}, (312) \cite{2006ApJ...638..963K}, (313) \cite{2007AstL...33..159K}, (314) \cite{2008AstL...34..753K}, (315) \cite{2019MNRAS.489.3031X}, (316) \cite{2018ATel11341....1S}, (317) \cite{2018ATel11787....1H}, (318) \cite{2018AJ....155..247H}, (319) \cite{2003IAUC.8088....4H}, (320) \cite{2011A&A...526A.122P}, (321) \cite{2010AandA...523A..61K}, (322) \cite{2005ATel..669....1T}, (323) \cite{2011ATel.3741....1D}, (324) \cite{2008ATel.1653....1S}, (325) \cite{2018MNRAS.476.2110R}, (326) \cite{2006ATel..862....1T}, (327) \cite{2019ApJ...878...15H}, (328) \cite{2017A&A...608A..36T}, (329) \cite{2007A&A...464..277M}, (330) \cite{2006ATel..847....1H}, (331) \cite{2014A&A...561A.108B}, (332) \cite{2008ATel.1623....1G}, (333) \cite{2008AstL...34..653B}, (334) \cite{2006A&A...455...11M}, (335) \cite{2009ApJ...701.1644W}, (336) \cite{2006ATel..967....1W}, (337) \cite{2011ATel.3272....1L}, (338) \cite{2013ApJS..206...17M}, (339) \cite{2011MNRAS.411.2137G}, (340) \cite{2017MNRAS.467..540L}, (341) \cite{2007ApJ...669L...1B}, (342) \cite{2012ATel.4248....1M}, (343) \cite{2013MNRAS.433.2028E}, (344) \cite{2013AstL...39..513L}, (345) \cite{2007ATel.1288....1L}, (346) \cite{2012MNRAS.420.2087D}.


\bsp	
\label{lastpage}
\end{document}